\documentclass[preprint]{elsarticle}

\usepackage{amsfonts}
\usepackage{amsmath}
\usepackage{amssymb}
\usepackage{graphicx} 
\usepackage{graphics} 
\usepackage{epsfig,epstopdf}
\usepackage{color} 
\usepackage[dvips]{rotating}
\usepackage{subcaption}
\usepackage{epsfig}
\usepackage{stmaryrd}
\usepackage[percent]{overpic}
\usepackage[]{algorithm2e}

\newcommand\solidrule[1][2mm]{\rule[0.5ex]{#1}{.9pt}}
\newcommand\dashedrule{\mbox{\solidrule[1mm]\hspace{1mm}\solidrule[1mm]}}
\newcommand\dashdotrule{\mbox{\solidrule[1mm]\hspace{1mm}\solidrule[0.9pt]\hspace{1mm}\solidrule[1mm]}}
\newcommand\dotsrule{\mbox{\solidrule[0.9pt]\hspace{1mm}\solidrule[0.9pt]\hspace{1mm}\solidrule[0.9pt]}}
\newcommand\Smblacksquare{\protect\scalebox{0.6}{$\blacksquare$}}

\definecolor{red}{RGB}{248, 118, 109}
\definecolor{green}{RGB}{125, 174, 0}
\definecolor{purple}{RGB}{199, 123, 255}
\definecolor{cyan}{RGB}{0, 191, 196}
\definecolor{blue}{RGB}{95, 155, 255}
\definecolor{green_darker}{RGB}{6, 186, 51}
\definecolor{yellow}{RGB}{163, 165, 0}

\journal{Journal of Computational Physics}

\begin{document}
\begin{frontmatter}

\title{Direct numerical simulation of variable surface tension flows using a Volume-of-Fluid method}
\author[]{Ivana Seric}
\author[]{Shahriar Afkhami\corref{cor}}
\ead{shahriar.afkhami@njit.edu}
\author[]{Lou Kondic}

\cortext[cor]{Corresponding author}
\address{Department of Mathematical Sciences, New Jersey Institute of Technology, Newark, NJ, USA}



\begin{abstract}
We develop a general methodology for the inclusion of variable surface tension into 
a Volume-of-Fluid based Navier-Stokes solver.
This new numerical model provides a robust and accurate method for computing the
surface gradients directly by finding the tangent directions on the interface 
using height functions. 
The implementation is applicable to both temperature and concentration dependent
surface tension, along with the setups involving a large jump in the temperature 
between the fluid and its surrounding, as well as the situations where the
concentration should be strictly confined to the fluid domain, such as the mixing 
of fluids with different surface tension coefficients. 
We demonstrate the applicability of our method to thermocapillary migration of 
bubbles and coalescence of drops characterized by different surface tension.
\end{abstract}

\end{frontmatter}

\section{Introduction}

Flows induced by the spatial variations in the surface tension, also known as
Marangoni effect \citep{scriven1960}, can be caused by surfactants, temperature 
or concentration gradients, or a combination of these effects. Understanding 
these flows is important since they are relevant in microfluidics 
\citep{Farahi2004}, heat pipe flows \citep{Kundan2015}, motion of drops or 
bubbles in materials processing applications that include heating or cooling 
\citep{subramanian2002}, evolution of metal films of nanoscale thickness 
melted by laser pulses \citep{trice_prl08,dong2016}, and in a variety of other 
thin film flows, see~\citep{Davis1987,cm_rmp09} for reviews. 

Numerical methods for studying variable surface tension flows include front 
tracking \citep{Muradoglu2014}, level set \citep{Xu2006}, diffuse interface 
\citep{Teigen2011}, marker particle \citep{Blanchette2009, Blanchette2012}, 
immersed boundary \citep{Lai2008}, boundary integral \citep{Booty2010}, 
interface-interaction \citep{Schranner2016}, and Volume-of-Fluid (VOF) 
\citep{Renardy2004,James2004,Ma2011} methods. 
The VOF method is efficient and robust for tracking topologically complex evolving 
interfaces. The improvements in recent years in the computation of the surface
tension have empowered the VOF method to become a widespread 
method for modeling interfacial flows \citep{Francois2010, Lopez2010}.
However, an accurate implementation of the variable surface tension 
in the VOF formulation is still lacking a general treatment.

A challenge of including variable surface tension effects into the VOF method is 
that the surface tension is not known exactly at the interface - only the value 
averaged over a computational cell containing the interface is known.   To obtain 
the surface tension at the interface, an approximation from the values near the 
interface, usually calculated at the center of each adjacent computational cell, 
is necessary. As we outline below, the approximation of the interface values has 
been carried out in the literature differently, depending on the physics of the 
problem studied.  Additional major issue concerns computing the surface gradients 
of the surface tension. 
  
In \citet{Alexeev2005} and \citet{Ma2011}, the VOF method is used to study flows 
involving temperature dependent surface tension. The implementation in 
\citet{Alexeev2005} solves the heat equation in fluids on the both sides of the 
interface, and then imposes the continuity of the temperature and flux at the
interface, 
and conservation of energy in the cell containing the interface to approximate 
the temperature in the fluid and air in the cell. These temperature values are 
then used to calculate surface gradients of the temperature from nearby cells  
that are not cut by the interface; these gradients are then exponentially 
extrapolated to the interface. 
In the work by \citet{Ma2011}, the temperature at the interface is approximated 
from the temperatures in the liquid and the gas by imposing the continuity of 
heat flux at the interface. The surface gradients of the temperature are 
approximated by computing the derivatives in each coordinate direction using 
finite differences, and then projecting them onto the tangential direction.
If the interface is not contained in all cells of the finite difference stencil,
then one sided differences are used. Hence, this method requires temperature 
solution on both sides of the interface and therefore cannot be used for the setups
involving a large difference in thermal conductivity of the two fluids, 
since the fluids may have a large difference in the temperature. 
Furthermore, both of these methods are not applicable to setups where the surface
tension only depends on the concentration, such as mixing of miscible liquids with 
different surface tension. 
In the work by \citet{James2004}, the VOF method is used to study the flows 
induced by the surfactant concentration gradient. In their method, the concentration
values at the interface are obtained by imposing the condition that the average
concentration at the interface is equal to the average concentration in the cell
containing the interface. Then, the surface gradients are computed using the 
cell-center interfacial concentration in the two adjacent cells.

Here, we develop a method that can be applied to both temperature and concentration 
dependent surface tension, with the surface gradients computed using the cell-center 
values in the interfacial cells only. We find the tangential gradients directly 
by computing the tangent directions on the interface using height functions
\citep{Popinet2009a}. 
This method can be applied to the setups such that the concentration is confined 
to the fluid domain, e.g.\ mixing of liquids with different surface tension 
coefficients, as well as the configurations involving large jump of the
temperature between the liquid and the surrounding. 
Since our method does not depend on whether we consider temperature or concentration 
gradients, we will use them interchangeably in the remaining part of the paper.

Our numerical method is implemented using {\sc Gerris}: an open source adaptive 
Navier-Stokes solver \citep{Popinet2003,Popinet2009a}. The current version includes 
Continuum Surface Force (CSF) \citep{Brackbill1992} implementation of the surface 
tension force with height function algorithm for computing interfacial normal and 
curvature \citep{Popinet2009a}.
Here, we present the method for extending this formulation to include variable 
surface tension, allowing to consider the surface force in the direction tangential 
to the interface.
As far as we are aware, this is the first implementation of the variable surface
tension combined with the accurate implementation of the CSF method, such that 
the curvature and interface normals are computed using generalized height functions
\citep{Popinet2009a}. Our extension is a step closer to cover all aspects of the 
variable surface tension flows; the remaining one is the implementation of the 
surfactant transport and surface tension gradients due to the presence of the surfactants.
This will be the topic of our future work.

The rest of this paper is organized as follows: Section \ref{sec:VOF} gives an 
overview of the VOF method, including the CSF method for the computation of the 
surface tension; Section \ref{sec:method} describes in detail the implementation 
of the variable surface tension in two and three dimensions; and Section 
\ref{sec:results} illustrates the performance of our method for various test cases, 
including temperature and concentration dependent surface tension.

\section{Governing equations }
\label{sec:VOF}

\begin{figure}
\includegraphics[width=\textwidth]{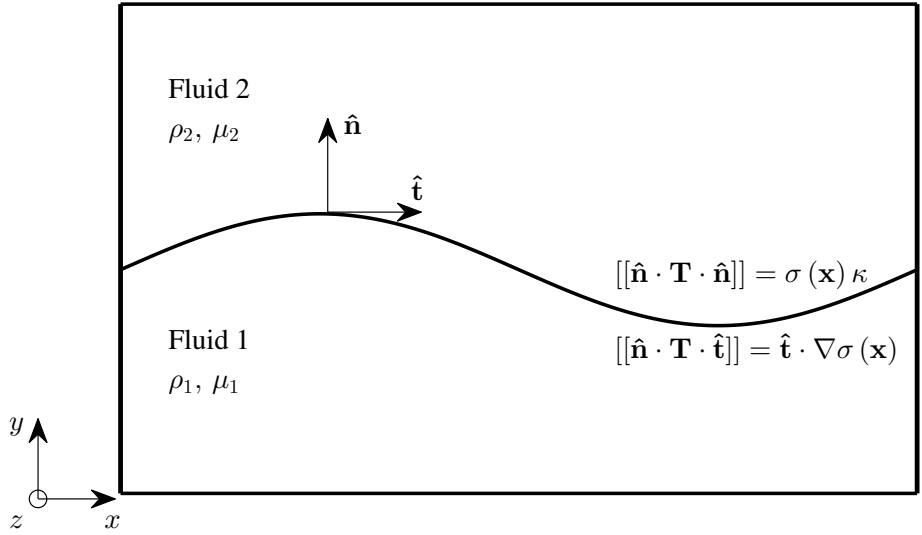}
\caption{Schematic of a system with two immiscible fluids and the corresponding boundary conditions.}
\label{fig:1}
\end{figure}

We consider  incompressible two-phase flow described by Navier-Stokes equations
\begin{gather}
\label{eq:Fma}
  \rho(\partial_t \textbf{u} + \textbf{u}\cdot \nabla \textbf{u}) = 
     - \nabla p + \nabla \cdot \left( 2 \mu D\right)
     + \textbf{F}
     , \\
  \nabla\cdot\textbf{u} = 0,
\end{gather}
and the advection of the phase-dependent density $\rho\left( \chi \right)$
\begin{equation}
  \label{eq:adv_rho}
  \partial_t \rho + (\textbf{u} \cdot \nabla)\rho = 0, \,
\end{equation}
where $\textbf{u} = (u, v, w)$ is the fluid velocity, $p$ is the pressure, $ \rho
(\chi) = \chi \rho_1 + (1-\chi)\rho_2$ and  $\mu (\chi) = \chi \mu_1 + 
(1-\chi)\mu_2$  are the phase dependent density and viscosity respectively, and 
$D$ is the rate of deformation tensor $D = \left(\nabla \textbf{u} + 
\nabla\textbf{u}^T\right)/2$.
Subscripts $1$ and $2$ correspond to the fluids $1$ and $2$, respectively (see Figure \ref{fig:1}).
Here, $\chi$ is the characteristic function, such that $\chi = 1$ in 
the fluid $1$, and $\chi = 0$ in the fluid $2$. Note that any body force can be 
included in $\textbf{F}$. The characteristic 
function is advected with the flow, thus
\begin{equation}
 \partial_t \chi + (\textbf{u} \cdot \nabla)\chi = 0.
 \label{chi}
\end{equation}
Note that solving equation \eqref{chi} is equivalent to solving equation \eqref{eq:adv_rho}.

The presence of an interface gives rise to the stress boundary conditions, see 
Figure \ref{fig:1}. The normal stress boundary condition at the interface defines the stress jump \citep{landau1987, levich1969}
\begin{gather}
 \llbracket \hat{\mathbf{n}}\cdot \mathbf{T}\cdot \hat{\mathbf{n}}\rrbracket = \sigma \left( \mathbf{x} \right) \kappa,
\end{gather} 
where $ \mathbf{T} = -p \mathbf{I} + \mu \left(\nabla \textbf{u} + \nabla\textbf{u}^T\right)$
is the total stress tensor, $\sigma \left( \mathbf{x} \right)$ is the surface tension
coefficient, $\kappa$ is the curvature of the interface, and $\hat{\mathbf{n}}$ 
is the unit normal at the interface pointing out of the fluid 1.
The variation of surface tension coefficient results in the tangential 
stress jump at the interface
\begin{equation}
\llbracket \hat{\mathbf{n}}\cdot \mathbf{T}\cdot \hat{\mathbf{t}}\rrbracket = 
		 \hat{\mathbf{t}}\cdot\mathbf{\nabla} \sigma \left( \mathbf{x} \right),
\end{equation}
which drives the flow from the regions of low surface tension to the ones with high surface tension.
Here, $\hat{\mathbf{t}}$ is the unit tangent vector in two dimensions (2D); 
in three dimensions (3D) there are two linearly independent unit tangent vectors.
Using the Continuum Surface Force (CSF) method \citep{Brackbill1992}, the forces
resulting from the normal and tangential stress jump at the interface can be 
included in the body force $\textbf{F} = \textbf{F}_{sn} + \textbf{F}_{st}$, 
defined as  
\begin{equation}
\label{eq:surf_fn}
 \textbf{F}_{sn} = \sigma \left( \mathbf{x} \right) \kappa \delta_s \hat{\mathbf{n}},
\end{equation}
and 
\begin{equation}
\label{eq:surf_ft_nabla}
 \textbf{F}_{st} =\mathbf{\nabla}_s \sigma \left( \mathbf{x} \right) \delta_s,
\end{equation}
where $\delta_s$ is the Dirac delta function centered at the interface, 
$ \delta_s \hat{\mathbf{n}} = \nabla \chi $, and $\mathbf{\nabla}_s$ is the surface gradient.
Substituting equations \eqref{eq:surf_fn} and \eqref{eq:surf_ft_nabla} in 
the momentum equation \eqref{eq:Fma}
gives 
\begin{equation}
\label{eq:full_NS}
 \rho(\partial_t \textbf{u} + \textbf{u}\cdot \nabla \textbf{u}) = 
     - \nabla p + \nabla \cdot \left( 2 \mu D\right)
     + \sigma \left( \mathbf{x} \right) \kappa \delta_s \hat{\mathbf{n}} 
     + \mathbf{\nabla}_s \sigma \left( \mathbf{x} \right) \delta_s.
\end{equation}
We define the nondimensional variables, denoted with a superscript ``*'', as
\begin{gather*}
\begin{aligned}
  x^* = \frac{x}{a}, \,\,\,\,
  t^* = \frac{t}{t_r}, \,\,\,\, 
  \mathbf{u}^* = \frac{\mathbf{u}}{U_0}, \,\,\,\,
  p^* = \frac{p}{p_0},   \\
  \rho^*\left( \chi \right)  = \frac{\rho\left( \chi \right) }{\rho_1}, \,\,\,\, 
  \mu^*\left( \chi \right)  = \frac{\mu\left( \chi \right) }{\mu_1}, \,\,\,\,
  \sigma^* = \frac{\sigma}{\sigma_0},
\end{aligned}
\end{gather*}
where the scales $a$, $p_0$, $t_c$, $U_0$ and $\sigma_0$ are chosen based on the problem
studied. Hence the dimensionless equation \eqref{eq:full_NS} is
\begin{multline}
\label{eq:nondim_NS}
  \text{Re} \rho^*
  (\partial_t^* \mathbf{u}^* + \mathbf{u}^*\cdot \nabla^* \mathbf{u}^*) 
  = 
  - \nabla^* p^* + \nabla^* \cdot \left( 2 \mu^* D^* \right) + \\
  + \text{Ca}^{-1} \sigma^* \kappa^* \delta_s^*\hat{\mathbf{n}} 
  +  \frac{\sigma_0}{U_0 \mu_1} \nabla_s^* \sigma^* \delta_s^*\hat{\mathbf{t}},  
  \end{multline}
where $\text{Re}$ and $ \text{Ca}$ are the Reynolds and Capillary numbers respectively,
defined as
\begin{gather}
 \text{Re} = \frac{\rho_1 U_0 a}{\mu_1} , \,\,\,\,\,
 \text{Ca} = \frac{U_0 \mu_1}{\sigma_0}. 
\end{gather}
The surface tension is a function of temperature, $T$, or concentration, $C$, which
satisfy advection diffusion equation
 \begin{align}
 \label{eq:advec_diff1}
 \rho \left( \chi \right) C_p \left( \chi \right) 
 \left( \partial_t T + (\textbf{u} \cdot \nabla) T \right)
	&= \nabla \cdot \left( k \left( \chi \right) \nabla T \right), \\
\label{eq:advec_diff2}
\partial_t C + (\textbf{u} \cdot \nabla) C 
	&= \nabla \cdot \left( \alpha \left( \chi \right) \nabla C \right),
 \end{align}
where $C_{p} \left( \chi \right)$, $k \left( \chi \right)$ and 
$\alpha \left( \chi \right)$ are the phase dependent heat capacity, conductivity
and diffusivity, respectively.
Along with the scales given above, equations
\eqref{eq:advec_diff1} and \eqref{eq:advec_diff2} are nondimensionalized using the following scales
\begin{gather}
  k^*\left( \chi \right) = \frac{k\left( \chi \right)}{k_1}, \,\,\,\, 
  C_p^*\left( \chi \right) = \frac{C_{p}\left( \chi \right)}{C_{p_1}}, \,\,\,\, 
  T^* = \frac{T}{T_0},  \,\,\,\,
  \alpha^*\left( \chi \right) = \frac{\alpha\left( \chi \right)}{\alpha_1},
\end{gather}
where $T_0$ is chosen based on the physics of the system. Hence the dimensionless equations
\eqref{eq:advec_diff1} and \eqref{eq:advec_diff2} are
\begin{gather}
\label{eq:advec_diff_nd1}
  \text{Ma} \rho^* C_p^* 
   \left( \partial_t^* T^* + (\mathbf{u}^*\cdot \nabla^*) T^* \right) 
	  = \nabla^* \cdot \left( k^* \nabla^* T^* \right), \\
	  \label{eq:advec_diff_nd2}
	  \text{Ma}
   \left( \partial_t^* C^* + (\mathbf{u}^*\cdot \nabla^*) C^* \right) 
	  = \nabla^* \cdot \left( \alpha^* \nabla^* C^* \right), 
\end{gather}
where $\text{Ma}$ is the Marangoni number defined as
\begin{gather}
 \text{Ma} = \frac{U_0 a}{\alpha_1}.
\end{gather} 
The diffusivity, $\alpha_1$, in the heat equation is 
$\alpha_1 = {k_1}/({\rho_1 C_{p_1}})$.
Surface tension can have linear or nonlinear dependence on temperature or 
concentration.  
In many applications the surface tension depends on the temperature linearly, i.e.
\begin{equation}
\label{eq:linSigma}
 \sigma = \sigma_0 + \sigma_T \left( T - T_R\right),
\end{equation}
where $\sigma_0$ is the surface tension at a reference temperature $T_R$, and 
$\sigma_T$ is a constant.
Then, we can write
\begin{equation}
\mathbf{\nabla}_s \sigma = \sigma_T \mathbf{\nabla}_s T,
\end{equation}
and compute $\mathbf{\nabla}_s T$ in the same manner as $\mathbf{\nabla}_s \sigma$. 
Using the scales given above, the dimensionless equation \eqref{eq:linSigma} is 
\begin{equation}
\label{eq:sigma_nd}
\sigma^* =  1 + \frac{\sigma_T T_0}{\sigma_0} \left( T^* - T_R^* \right).
\end{equation}
In the following section, we describe a method for computing 
$\nabla_s \sigma$ in general, regardless of the dependence
on the temperature or concentration.

\section{Numerical method}
\label{sec:method}

The proposed numerical method is implemented into {\sc Gerris}, 
which numerically solves equations
\eqref{eq:Fma} to \eqref{eq:adv_rho} using the VOF interface tracking method
with implicit treatment of the viscous forces \citep{Popinet2009a,Popinet2003,Gerris}. 
The CSF method is used for the 
implementation of the surface tension force with curvatures computed using the 
height function method \citep{Popinet2009a,Afkhami2009a}. 
The {\sc Gerris} code uses octree (3D) and quadtree (2D) grids, allowing to 
adaptively refine the grid in the immediate neighborhood of the interface. 
While we describe our implementation of the variable surface tension for uniform
meshes, the extension to adaptively refined meshes is straightforward, following 
the implementation details described by \citet{Popinet2009a,Popinet2003}.

The surface gradient of any scalar field $Q$ is defined as the projection of the
gradient onto the surface, i.e.\ 
\begin{equation}
\label{eq:nabla_s_def}
    \nabla_s Q = \nabla Q - \hat{\mathbf{n}}\left(\hat{\mathbf{n}}\cdot  \nabla Q  \right) 
\end{equation}
where $\hat{\mathbf{n}}$ is the unit normal vector at the surface.
However, this definition of the surface gradient can result in inaccuracies when
implemented in the VOF method for general variable surface tension for two reasons. 
First,  the discontinuities of the material properties across the
interface can result in $Q$ having a large jump across the interface: for example,
in the case of surface tension dependence on the temperature where the fluids on 
each side of the interface have large difference in the conductivity. 
The second reason is that, in general, surface tension can depend on the 
concentration: for example, in the case of the mixing of two liquids with different 
surface tension, or in the case of surface tension dependent on the surfactant
concentration. 

Here, we propose a numerical method for implementing the general variable
surface tension. We compute the surface gradient as
\begin{equation}
\label{eq:surf_ft}
 \textbf{F}_{st} = \frac{\partial \sigma}{\partial s_1} \delta_s \hat{\mathbf{t}}_1 
 +\frac{\partial \sigma}{\partial s_2} \delta_s \hat{\mathbf{t}}_2,
\end{equation}
where $\hat{\mathbf{t}}_1$ and $\hat{\mathbf{t}}_2$ are the unit tangent 
vectors at the interface, pointing in the $s_1$ and $s_2$ directions, respectively.
In our approach, we first define surface tension values at the interface, then
compute the derivatives of $\sigma$ along the interface, and finally project the
derivatives onto the tangent space defined by $\hat{\mathbf{t}}_1$ and $\hat{\mathbf{t}}_2$.  
In the following Sections we present the details of the implementation. 
In Section \ref{sec:sigma_tilde}, we show how to approximate the surface 
tension value on the interface using the cell-center values.  Then in Section
\ref{sec:dsigma_ds}, we show how ${\partial \sigma}/{\partial s_d}$, for $d = 1, 2$,
are evaluated, along with the choice of the tangent vectors and addition of the 
tangential surface force using CSF method.  This is done first for 2D is Section 
\ref{sec:2D_derivative}, and then for 3D in Section \ref{sec:3D_derivative}.

\subsection{Approximation of interfacial values of surface tension}
\label{sec:sigma_tilde}

The algorithm for implementing $\nabla_s\sigma\left( \mathbf{x} \right)$ in the 
VOF method starts with the approximation of the interfacial values of the surface 
tension in each cell containing an interface segment. More precisely, we use the idea of
constructing the columns of cells inspired by the computation of interfacial 
curvature and normals using height functions \citep{Popinet2009a} (see \ref{sec:HF}). 

\begin{figure}[t]
\begin{subfigure}{0.5\textwidth}
\includegraphics[width=1\textwidth]{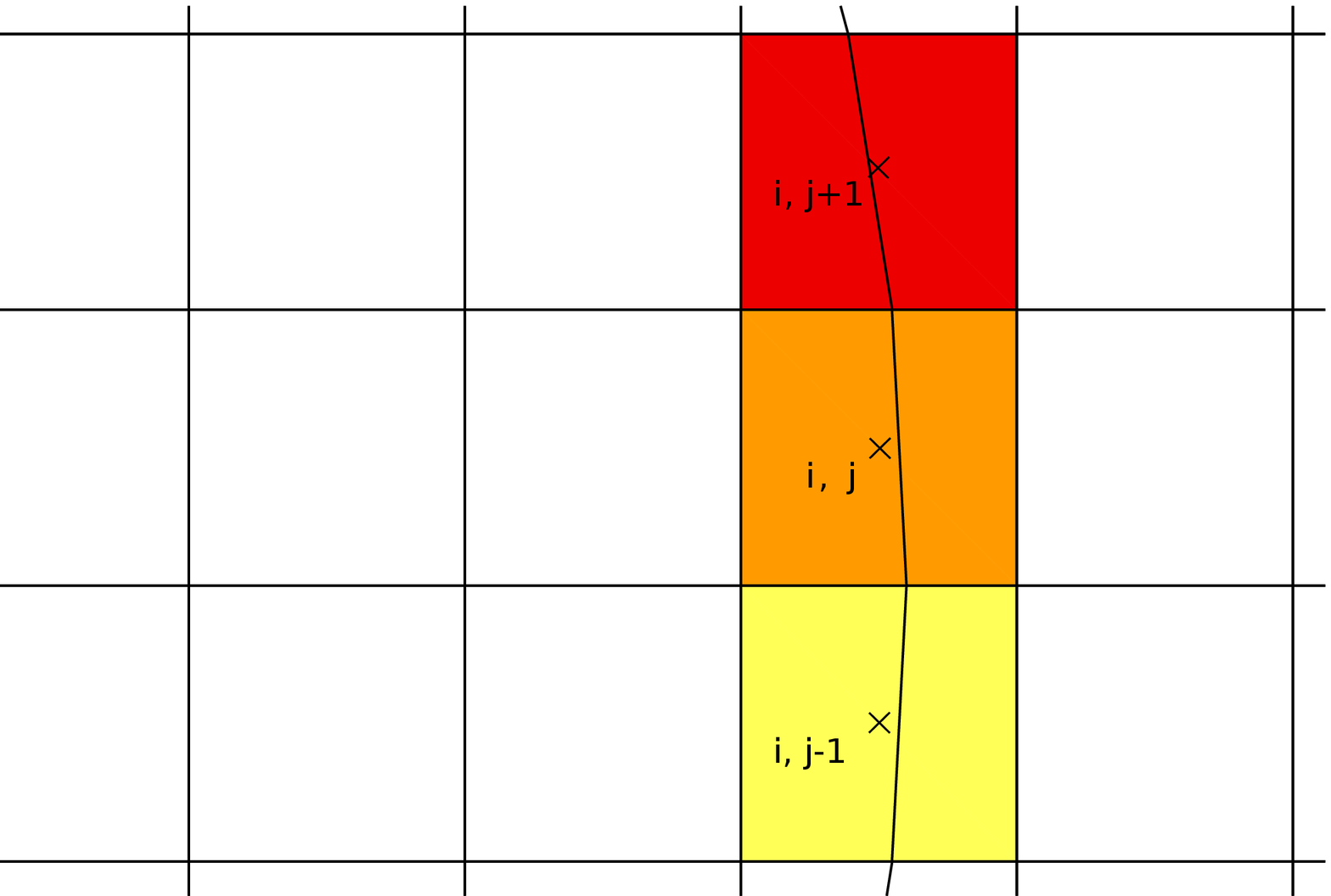}
 \caption{}
\end{subfigure}
\begin{subfigure}{0.5\textwidth}
 \includegraphics[ width=1\textwidth]{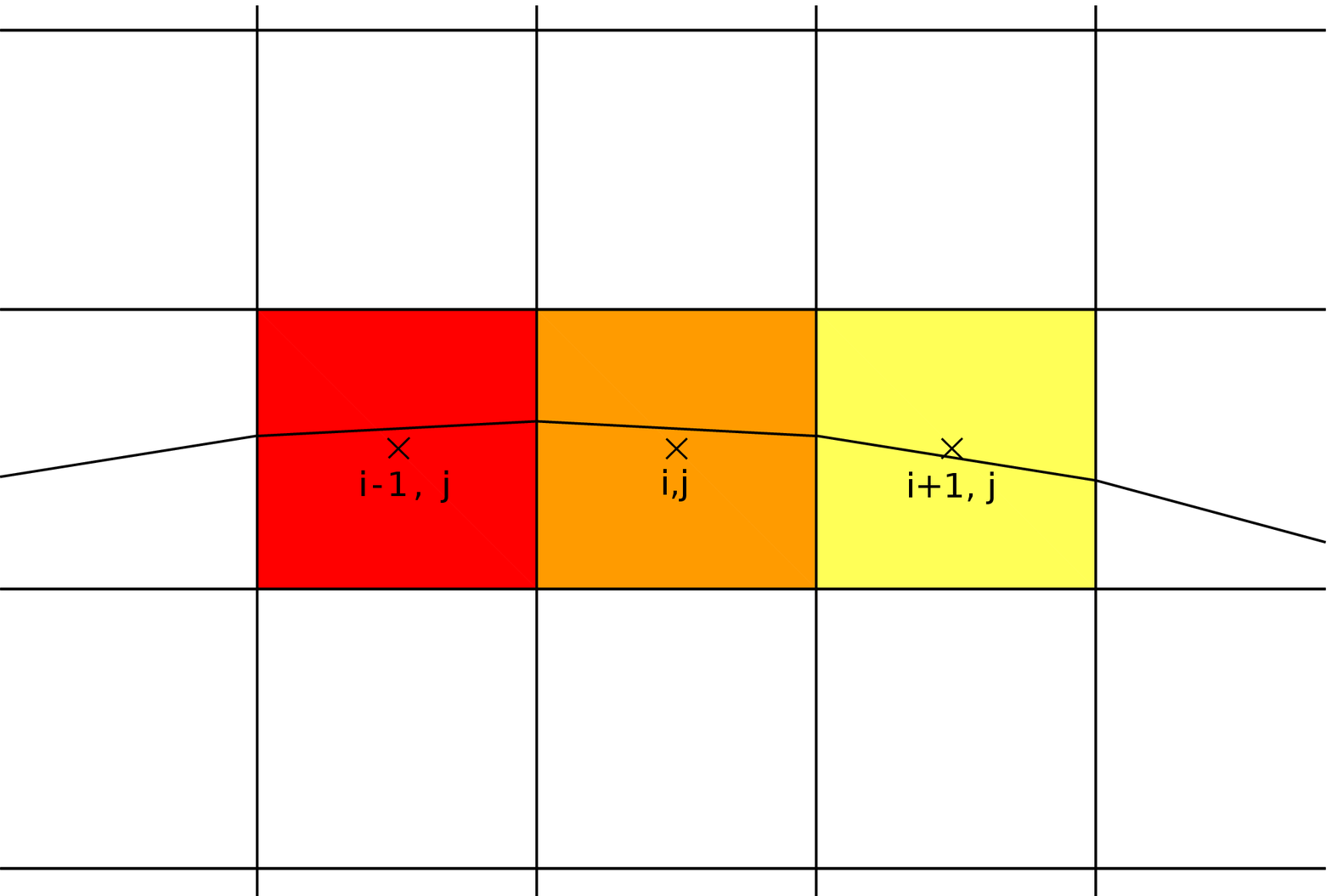}
 \caption{}
\end{subfigure}
  \caption{ 
  An example of the interface orientation, where columns in $x$ (a) 
  and $y$ (b) direction for computing $\tilde{\sigma}^c\left( \mathbf{x} \right)$
  contain one interfacial cell. Each color shows a different column in 
  interfacial cells $\mathcal{C}$. }
\label{fig:genConc_flat}
\end{figure}

\begin{figure}[t]
\begin{subfigure}{0.5\textwidth}
  \includegraphics[width=1\textwidth]{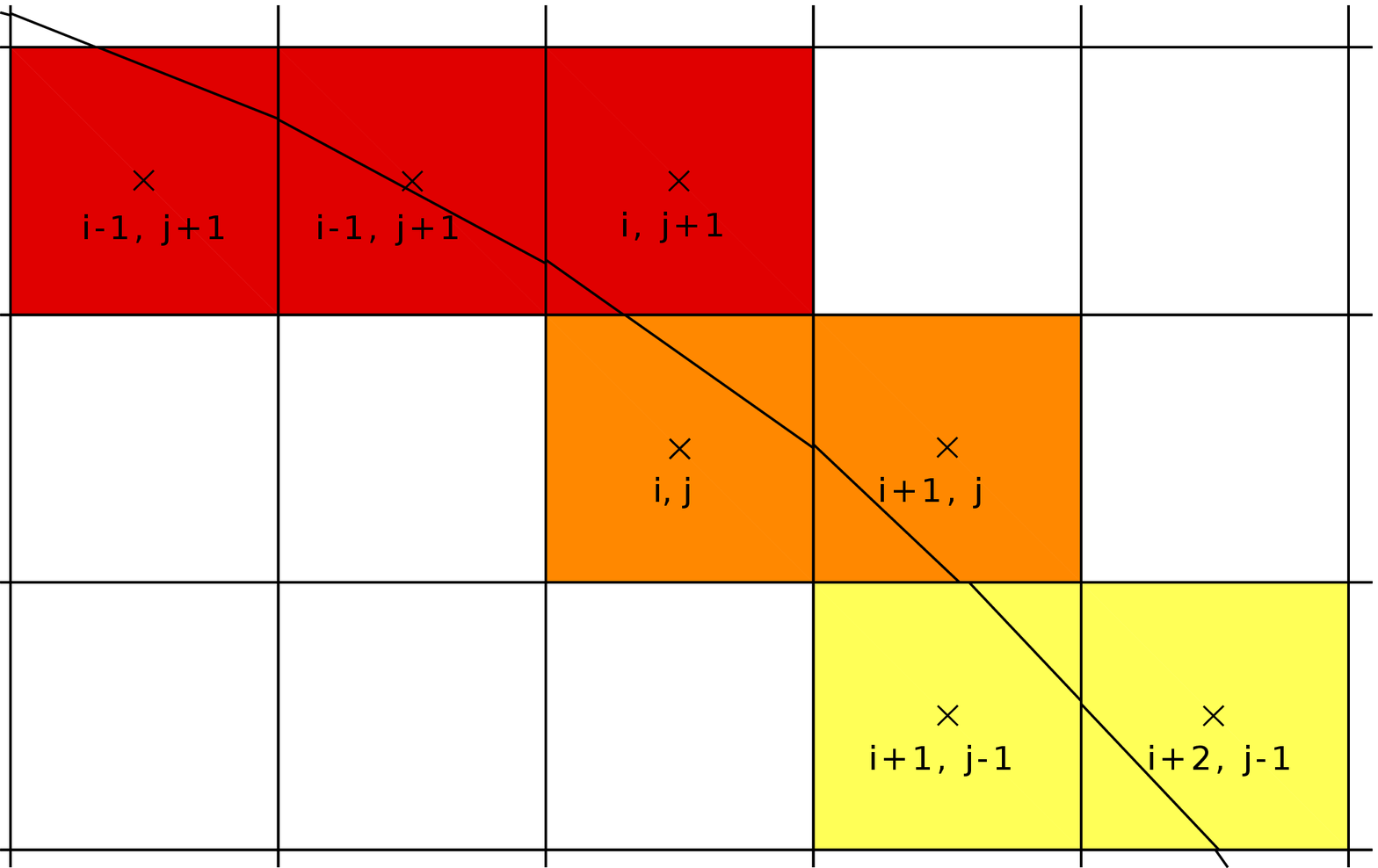}
 \caption{}
\end{subfigure}
\begin{subfigure}{0.5\textwidth}
  \includegraphics[ width=1\textwidth]{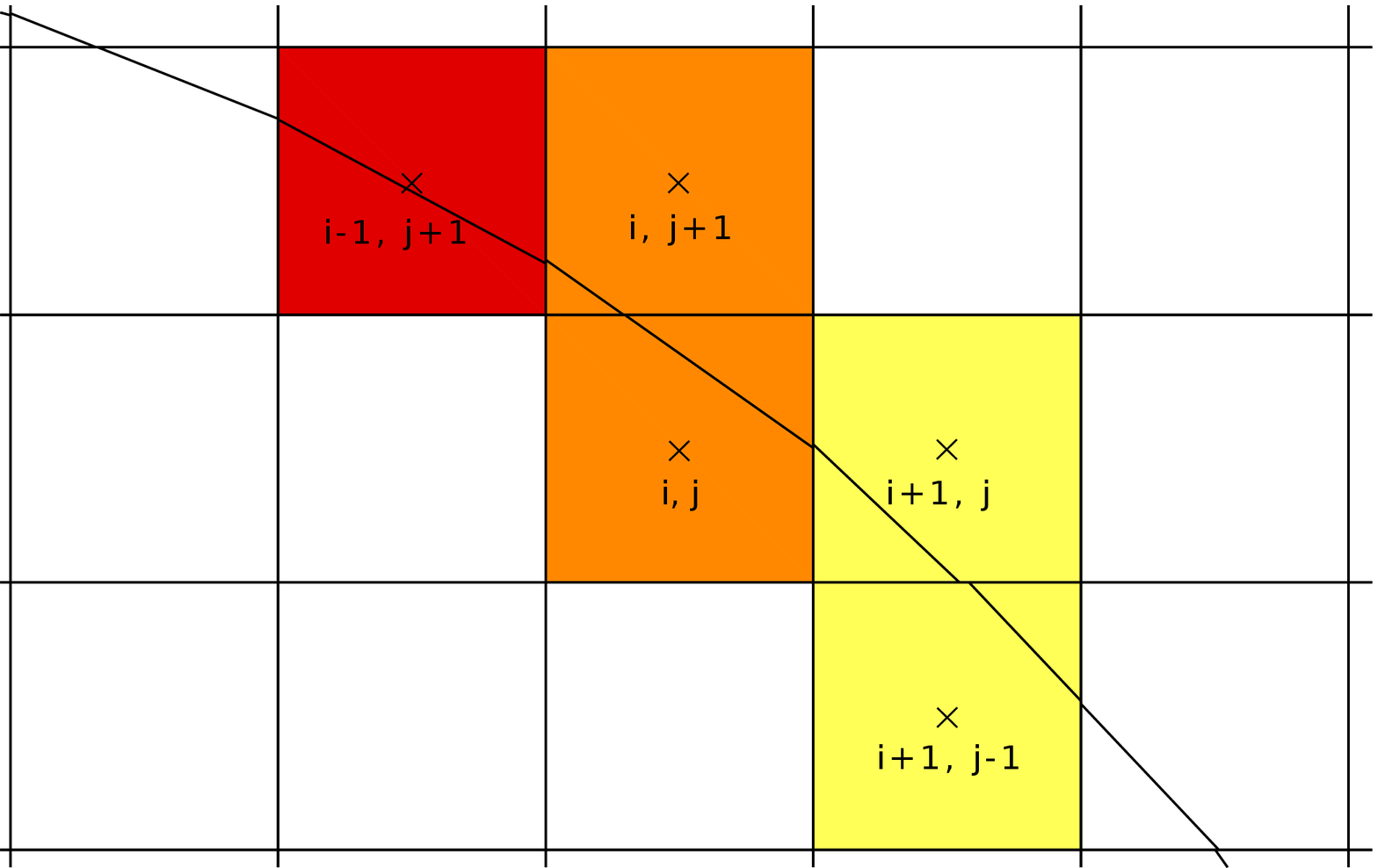}
 \caption{}
\end{subfigure}
  \caption{  An example of the interface orientation where columns in $x$ (a) 
  and $y$ (b) direction for computing $\tilde{\sigma}^c$ contain more than one 
  interfacial cell. The cells with the same color belong to the same column. }
  \label{fig:genConc_dirs}
\end{figure}

Let $\sigma({\mathcal{C}})$ be the surface tension evaluated from the temperature
or concentration at the center of all interfacial cells $\mathcal{C}$, with the
volume fraction $\chi(\mathcal{C})$.
The surface tension in each column, denoted by $\tilde{\sigma}^c\left(\mathbf{x}
\right)$, is defined so that it has only one value in each column, regardless of
how many interfacial cells are contained in that column. 
For illustration, Figure \ref{fig:genConc_flat} shows columns that contain only 
one interfacial cell, and Figure \ref{fig:genConc_dirs} shows columns that contain
more than one interfacial cell, where the same color denotes cells in the same 
column. The superscript, $c = x,y,z$, represents the column direction. For 
simplicity, here we show examples of the implementation in 2D, however, the 
algorithm extends trivially to 3D. 

For columns with only one interfacial cell (see Figure \ref{fig:genConc_flat}(a) 
and (b) for the columns in $x$ and $y$ direction, respectively), the surface 
tension of the interfacial cells, $\tilde{\sigma}^c$, is equal to the surface
tension $\sigma ({\mathcal{C}})$ in the same cells. If there is more than one 
interfacial cell in the column, then $\tilde{\sigma}^c$ is approximated by the 
volume weighted average of the $\sigma ({\mathcal{C}})$ values. In Figure 
\ref{fig:genConc_dirs}, the cells labeled with cell indices will be used for 
computing $\tilde{\sigma}^c$ for columns in the $x$ and $y$ directions -- Figures 
\ref{fig:genConc_dirs}(a) and (b) respectively. For example, in Figure 
\ref{fig:genConc_dirs}(a), the $\tilde{\sigma}^x$ is computed using the columns in 
the $x$ direction, and the value of $\tilde{\sigma}^x$ in the column containing cell 
$\mathcal{C}_{i,j}$, denoted $\tilde{\sigma}_{j}^x$, is 
\begin{equation}
\label{eq:interfaceTemp}
 \tilde{\sigma}_{j}^x = \frac{\chi_{i,j} \sigma_{i,j} + \chi_{i+1,j} \sigma_{i+1,j}}
 {\sum \chi_i}.
\end{equation} 
Note that the cells in the same column, in this particular example cells  
$\mathcal{C}_{i,j}$ and $\mathcal{C}_{i+1,j}$, have the same value of 
$\tilde{\sigma}^{x}$. For the columns in the $y$ direction, as in Figure 
\ref{fig:genConc_dirs}(b), $\tilde{\sigma}^y$ in the column containing  cell 
$\mathcal{C}_{i,j}$, denoted $\tilde{\sigma}_{i}^y$, is computed as
\begin{equation}
\label{eq:interfaceTemp_y}
 \tilde{\sigma}_{i}^y = \frac{\chi_{i,j} \sigma_{i,j} + \chi_{i,j+1} \sigma_{i,j+1}}{\sum \chi_i}.
\end{equation}
Again, the cells in the same column, in this case $\mathcal{C}_{i,j}$ and 
$\mathcal{C}_{i,j+1}$ have the same value of $\tilde{\sigma}^{y}$.

In our implementation, we first define $\tilde{\sigma}^c$ for all $c$ in all 
interfacial cells. For certain interface orientations, it is possible to define
$\tilde{\sigma}^c$ for columns in more than one direction, e.g.\ the interface 
in Figure \ref{fig:genConc_dirs}. However, this is not always the case, e.g. in 
Figure \ref{fig:genConc_flat}(a) we can only compute $\tilde{\sigma}^x$, and in 
Figure \ref{fig:genConc_flat}(b) we can only compute $\tilde{\sigma}^y$. For the 
former case, in the following sections we describe how the direction of the 
columns is chosen along with the discussion of the computation of the surface forces.

\subsection{Computation of the surface forces}
\label{sec:dsigma_ds}
The next step in the variable surface force implementation is the evaluation of the 
derivatives along the interface, ${\partial \sigma}/{\partial s_d}$ in equation
\eqref{eq:surf_ft}. In 2D, we only need to compute the derivative in one direction,
since the basis for a tangent line consists of only one vector. However, in 3D, we
need two tangent vectors to form a basis for the tangent space, hence we need to 
evaluate the derivative in two directions. We now discuss the implementation of the method 
for 2D and 3D.

\subsubsection{Surface force in 2D}
\label{sec:2D_derivative}

In 2D, equation \eqref{eq:surf_ft} simplifies to
\begin{equation}
\label{eq:surf_ft_2D}
 \textbf{F}_{st} = \frac{\partial \sigma}{\partial s^c} \delta_s \hat{\mathbf{t}},
\end{equation}
since we only have one tangential direction. We remind the reader that in this
case, $c = x,y$. 
The derivative of the surface tension along the interface, ${\partial \sigma}/{\partial s^c}$,
is approximated by the derivative of the interfacial value, $\tilde{\sigma}^c$ in 
the column which is formed in the direction $c$.  
The choice of the direction, $c$, is based on the interface orientation: $c$ is 
chosen to be the same as the largest component of the normal vector to the
interface. The same choice is made for computing curvature and the interface normal 
using height functions \citep{Popinet2009a}.

In each interfacial cell, we compute the derivative along the interface 
using center difference, i.e.\ the finite difference of the $\tilde{\sigma}^c$
in the two neighboring columns. For example, in Figure \ref{fig:genConc_flat}(a) 
and \ref{fig:genConc_dirs}(a), the derivative is computed with respect to the 
$y$ direction, as
\begin{equation}
 \label{eq:generalGrads}
	\left( \frac{\partial \sigma}{\partial s^x} \right)_{i,j} = 
	\frac{\tilde{\sigma}_{j+1}^x - \tilde{\sigma}_{j-1}^x }{\text{d}s}.
\end{equation}
As a reminder,  $\tilde{\sigma}_{j}^x$ is the interfacial value of the surface
tension in the column $j$ constructed in the $x$ direction. The arc length, 
$\text{d} s$, is computed from the height function in the same direction as 
${\partial \sigma}/{\partial s^c}$. For the example given in equation 
\eqref{eq:generalGrads}, the arc length is 
\begin{equation}
 \text{d}s = 2\Delta \sqrt{1 + h_y },
\end{equation}
where $h_y$ is the derivative of the height function (see \ref{sec:HF}) and 
$\Delta$ is the cell size.

The next part of the surface gradient implementation is the choice of the 
tangent vector, $\hat{\mathbf{t}}$, which is computed so that it satisfies
$ \hat{\mathbf{t}} \cdot \hat{\mathbf{n}} = 0 $, where $\hat{\mathbf{n}}$ is 
found using Mixed Young's Center method by \citet{aulisa2007}. The direction of 
$\hat{\mathbf{t}}$ depends on the direction used for computing 
${\partial \sigma}/{\partial s^c}$: $\hat{\mathbf{t}}$ points in the direction of 
the positive component orthogonal to the $c$ direction. For example, $\hat{\mathbf{t}}$ points in 
the positive $x$ direction if we construct columns in the $y$ direction. 

\begin{figure}[t]
\begin{subfigure}{0.325\textwidth}
 {\includegraphics[width=\textwidth]{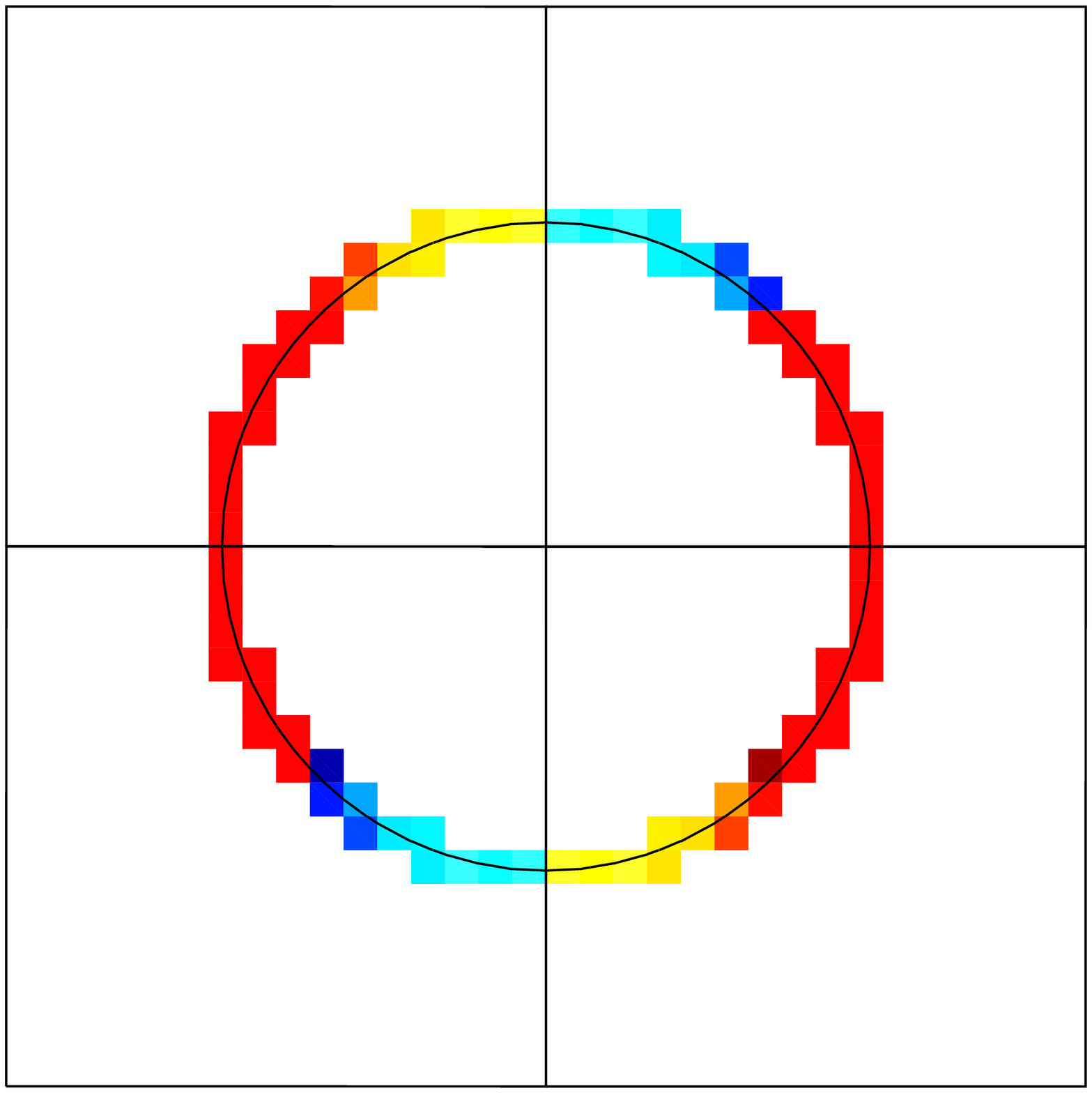}}
 \caption{}
\end{subfigure}
\begin{subfigure}{0.325\textwidth}
 {\includegraphics[width=\textwidth]{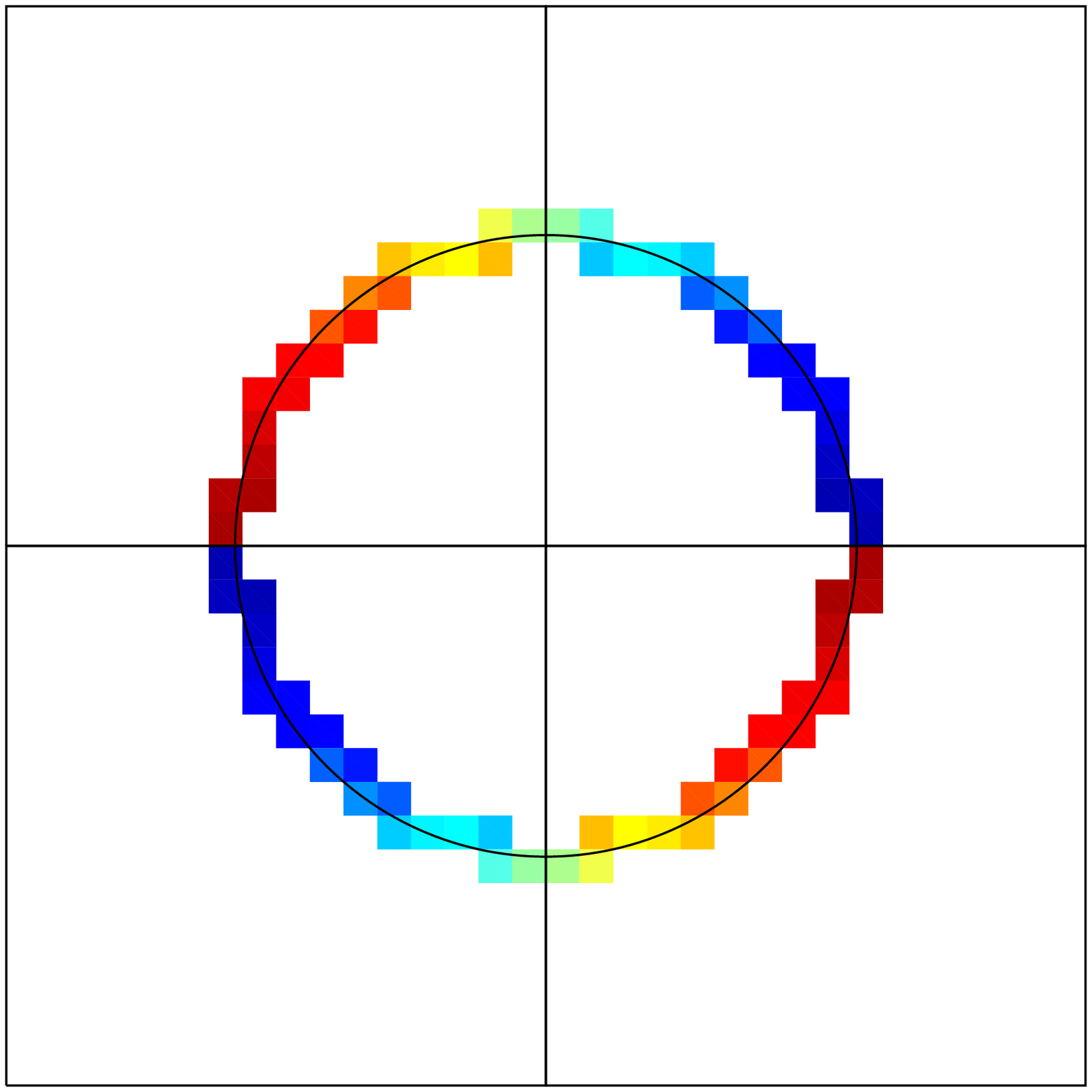}}
 \caption{}
\end{subfigure}
\begin{subfigure}{0.325\textwidth}
 {\includegraphics[width=\textwidth]{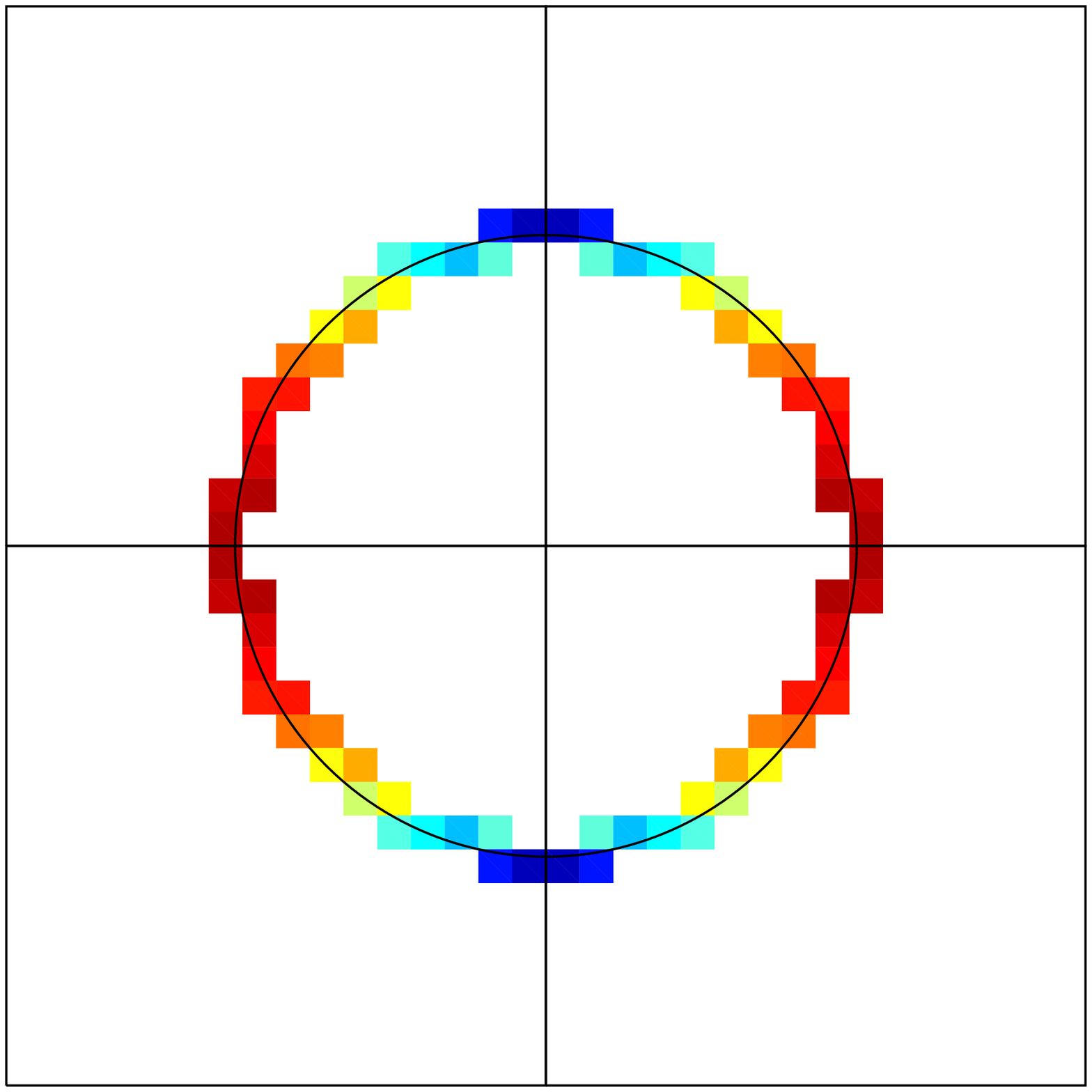}}
 \caption{}
\end{subfigure}
  \caption{(a) Example of the gradient ${\partial \sigma}/{\partial s^c}$, computed 
  on a circular interface, where $\sigma$ depends on the $y$ direction linearly.
  The dark red and dark blue colors are the most positive and negative values respectively. 
  The $x$ (b) and $y$ (c) components of $\mathbf{G}$. }
  \label{fig:sign_eg} 
\end{figure}

To illustrate the importance of the choice of the tangent vector, consider 
an intermediate value of the surface force, $\mathbf{G}$, defined as
\begin{align}
    \label{eq:G_2D_x}
    G_x = \frac{\partial \sigma}{\partial s^c} \text{sgn}(t_x), \\
    \label{eq:G_2D_y}
    G_y = \frac{\partial \sigma}{\partial s^c} \text{sgn}(t_y).
\end{align}
Figure \ref{fig:sign_eg}(a), (b) and (c) show an examples of 
${\partial \sigma}/{\partial s^c}$, $G_x$ and $G_y$ respectively, computed in 
all interfacial cells, where we impose a positive uniform gradient of the surface 
tension in the $y$ direction. In figure \ref{fig:sign_eg}(a), 
${\partial \sigma}/{\partial s^c}$ changes sign in the first and third quadrant 
at the angles, defined from the positive $x$ axis, of $\pi/4$ and $5\pi/4$, respectively.
At these points the direction of the columns used in gradient computation changes.
Hence, the two neighboring cells have opposite sign of 
${\partial \sigma}/{\partial s^c}$. However, once we include the correct sign of the 
tangent vector components and consider each component separately, as in equations
\eqref{eq:G_2D_x} and \eqref{eq:G_2D_y}, this inconsistency in the sign is
corrected; see Figure \ref{fig:sign_eg}(b) and (c) for illustration.

\begin{figure}[t]
\centering
\begin{subfigure}{0.35\textwidth}
 {\includegraphics[width=\textwidth]{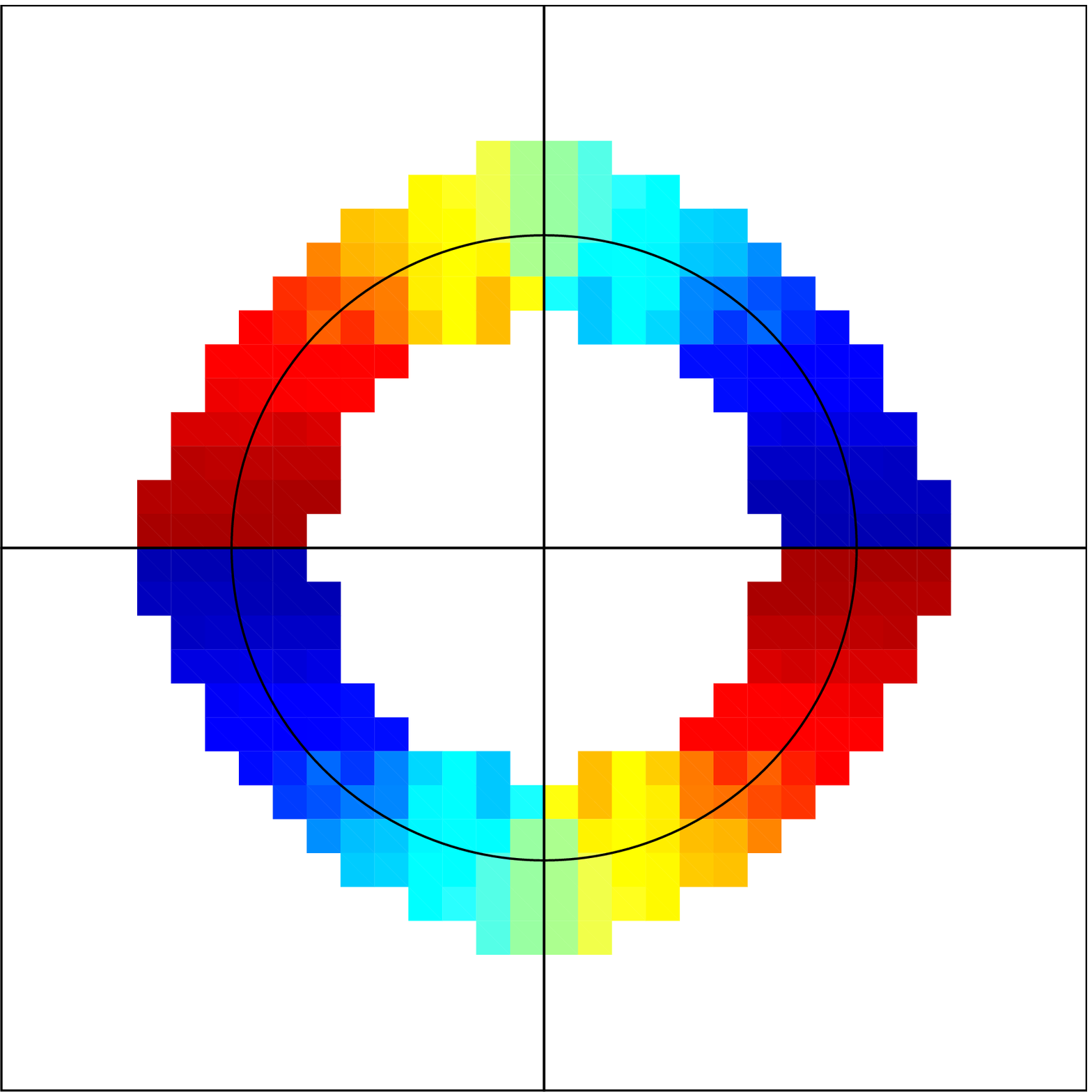}}
 \caption{}
\end{subfigure}
\begin{subfigure}{0.35\textwidth}
 {\includegraphics[width=\textwidth]{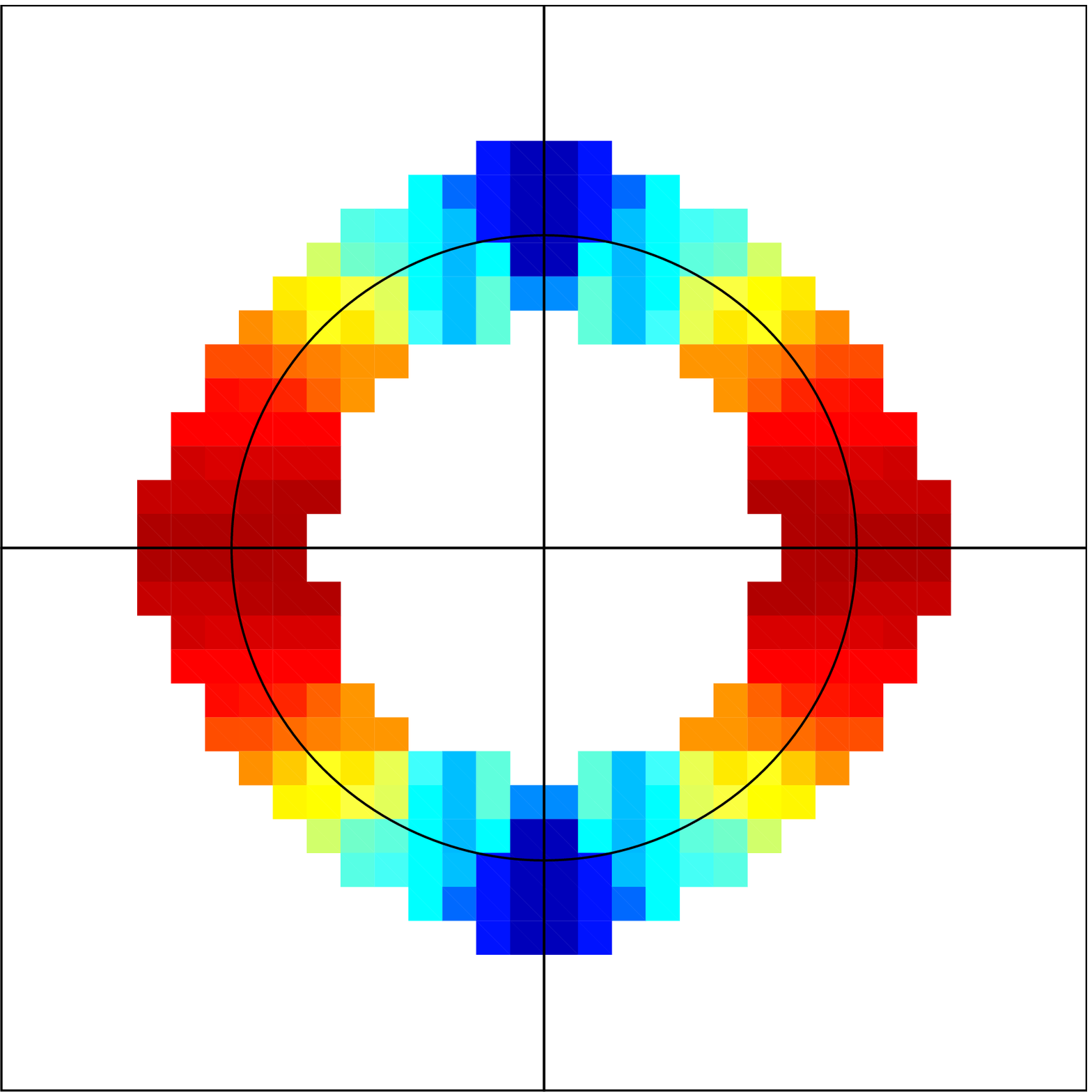}}
 \caption{}
\end{subfigure}
  \caption{ The $x$ (a) and  $y$ (b) component of $\mathbf{G}$, with values 
  around the interfacial cells defined by averaging the neighboring cells.}
  \label{fig:smeared} 
\end{figure}

The complete surface force defined in equation \eqref{eq:surf_ft_2D}, 
given in the component form, is
\begin{align}
    \label{eq:Force_2D}
    F_x = G_x | t_x  |\, \delta_s, \\ 
    \label{eq:Force_2D_2}
    F_y = G_y | t_y  |\, \delta_s. 
\end{align}
where $\delta_s = || \nabla \chi  ||_2 $.
In the CSF method \citep{Brackbill1992}, we need to know $\mathbf{G}$ in the cells 
around the interface, i.e.\ in all the cells where $|| \nabla \chi  ||_2 $ is 
nonzero. We proceed by using the same approach as for defining the curvature
in the cells around the interface \citep{Gerris}, i.e.\ the values in the cells neighboring the 
interfacial cells are defined by averaging the values in the direct neighbors that 
already have the curvature value defined. This procedure is repeated twice, insuring that the 
curvature values for the corner neighbors to the interfacial cells are defined as 
well. We use an identical approach for defining the $x$ and $y$ components of 
$\mathbf{G}$ in the cells around the interface which are subsequently used in equations
\eqref{eq:Force_2D} and \eqref{eq:Force_2D_2}. Figure \ref{fig:smeared} shows the 
result of this procedure for the same example of the surface gradient as discussed 
in Figure \ref{fig:sign_eg}.

\subsubsection{Surface force in 3D}
\label{sec:3D_derivative}
\begin{figure}[t]
\centering
 \includegraphics[width = 0.8\textwidth]{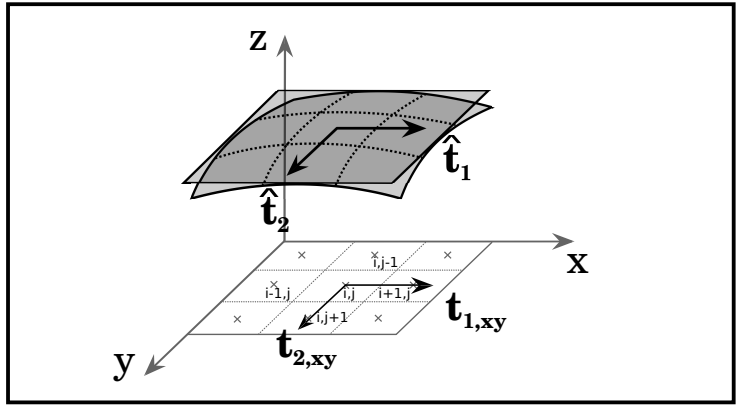}
 \caption{ A stencil used for computing surface gradient in the column containing the 
 cell $\mathcal{C}_{i,j}$, 
 with a tangent plane defined by the vectors $\hat{ \mathbf{t}}_1$ and $\hat{ \mathbf{t}}_2$.
 Vectors $\hat{ \mathbf{t}}_{1,xy}$ and $\hat{ \mathbf{t}}_{2,xy}$ are the projections
 of $\hat{ \mathbf{t}}_1$ and $\hat{ \mathbf{t}}_2$ onto the $xy$-plane respectively. }
 \label{fig:vectors_3D}
\end{figure}

The implementation of the surface gradient in 3D extends the 2D implementation 
by considering the second tangential direction as stated in equation \eqref{eq:surf_ft}.
Equivalently as in 2D, we first define the column values $\tilde{\sigma}^c$ of the 
surface tension $\sigma$. This part of the algorithm is identical to the 2D part, 
with the addition of one more direction.
After the column values, $\tilde{\sigma}^c$, are defined, we compute the gradients
along the two components orthogonal to the columns: for example, if the columns are 
constructed in the $z$ direction, see Figure \ref{fig:vectors_3D}, then the 
derivatives along the interface are computed in the $x$ and $y$ directions as
\begin{equation}
\begin{aligned}
\label{eq:z_cols}
\left(\frac{\partial \sigma}{\partial s^z_1}\right)_{i,j} &= \frac{ \tilde{\sigma}_{i+1,j}^z - \tilde{\sigma}_{i-1,j}^z  }{2\Delta \sqrt{ 1 + h_{x}^2 }},\\
\left(\frac{\partial \sigma}{\partial s^z_2}\right)_{i,j} &= \frac{ \tilde{\sigma}_{i,j+1}^z - \tilde{\sigma}_{i,j-1}^z  }{2\Delta \sqrt{ 1 + h_{y}^2 }}.
\end{aligned}
\end{equation}
As previously discussed in 2D, the direction, $c$, in which the columns are constructed, is
chosen based on the interface orientation,
where $c$ is the same as the direction of the largest component of the interface
normal vector. 

Next part of the surface gradient computation is the choice of the tangent vectors,
$\hat{\mathbf{t}}_d$, which are computed so that they satisfy 
$\hat{\mathbf{t}}_d \cdot \hat{\mathbf{n}} = 0$. Among all the possibilities for 
$\hat{\mathbf{t}}_d$, we choose the two whose projections onto the coordinate plane,
defined by all points with $c$ coordinate equal to zero, are parallel to the axes.
Figure \ref{fig:vectors_3D} illustrates this procedure by an example where the columns are
constructed in the $z$ direction and the projections of the tangent vectors 
$\hat{\mathbf{t}}_1$ and $\hat{\mathbf{t}}_2$ onto the $x$-$y$ plane
are parallel to the $x$ and $y$ axes and denoted by
$\hat{ \mathbf{t}}_{1,xy}$ and $\hat{ \mathbf{t}}_{2,xy}$, respectively.
In this particular example, the tangent vectors will be of the form 
\begin{gather}
\label{eq:tangent_form}
\hat{\mathbf{t}}_1 = ( t_{1x}, 0, t_{1z} ),\\
\hat{\mathbf{t}}_2 = ( 0, t_{2y}, t_{2z} ). 
\end{gather} 
The signs of the components of the tangential vectors are chosen so that their
projections onto the coordinate plane point in the positive direction of the
coordinate axes (see e.g.\ Figure \ref{fig:vectors_3D}). 

Finally, we compute the surface force, $ \textbf{F}_{st}=(F_x, F_y,F_z)$. In the 
case such that the columns are constructed in the $z$ direction, the components
of $ \textbf{F}_{st}$ are
\begin{align}
\label{eq:force_3D}
 F_x &= \frac{\partial \sigma}{\partial s^z_1} t_{1x}, \\
 F_y &= \frac{\partial \sigma}{\partial s^z_2} t_{2y}, \\
\label{eq:force_3D_z}
 F_z  &= \frac{\partial \sigma}{\partial s^z_1} t_{1z} 
 + \frac{\partial \sigma}{\partial s^z_2} t_{2z}.
\end{align}

Similarly as in the 2D case, in order to use the CSF formulation, the components
of the tangential force need to be defined in the cells around 
the interface. 
This is done equivalently as in 2D, using the neighbor averaging procedure,
see Section \ref{sec:2D_derivative}. 
However, in 3D, there is one extra step due to one of the components containing 
an addition of two terms, e.g.~as in equation \eqref{eq:force_3D_z}. 
In order to illustrate this, consider the general form of the $x$ component of the 
tangential force
\begin{align}
\label{eq:F_x}
F_x = 
 \begin{cases} 
	(\partial \sigma/\partial s^x_1) t_{1x} +
	(\partial \sigma/\partial s^x_2) t_{2x} 
	& \mbox{ if } c = x, \\
	(\partial \sigma/\partial s^y_1) t_{1x} 
	& \mbox{ if } c = y,\\
	(\partial \sigma/\partial s^z_1) t_{1x} 
	& \mbox{ if } c = z.
 \end{cases}
\end{align}
Similarly as in 2D, the differences in the sign in the derivatives, 
$\partial \sigma/\partial s^c_d$, may arise from the choice of the column
directions. We proceed by defining the intermediate value of the surface force,
$\mathbf{G}$. The components of $\mathbf{G}$ are computed equivalently as in 2D,
except for the $c$ component which is defined as 
\begin{equation}
 G_c = \frac{(\partial \sigma/\partial s^c_1) t_{1c} +
(\partial \sigma/\partial s^c_2) t_{2c} }{\sqrt{ t_{1c}^2 + t_{2c}^2  }},
\end{equation}
where $c$ is the direction of the columns. Now we can carry out the averaging
procedure for each component of $\mathbf{G}$.
Finally, the component of the force in the direction $c$ is
\begin{equation}
 F_c =   G_c |t_{1c}| \delta_s + G_c |t_{2c}| \delta_s. 
\end{equation}
The other components are computed equivalently as in the 2D case.

\section{Results}
\label{sec:results}

\subsection{Surface gradient computation}

\begin{figure}[t]
\centering
 \includegraphics[width = 0.4\textwidth]{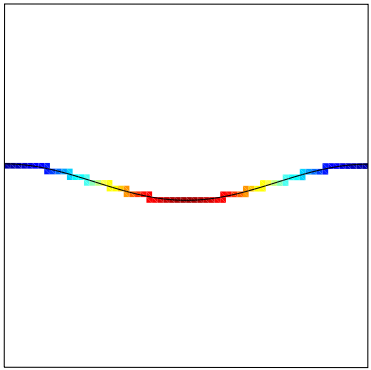}
 \caption{ The setup of the perturbed interface with surface tension dependent
 on the interface profile $h \left( x \right) = h_0 + \varepsilon \cos(2 \pi x)$.
 The color represents the surface tension at the interface,  with dark red and 
 dark blue being the maximum and minimum values respectively. }
 \label{fig:sigma_sketch}
\end{figure}

We first present the validation of our methodology for computing the surface
gradient in 2D geometry where we can compute the gradient exactly. 
The simplest geometry that we consider is a flat perturbed interface, i.e.\
let the interface be a function of $x$ as
\begin{gather}
 h \left( x\right) = h_0 + \varepsilon \cos(2 \pi x).
\end{gather}
Let the surface tension be a function of the interface position as
\begin{gather}
 \sigma \left( h \right) = 1 + \sigma_h h \left( x \right),
\end{gather}
Figure \ref{fig:sigma_sketch} shows the interface profile and surface tension at
the interface, for $h_0 = 0.5$,  $\varepsilon = 0.05$, and 
$\sigma_h = 10^{-2}$ in a computational domain of $1\times1$, 
with symmetry boundary conditions imposed on all sides.
In this case, apart from using the definition of the surface gradient  given in 
equation \eqref{eq:nabla_s_def}, we can also compute the exact surface gradient 
using the chain rule as
\begin{equation}
\label{eq:grad_sigma_film}
    \nabla_s \sigma = \frac{\sigma_h h_{x}}{\sqrt{1 + h_{x}^{2} } } \, \hat{\mathbf{t}},
\end{equation}
where the unit tangent vector, $\hat{\mathbf{t}}$, is defined to point in the 
positive $x$ direction as
\begin{equation}
     \hat{\mathbf{t}} = \left( 1, - h_{x} \right)/\sqrt{1 + h_{x}^{2}},
\end{equation}
Note that the numerator in equation \eqref{eq:grad_sigma_film}, $\sigma_h h_{x}$,
is equivalent to the numerator of equation \eqref{eq:generalGrads}; hence, we can 
compare their computed values to the exact ones directly.  
We present the errors associated with computing $\sigma_h h_{x}$ and $h_{x}$, separately,
as well as each component of the surface gradient in equation \eqref{eq:grad_sigma_film}. 
We test the convergence as a function of the mesh size, $\Delta$, using $L_1$ and 
$L_\infty$ norms to define $E_1$ and $E_\infty$ errors respectively as
\begin{gather}
\label{eq:L_1_norm}
    E_1 (f) = \frac{\sum\limits_j^N |f_{\text{approx}} - f_{\text{exact}}|}{N}, \\
\label{eq:L_infty_norm}
    E_\infty (f) = \max{|f_{\text{approx}} - f_{\text{exact}}|},
\end{gather}
where the summation is over all interfacial cells and $N$ is the number of 
interfacial cells. The interface position in each cell can influence the errors 
obtained in constructing the columns for the computation of both surface gradients, 
$\partial \sigma/\partial s^y$, and the derivative of the height function, $h_x$.
To avoid this error bias, we average the errors from 100 simulations 
where $h_0$ was modified to $\tilde{h_0} = h_0 + h_r$, where $h_r$ is a random 
number in the interval $[0, \Delta]$ with uniform distribution. 

\begin{figure}[t]
\centering
    \includegraphics[width = 0.6\textwidth,trim=0 0 40mm 0,clip=true]{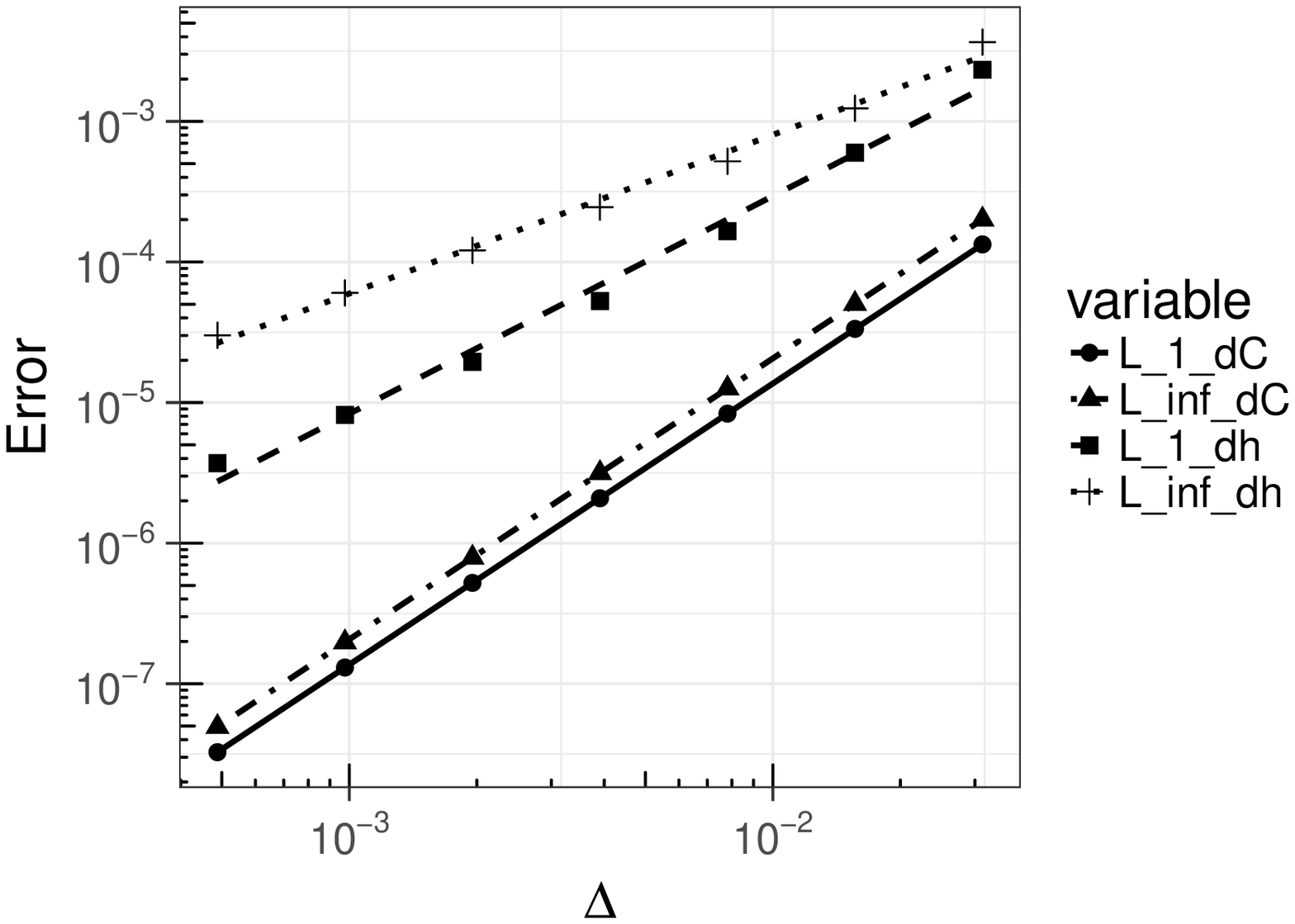}
    \caption{The computed errors for 
     $E_1(\sigma_h h_{x})$ $({\color{black} \bullet})$, 
     $E_{\infty}(\sigma_h h_{x})$ $({\color{black} \blacktriangle})$,
     $E_1(h_{x})$ (\Smblacksquare), and
     $E_{\infty}(h_{x})$ $({\color{black} +})$. 
    The order of convergence for $\sigma_h h_{x}$ for both 
    $E_1$ (\solidrule) and $E_\infty$ (\dashdotrule) errors is $2$, 
    and the order of convergence for $h_x$ for 
    $E_1$ (\dashedrule ) and $E_\infty$ (\dotsrule) errors is $1.55$ and $1.13$,
    respectively. The symbols represent the errors from the computations and the 
    lines show the linear fits.}
     \label{fig:98_numerator}
\end{figure}

We initialize the surface tension, $\sigma$, directly as a function of $x$, i.e. 
$\sigma\left( h \right) = 1 + \sigma_h (\tilde{h_0} + \varepsilon \cos(2 \pi x))$.
Figure \ref{fig:98_numerator} shows the convergence of the computed
$\sigma_h h_{x}$ as a function of mesh refinement. As shown, the order of convergence 
is $2$ for both $E_1$ and $E_\infty$ errors. In this test case, the interfacial value of 
the surface tension $\tilde{\sigma}^y$ is computed in the $y$ direction for all cells. 
Figure \ref{fig:98_numerator} also shows the order of convergence of $h_{x}$, computed
using height functions (see \ref{sec:HF}), that is $1.55$ and $1.13$ for $E_1$ and 
$E_\infty$ errors, respectively. The lower order of convergence is contributed to 
the errors in initializing the volume fractions, and their convergence to the
prescribed initial condition.

\begin{figure}[t]
\centering
    \includegraphics[width = 0.6\textwidth,trim=0 0 35mm 0,clip=true]{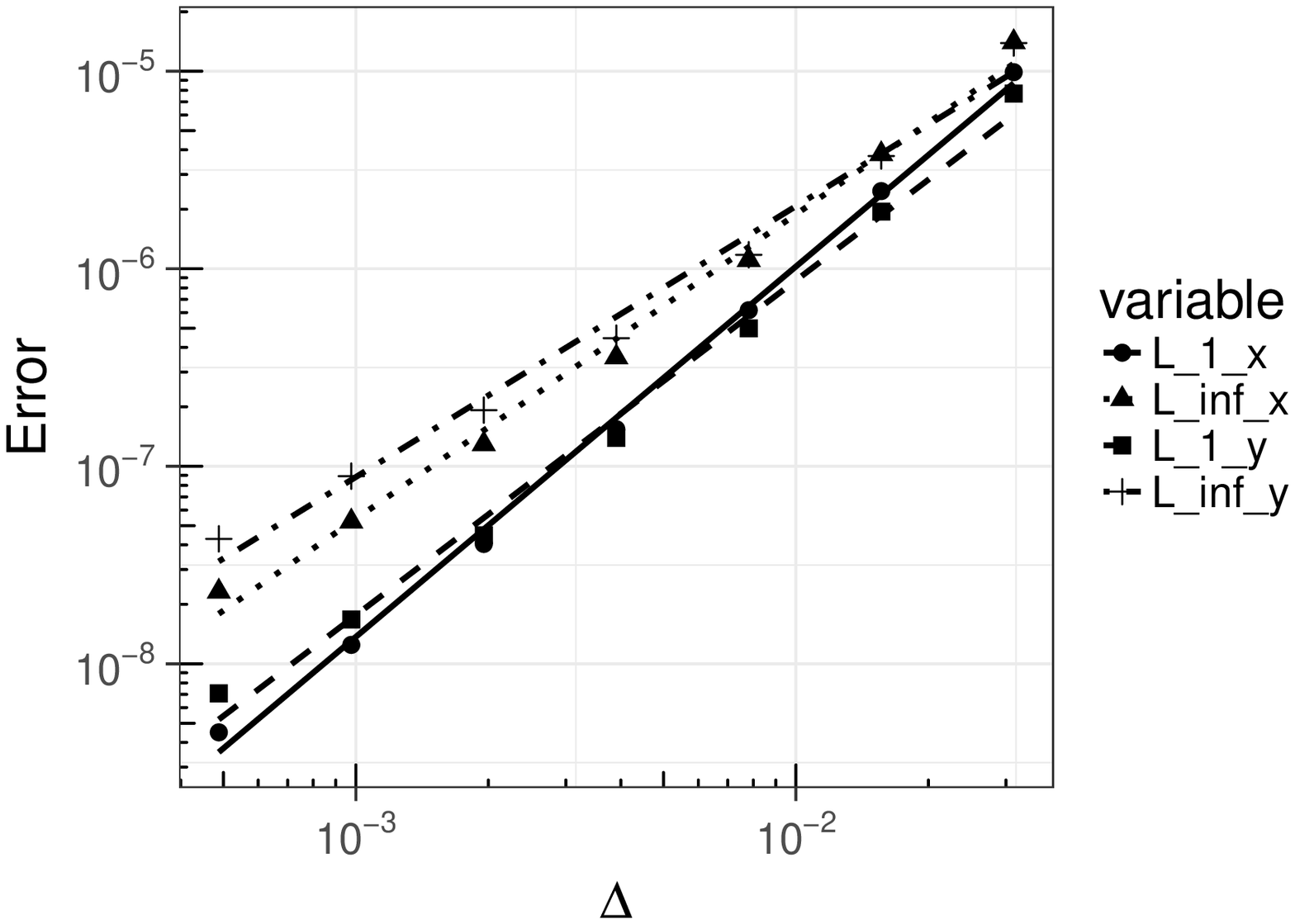}
\caption{The computed errors for 
     $E_1((\nabla_s\sigma)_x)$ $({\color{black} \bullet})$, 
     $E_{\infty}((\nabla_s\sigma)_x)$ $({\color{black} \blacktriangle})$,
     $E_1((\nabla_s\sigma)_y)$ (\Smblacksquare), and
     $E_{\infty}((\nabla_s\sigma)_y)$ $({\color{black} +})$. 
    The order of convergence for $(\nabla_s\sigma)_x$ is $1.9$ and $1.5$
    for $E_1$ (\solidrule) and $E_\infty$ (\dotsrule) errors respectively, 
    and the order of convergence for $(\nabla_s\sigma)_y$ is $1.7$ and $1.4$ 
    for $E_1$ (\dashedrule) and $E_\infty$ (\dashdotrule) errors respectively.
    The symbols represent the errors from the computations and the lines show the
    linear fits.}
    \label{fig:98_forces}
\end{figure}

Next we investigate the accuracy of the computed surface gradient 
$$\nabla_s\sigma\left( \mathbf{x} \right) = 
    (\nabla_s\sigma)_x\,\hat{i} + (\nabla_s\sigma)_y\,\hat{j},$$
where 
$\left((\nabla_s\sigma)_x, (\nabla_s\sigma)_x\right)=\left(G_x|t_x|, G_y|t_y|\right)$. 
Figure \ref{fig:98_forces} compares the $x$ and $y$ components of the surface 
gradient with the exact solution. As shown, the $x$ component converges with order
$1.9$ and $1.5$ for $E_1$ and $E_\infty$ errors, respectively, and the $y$ 
component converges with order $1.7$ and $1.4$ for $E_1$ and  $E_\infty$ errors,
respectively. The difference in the order of convergence is due to the interface
orientation being in the horizontal direction and the gradient being imposed in 
the $x$ direction. Hence, the columns are always constructed in the $y$ direction, 
and the derivative along the interface is computed as 
${\partial \sigma}/{\partial s^x}$, which captures the gradient in the horizontal
direction more accurately.

\begin{figure}[htb]
\centering
 \begin{overpic}[width = 0.4\textwidth]{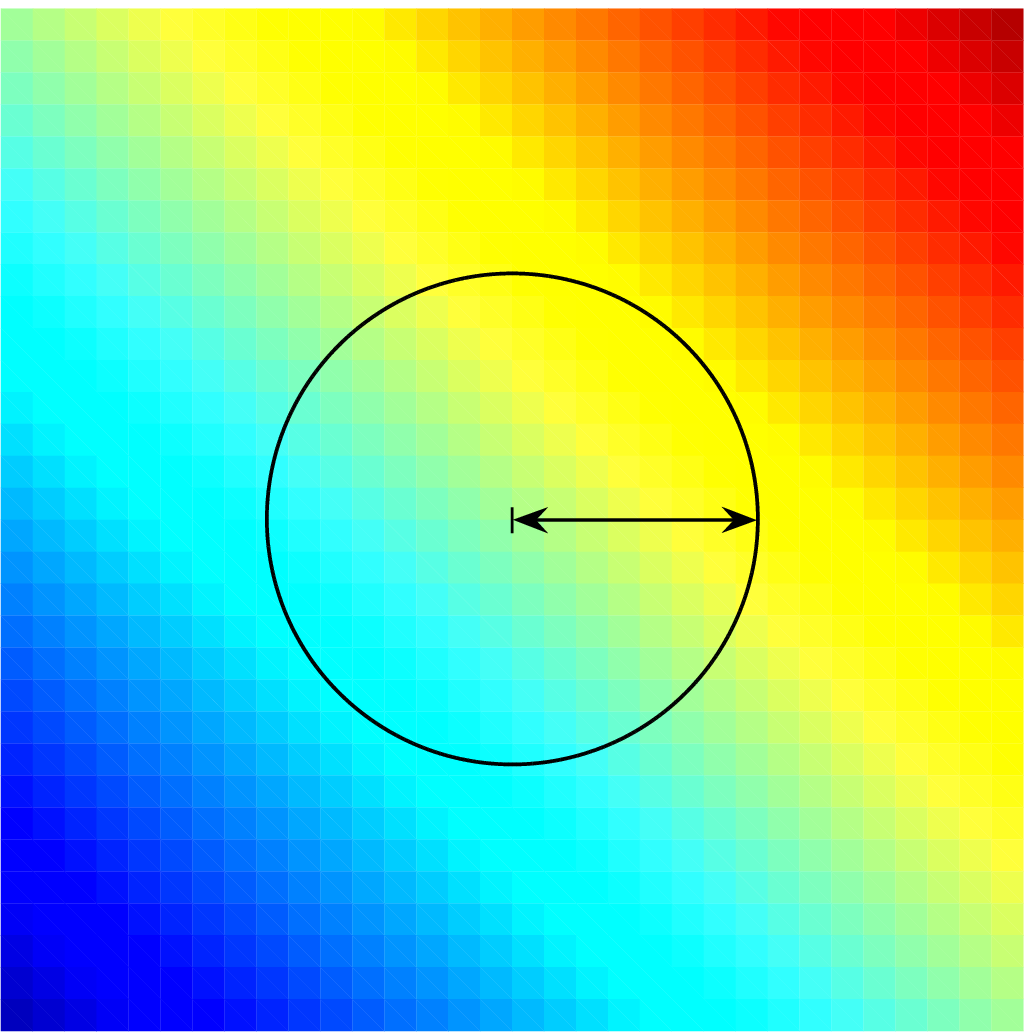}
   \put (5, 70) {$ \rho_1, \mu_1$}
   \put (5, 60) {$ k_1, C_{p1} $}
   \put (60,45 ) {$ a $ }
  \end{overpic}
  \caption{ The initial setup of a circular drop with an imposed uniform temperature
  gradient. The color shows the temperature with dark blue and dark red being the
  minimum and maximum values respectively.}
  \label{fig:temp_setup_xy}
\end{figure}

Next we test the convergence for a more general interfacial geometry where the 
interfacial values of $\tilde{\sigma}^c$ are computed using columns in both $x$ and 
$y$ directions (see Section \ref{sec:method}). We consider a circle of radius 
$a = 0.25$ positioned at $(0.5, 0.5)$ in a $1 \times 1$ domain with an imposed 
temperature distribution 
 \begin{gather}
 \label{eq:init_temp}
    T(x, y) = \Delta T \left( x + y \right),
\end{gather}
where $\Delta T$ is a constant. We assume that the thermal diffusivity is equal 
for the fluid inside and outside of the circle, i.e.~$
 k_1 = k_2,  C_{p1} = C_{p2}, 
 \rho_1 = \rho_2,  \mu_1 = \mu_2,
$
where the subscripts $1$ and $2$ denote surrounding and the fluid inside of the
drop, respectively. Figure \ref{fig:temp_setup_xy} shows the setup with color 
representing the temperature field. Here we choose 
$\Delta T = 0.1$, $k_1 = 1$, $C_{p1} = 1$ and $\rho_1 = 1$. 
For simplicity, we assume that the surface tension is a linear function of 
temperature, i.e.~$
    \sigma \left( T \right) = 1 + \sigma_T T ,
$
where we let $\sigma_T = -0.1$.
We set the velocity to zero, and knowing that the interface is exactly circular,
we can compute exact surface gradient from the definition
\begin{align}
\label{eq:gradient_xy_1}
  \nabla_s \sigma &= \nabla \sigma - \hat{\mathbf{n}} 
	\left( \hat{\mathbf{n}} \cdot \nabla \sigma \right), \\
\label{eq:gradient_xy_2}
    &= \frac{\sigma_T  \Delta T}{a^2} 
	\left\langle a^2 - x^2 \mp x \sqrt{a^2 - x^2}, 
			x^2 \mp x \sqrt{a^2 - x^2} \right\rangle, \\
\label{eq:gradient_xy_3}
    &= \frac{\sigma_T  \Delta T}{a^2}  
	\left\langle y^2 \mp y \sqrt{a^2 - y^2},
			a^2 - y^2 \mp y \sqrt{a^2 - y^2} \right\rangle
\end{align}
Equations \eqref{eq:gradient_xy_2} and \eqref{eq:gradient_xy_3} give the surface 
gradient as a function of $x$ and $y$, respectively.

We initialize the temperature following two approaches, and discuss their performance. 
First approach is to define the interface as a 
function of $x$ and $y$ depending on the more favorable interface orientation as
follows
\begin{equation}
\label{eq:temp45}
    T (x,y) = \begin{cases}  
		 \Delta T\left( x \pm \sqrt{ a^2 - x^{2} } \right) 
		 &\mbox{if } |x| < |y|, \\
                 \Delta T \left( y \pm \sqrt{ a^2 - y^{2} } \right) 
                 & \mbox{otherwise},
              \end{cases}
\end{equation}
where $x$ and $y$ are coordinates of the cell centers. 
Second approach is to use positions of the
centroid of the interface contained in each cell to initialize the temperature by
equation \eqref{eq:init_temp}.   We show below that the second approach leads
to more accurate results.

\begin{figure}[t] 
 \centering
    \includegraphics[width = 0.6\textwidth,trim=0 0 35mm 0,clip=true]{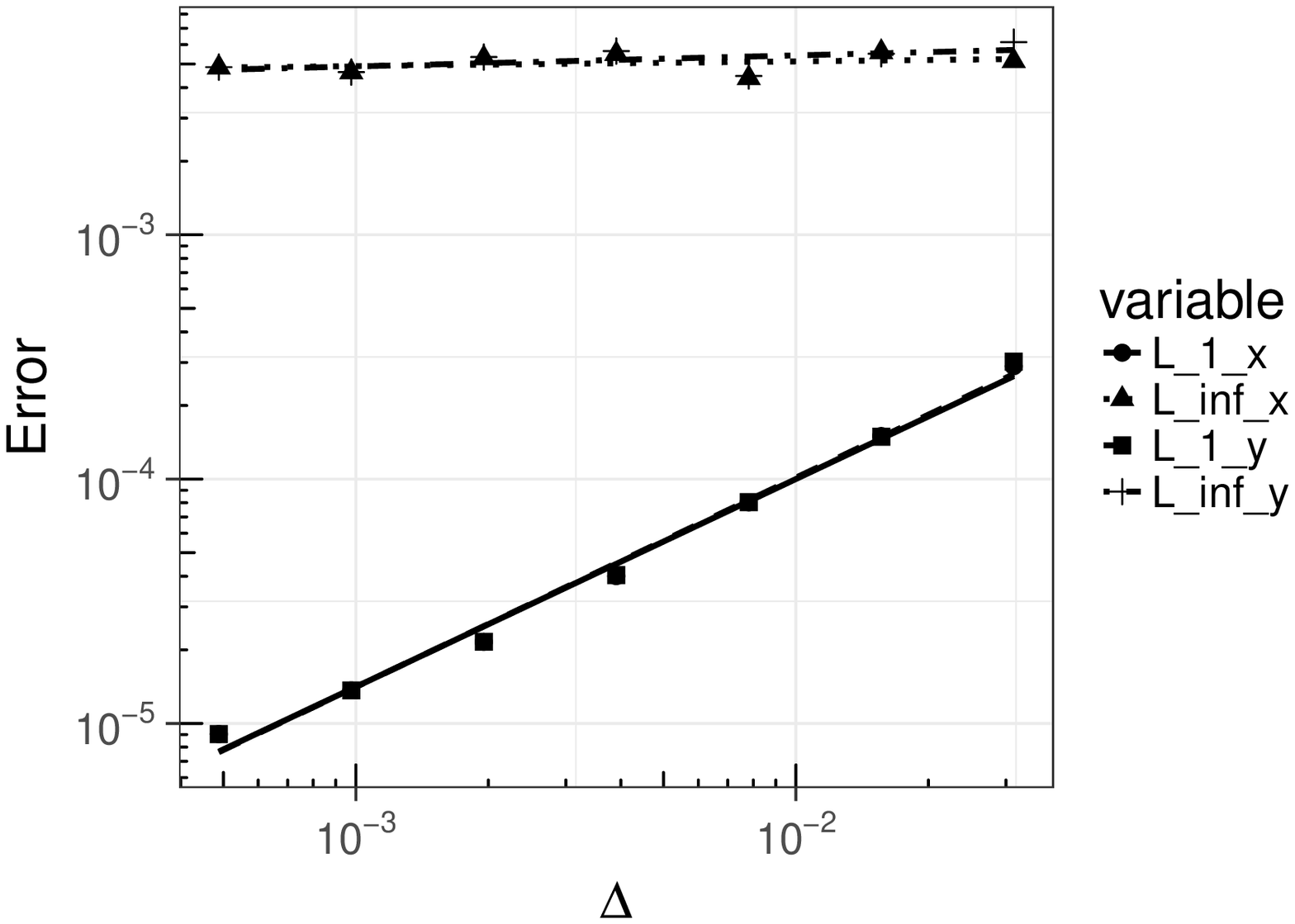}
    \caption{
    The computed error for $E_1((\nabla_s \sigma)_x)$ $({\color{black} \bullet})$, 
     $E_{\infty}((\nabla_s \sigma)_x)$ $({\color{black} \blacktriangle})$,
     $E_1((\nabla_s \sigma)_y)$ (\Smblacksquare), and
     $E_{\infty}((\nabla_s \sigma)_y)$ $({\color{black} +})$.  
     The order of convergence for $(\nabla_s \sigma)_x$ is $0.85$ and $0.019$  $E_1$ (\solidrule) and $E_\infty$ (\dotsrule)
    errors respectively, and the order of convergence for $(\nabla_s \sigma)_y$ is $0.86$ and $0.045$ 
    for $E_1$ (\dashedrule) and $E_\infty$ (\dashdotrule) errors, respectively.
    The symbols represent the errors from the computations, and the lines show the
    linear fit of those points.}
    \label{fig:errs_92}
\end{figure}

We compare the computed surface gradient with the exact solution by considering 
$E_1$ and $E_\infty$ errors defined in equations \eqref{eq:L_1_norm} and 
\eqref{eq:L_infty_norm}, respectively.
Similarly as in the previous example, in order to eliminate the dependence of the
errors on the interface position in the cell, the center of the drop is positioned
randomly in the interval $[0, \Delta]\times[0,\Delta]$, and the errors are 
averaged over $100$ random realizations. 
Figure \ref{fig:errs_92} shows the convergence to the exact solution for the $x$ and 
$y$ components of the gradient. The convergence of $E_1$ error is $0.85$ and $0.86$
for the $x$ and $y$ components, respectively.
The slow convergence of $E_\infty$ error is due to the errors in initializing the 
temperature at the lines $|x| = |y|$ from equation \eqref{eq:temp45}, demonstrated 
later. 

In order to reduce the influence of the initialization of $T$ on the convergence, 
we also compute the convergence of $L_1$ norm of each component of the surface gradient
\begin{gather}
\label{eq:L_1_solution}
    L_1 \left( f \right) = \frac{1}{N}\sum_i^N |f_i|, 
\end{gather}
where $f_i$ is the $x$ or $y$ component of the surface gradient, and $N$ is the 
number of interfacial points. 
 \begin{figure}[t]
 \centering
    \includegraphics[width = 0.8\textwidth,trim=0 0 35mm 0,clip=true]{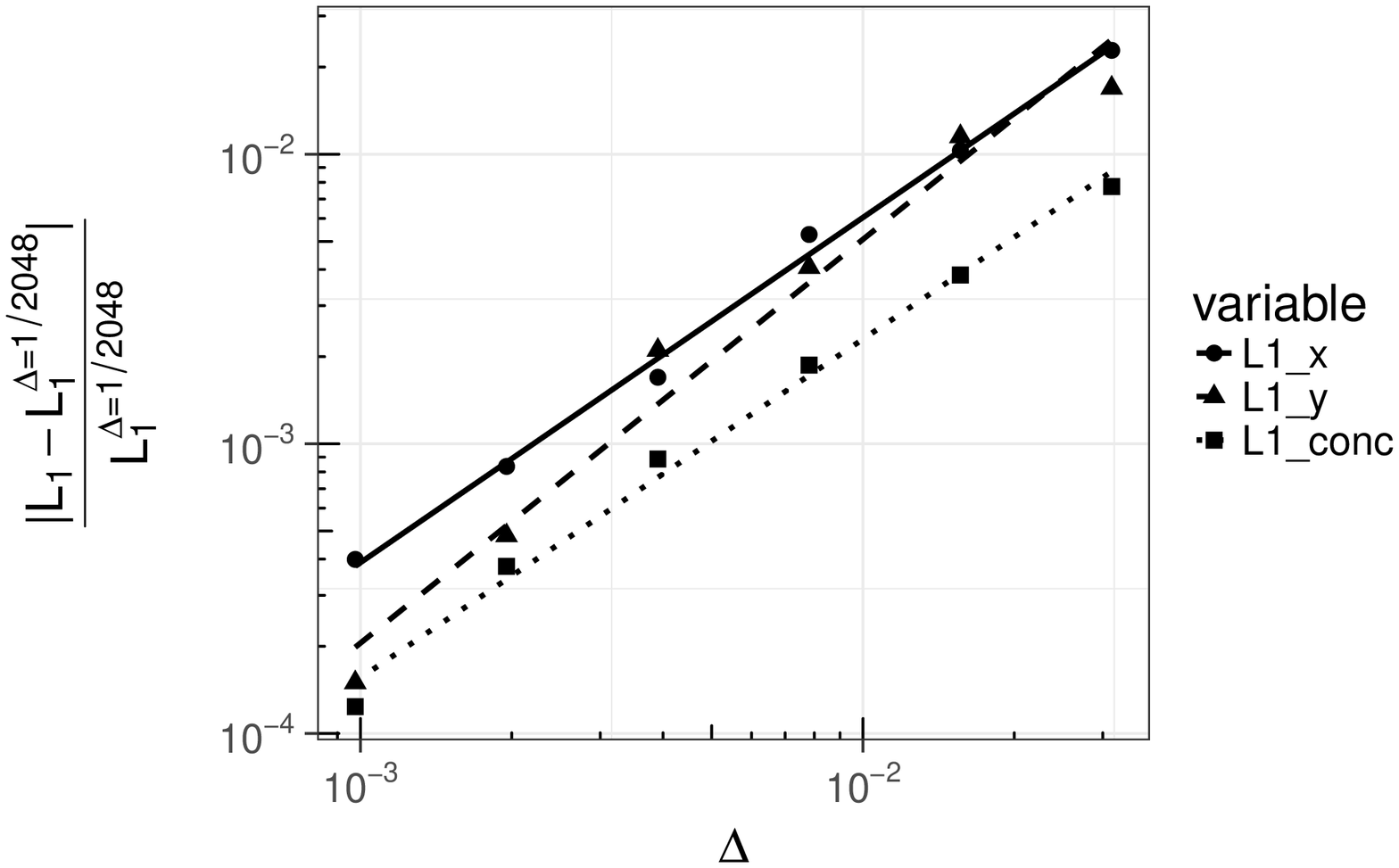}
    \caption{The computed errors for 
     $L_1((\nabla_s \sigma)_x)$ $({\color{black} \bullet})$, 
     $L_1((\nabla_s \sigma)_y)$ $({\color{black} \blacktriangle})$, and
     $L_1(T)$ (\Smblacksquare). 
     The order of convergence for $(\nabla_s \sigma)_x$ and $(\nabla_s \sigma)_y$
     is $1.2$ (\solidrule) and $1.4$ (\dashedrule), respectively. 
    The order of convergence of $T(x,y)$ at the interface is $1.2$ (\dotsrule). 
    The symbols represent the errors from the computations and the lines show the
    linear fit of those points.}
    \label{fig:92_conv} 
\end{figure}
Figure \ref{fig:92_conv} shows $L_1$ norm for the $x$ and $y$ 
components of the surface gradient and the order of convergence of the temperature
$T(x,y)$ in the interfacial points as a function of the mesh size, $\Delta$. We 
find the order of convergence of the $x$ and $y$ components of the surface 
gradient to be $1.2$ and $1.4$. The order of convergence of $T(x,y)$ along the 
interface is $1.2$. This indicates that the order of convergence of the surface 
gradient is limited by the order of convergence of the initial temperature at the
interface. 

\begin{figure}[t]
\centering
   \includegraphics[width = 0.49\textwidth]{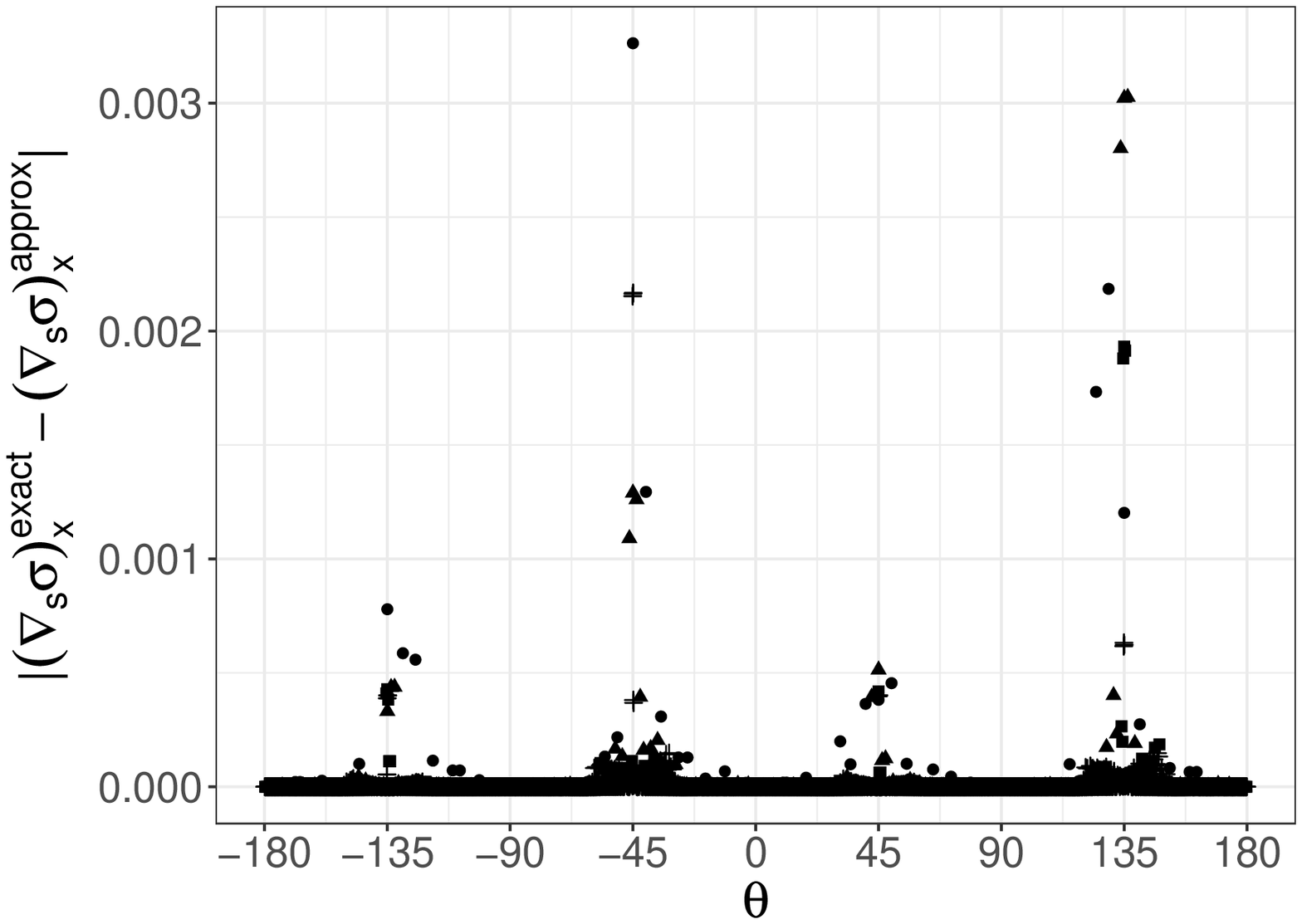}\,
   \includegraphics[width = 0.49\textwidth]{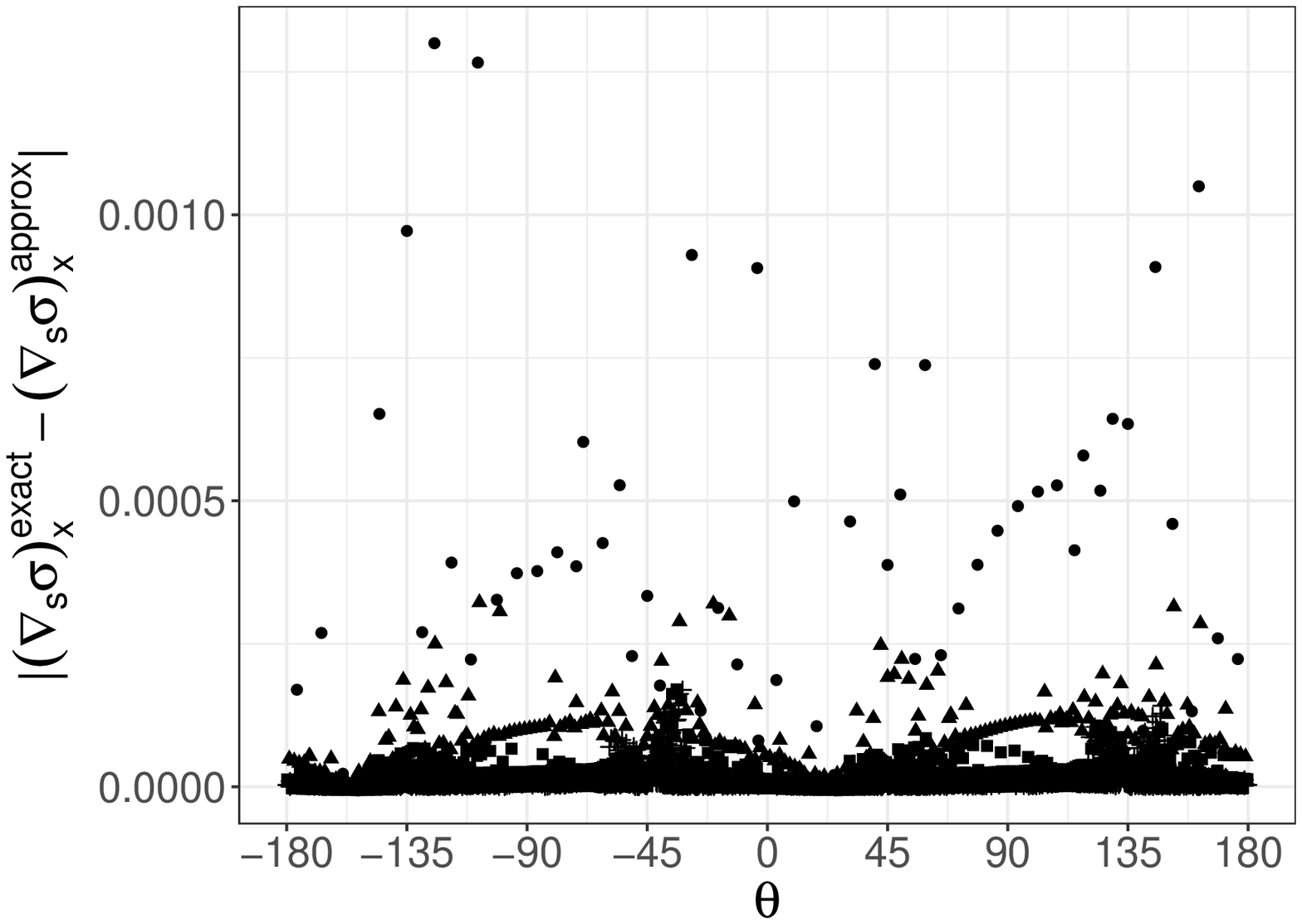}\\
\hspace{10mm}(a)\hspace{60mm}(b)   
    \caption{ Errors of the $x$ component of the surface gradient at the interfacial
    cells for $\Delta=a/8$ $({\color{black} \bullet})$, 
    $a/32$ $({\color{black} \blacktriangle})$, $a/128$ (\Smblacksquare),
    and $a/512$ $({\color{black} +})$. Initializing the temperature using 
    (a) equation  \eqref{eq:temp45}  and (b) equation \eqref{eq:temp_centroid}.
    $\theta$ is defined to be zero at the positive $x$ axis and increasing
    counterclockwise. }
    \label{fig:centroid_error}
\end{figure}

Figure \ref{fig:centroid_error}(a) shows the distribution of errors at the circular 
interface for one random realization. The largest errors appear around the lines 
$|x| = |y|$. Based on this we conclude that the lack of convergence of $E_\infty$ 
error is caused by the initialization of the temperature which changes the 
dependence on $x$ or $y$ variable at the lines $|x| = |y|$.

In order to initialize the temperature more accurately at the interface we use 
the centroid of the interface segment contained in each cell, $(x_c, y_c)$. 
Then the initial temperature is given by
\begin{equation}
\label{eq:temp_centroid}
    T(x,y) =  \Delta T  \left( x_c + y_c \right).
\end{equation}
This reduces the errors from initializing the temperature at the lines $|x|=|y|$ 
compared to using equation \eqref{eq:temp45}. 
Here, we also explore a different way of approximating interfacial temperature,
and use surface area weighted average instead of volume fraction weighted 
average (see Section \ref{sec:method}). 
The volume weighted average gives the temperature at the center of the mass of 
the fluid phase in the column, whereas the surface area weighted average gives 
the temperature at the center of the interface in the column. Hence, the latter 
is consistent with the initialization of the temperature using equation
\eqref{eq:temp_centroid}.

\begin{figure}[t]
 \centering
    \includegraphics[width = 0.6\textwidth,trim=0 0 35mm 0,clip=true]{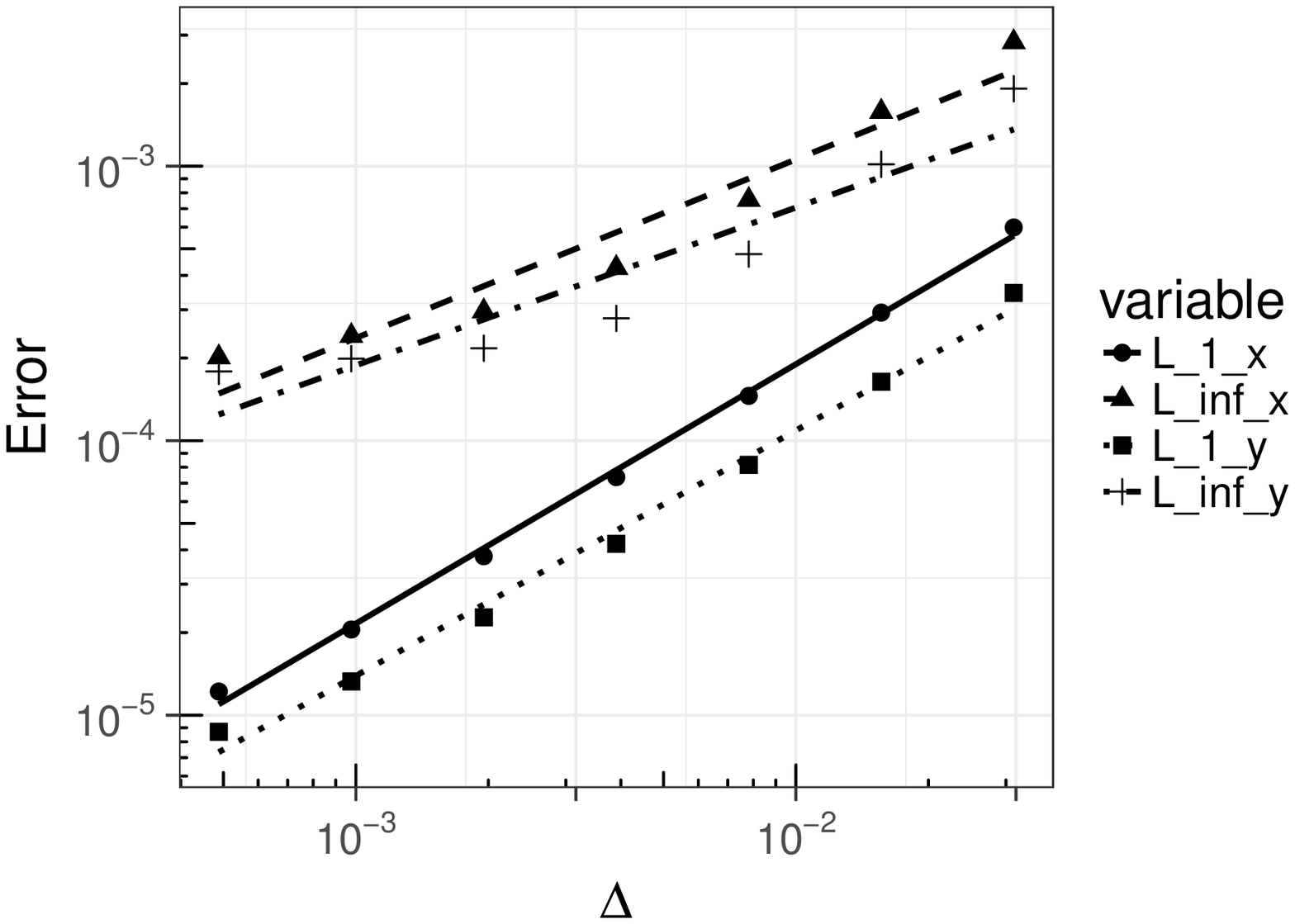}
    \caption{The computed errors for 
     $E_1((\nabla_s \sigma)_x)$ $({\color{black} \bullet})$, 
     $E_{\infty}((\nabla_s \sigma)_x)$ $({\color{black} \blacktriangle})$,
     $E_1((\nabla_s \sigma)_y)$ (\Smblacksquare), and
     $E_{\infty}((\nabla_s \sigma)_y)$ $({\color{black} +})$.  
     The order of convergence for for $(\nabla_s \sigma)_x$ is $0.94$ (\solidrule) 
     and $0.65$ (\dashedrule) for $E_1$ and $E_\infty$ errors, respectively, 
     and the order of convergence for for $(\nabla_s \sigma)_y$ is 
     $0.89$ (\dotsrule ) and $0.58$ (\dashdotrule), respectively.
    The symbols represent the errors from the computations, and the lines show the
    linear fit of those points.}
    \label{fig:errors_97} 
\end{figure}

Figure \ref{fig:centroid_error}(b) shows the errors of the $x$ component of the 
surface gradient at the interfacial cells if the temperature is initialized using 
equation \eqref{eq:temp_centroid}. The errors are still largest around $|x|=|y|$, 
however those are the usual ``weak'' spots of the height function construction.  
Figure \ref{fig:errors_97} shows the improvement in the convergence to the exact
solution using $L_1$ and $L_\infty$ norm for the $x$ and $y$ components of the surface
gradient as a function of mesh refinement. The order of convergence for $L_1$ norm
is $0.94$ and $0.89$ for the $x$ and $y$ components of the surface gradient, 
respectively, and the order of convergence for $L_\infty$ norm is $0.65$ and 
$0.58$ for the $x$ and $y$ components of the surface gradient, respectively. 
Hence, the second approach of initializing the temperature (using 
equation \eqref{eq:temp_centroid}) improves the convergence of the $L_\infty$ norm
significantly.

\subsection{Drop migration}
We further test our numerical implementation using a classical problem of the
thermocapillary drop migration (see the reviews \citep{subramanian2002,
wozniak1988}). A drop or a bubble placed in a fluid with an
imposed temperature gradient moves due to the variation in the surface tension as
a function of temperature.  
Several authors use this problem for benchmarking their numerical algorithms
for a temperature dependent surface tension \citep{Ma2011,nas2003,herrmann2008}. 
We show the comparison of our numerical results with the available work in the 
literature. We also show the comparison with the analytical solution of the drop
terminal velocity by \citet{Young1959}. 
\citet{Young1959} show that the nondimensional velocity of the drop in an unbounded
domain for an axysimmetric geometry in the limit of small Ma and Ca
numbers can be approximated as 
\begin{equation}
\label{eq:YGB}
 v^*_{ygb} = \frac{\mu_1}{\sigma_T a \Delta T} \frac{2}{(2 + k_r)(2 + 3 \mu_r)},
\end{equation}
where $\mu_1$ is the viscosity of the surrounding fluid, $\sigma_T$ is the 
(constant) gradient of the surface tension with respect to the temperature, $a$ 
is the drop/bubble radius, $\Delta T$ is the imposed temperature gradient, and 
$k_r$ and $\mu_r$ are the thermal conductivity and viscosity ratios, respectively,
for the drop/bubble compared to the surrounding fluid.

\begin{figure}[t]
\centering
 {\begin{overpic}[width = 0.45\textwidth]{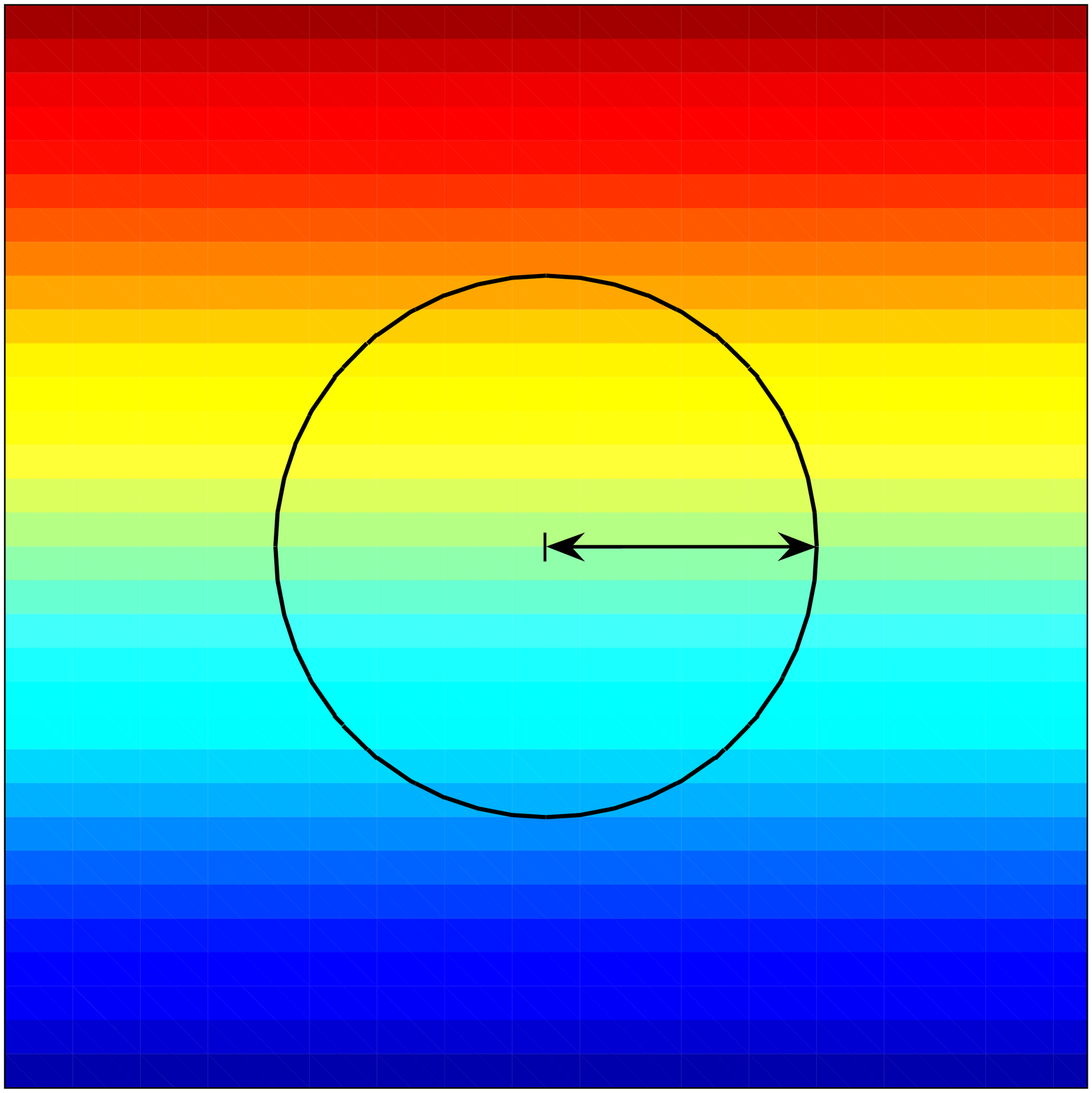}
   \put (40,3) {\color{white}$ T^* = T^*_1$}
   \put (39,93) {\color{white}$ T^* = T^*_2$}
   \put (5, 60) {$ \rho_1, \mu_1$}
   \put (5, 50) {$ k_1, C_{p_1} $}
   \put (60,45 ) {$ a $ }
  \end{overpic}
}
 \caption{ The initial setup of the drop migration problem. The color represents 
 the linear temperature distribution with imposed temperatures $T^*_1$ and $T^*_2$ at the horizontal boundaries.}
 \label{fig:temp_setup}
\end{figure}

Figure~\ref{fig:temp_setup} shows the considered setup: 
a drop or a bubble of radius $a$ is placed in an ambient fluid, with a linear temperature 
gradient  imposed in the $y$ direction. The 
temperature at the top and the bottom boundaries is set to constant values and a
zero heat flux boundary condition is imposed at the left and right boundaries. The
boundary conditions for the flow are no-slip and no penetration at the top and 
bottom boundaries and symmetry at the left and right boundaries.

We solve equations \eqref{eq:nondim_NS} and \eqref{eq:advec_diff_nd1} and consider the
following scales
$$
 p_0 = \frac{\mu_1 U_0}{a}, \,\,\,\,\, t_r = \frac{a}{U_0}, \,\,\,\,\,  U_0 = \frac{\sigma_T a \Delta T}{\mu_1}\,\,\,\,\, T_0 = a \Delta T,
$$
where the subscript $1$ denotes the properties of the ambient fluid. The surface
tension at the interface between the drop and the ambient fluid is assumed to 
depend linearly on temperature as given by equation \eqref{eq:sigma_nd}, which 
rescaled using the scales above yields
\begin{equation}
\sigma^* = 
1  +  \text{Ca}\left( T^* - T_R^* \right).
\end{equation}
Next we present the comparison of our results with the available studies in the literature.

\begin{figure}[t]
\centering
  \includegraphics[width = 0.45\textwidth]{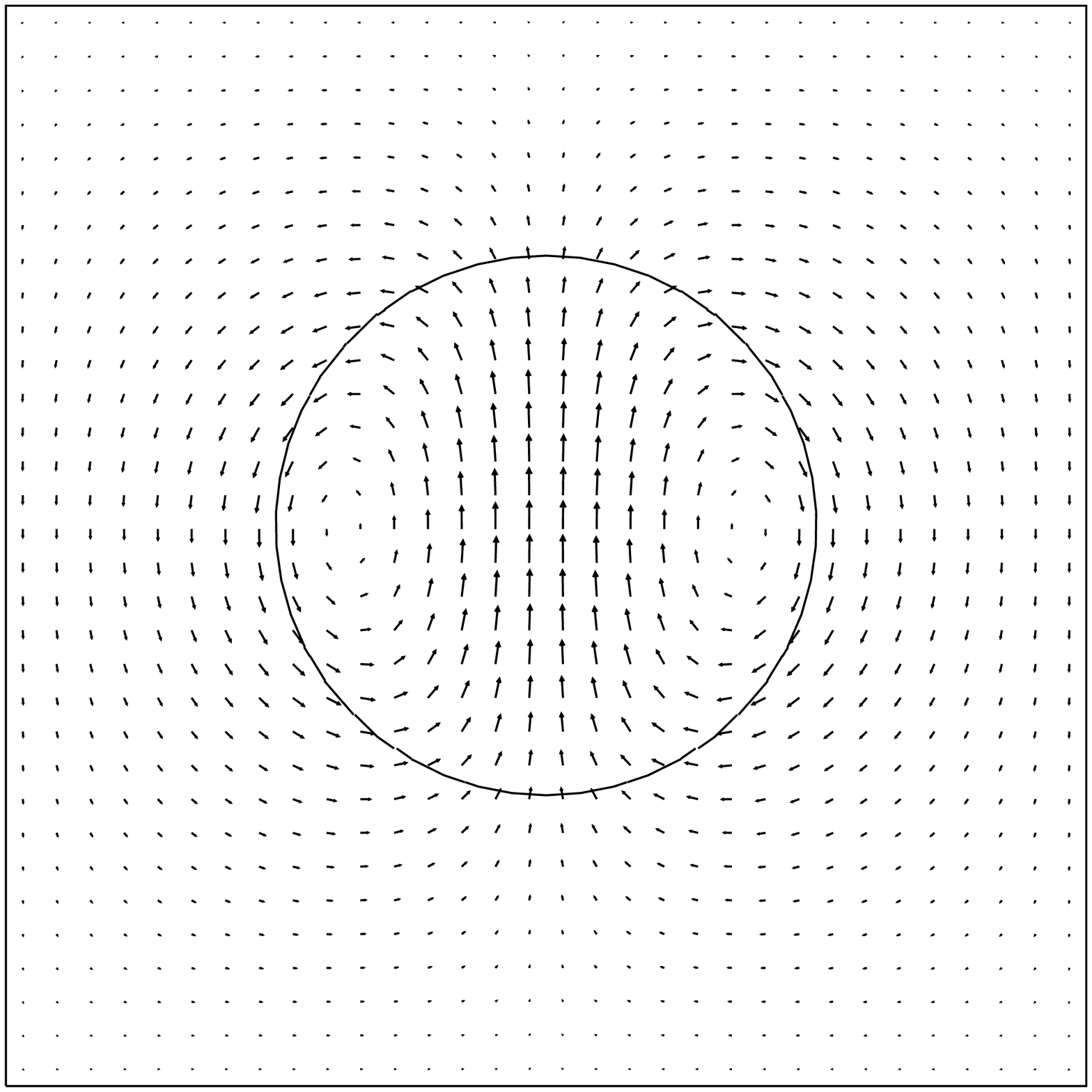}
  \caption{The velocity field in the drop and the surrounding fluid.}
  \label{fig:velocity_field} 
\end{figure}

We start by comparing our results with the ones by \citet{Ma2011}. The material 
properties are 
$\rho_1 = 500 \frac{kg}{m^3},\, \mu_1 = 0.024\, Pa \cdot s,\, 
\sigma_0 = 10^{-2} \frac{N}{m},\, \sigma_T = 2 \times 10^{-3} \frac{N}{m K},\,
k_1 = 2.4\times 10^{-6} \frac{W}{m K}, \, C_{p_1} = 10^{-4} \frac{J}{kg K}, \,
\Delta T = 200 \frac{K}{m},  \, T_2 = 290\, K,\, a = 1.44\times 10^{-3}\, m.$
The ratio of the material properties between the ambient fluid and the drop
is $2$. These physical properties give nondimensional parameters 
$\text{Re} = \text{Ma} = 0.72, \, \text{Ca} = 0.0576$,  and the velocity scale 
$U_0 = 0.024\, \frac{m}{s}$. Figure \ref{fig:velocity_field} shows the computed 
velocity field in the drop and the surrounding fluid. The surface tension gradient 
drives the flow from low surface tension region (top) to high surface tension 
region (bottom). This creates the flow inside the drop and as a result the drop 
moves in the positive $y$ direction. The drop velocity is computed using 
the following definition of the centroid velocity
$$
 v_c^* = \frac{ \sum\limits_{i,j} v^*_{i,j} \chi_{i,j} \, (\Delta_{i,j}^{*})^2}{\sum\limits_{i,j} \chi_{ij} \, (\Delta_{i,j}^{*})^2}
$$
where $v^*_{i,j}$ is the $y$ component of the cell-center velocity.

\begin{figure}[t]
\centering
 \includegraphics[width = 0.75\linewidth]{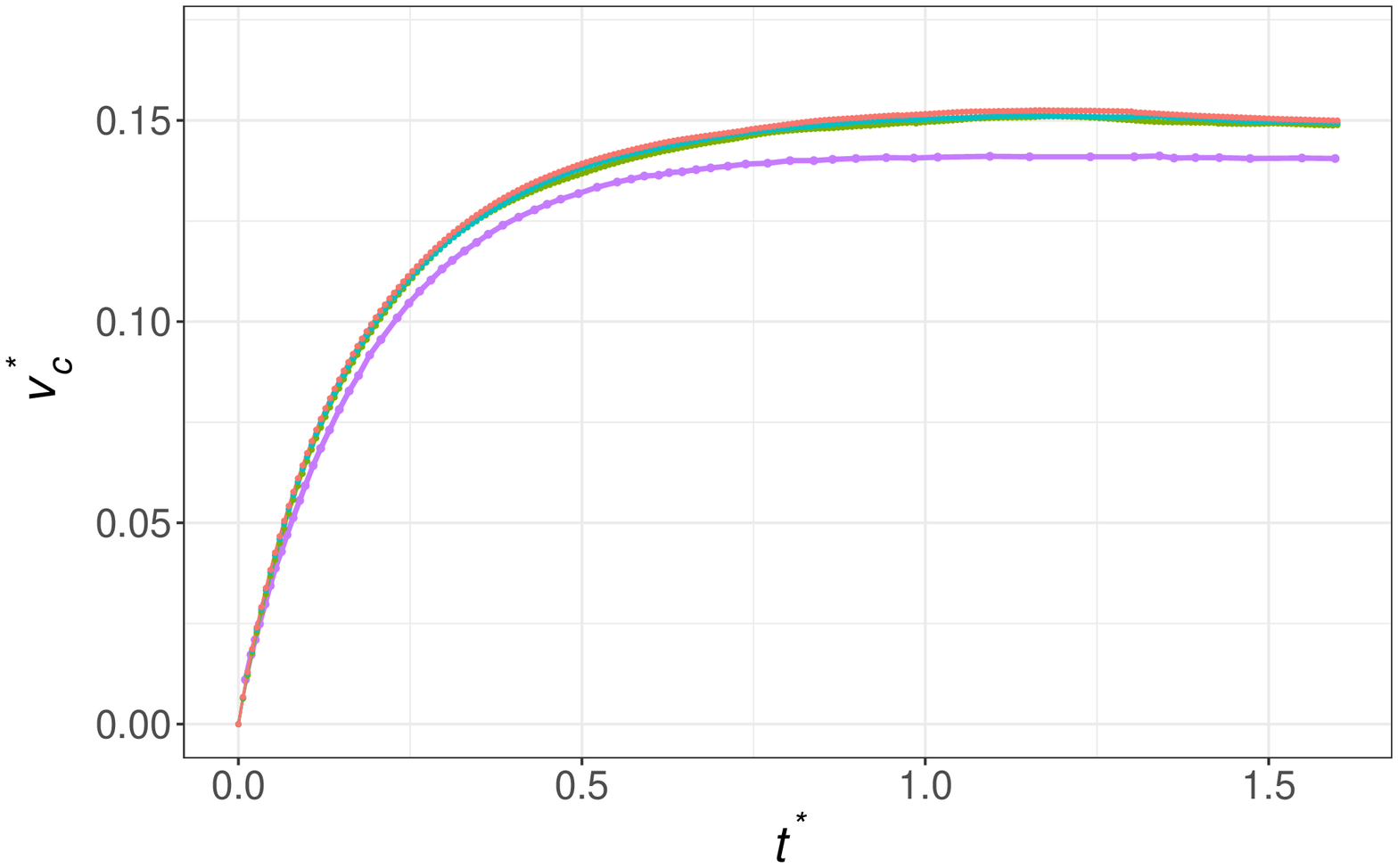}
 \caption{Drop migration velocity for $\text{Re} = \text{Ma} = 0.72$ and 
 $\text{Ca} = 0.0576$ for $\Delta=$ $1/64$ ({\color{red}\solidrule}), $1/256$
 ({\color{green_darker}\solidrule}), and $1/256$ ({\color{cyan}\solidrule})
 compared with the result given in \cite{Ma2011} ({\color{purple}\solidrule}) for
 2D simulations. }
   \label{fig:Bothe}  
\end{figure}

Figure \ref{fig:Bothe} shows the computed velocity of the drop compared to the
results in \citet{Ma2011}. In this test case, the computational domain is a square
box with a side length equal to  four times the drop radius;  the drop is 
initially placed at the center of the  domain. As shown, our results are in 
agreement with the previously obtained simulations in \citet{Ma2011}. 

\begin{figure}[t]
\centering
  \includegraphics[width = 0.75\linewidth]{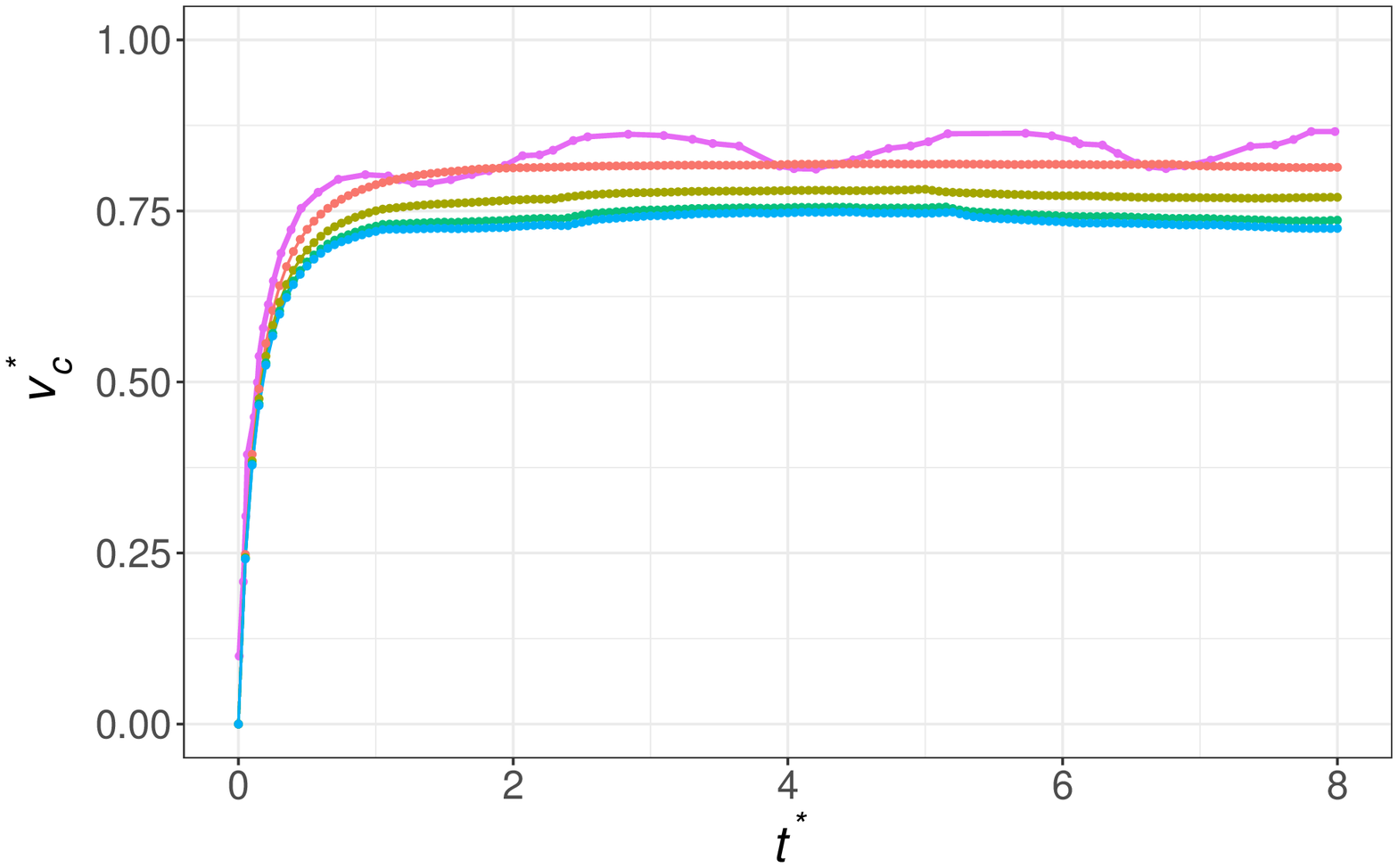}
  \caption{Convergence of the migration velocity as a function of the time step 
  for $\Delta t=$ $10^{-4}$ ({\color{red}\solidrule}), $5\times10^{-6}$
  ({\color{yellow}\solidrule}), $10^{-5}$ ({\color{green_darker}\solidrule}), 
  and $5\times10^{-6}$ ({\color{cyan}\solidrule}) compared with the results in 
  the results\cite{herrmann2008} ({\color{purple}\solidrule}) for 2D simulation, 
  for $\text{Re} =  \text{Ca} = 0.066$ and $\text{Ma} = 0 $. The velocity is 
  rescaled by $v^*_{ygb}$.}
  \label{fig:herrmann2D} 
\end{figure}

Next, we carry out another comparison for smaller value of Re and Ca numbers and
when $\text{Ma} = 0$; we choose $\text{Re} = \text{Ca} = 0.066$ in accordance with the 
results presented in \citep{herrmann2008} for the VOF method. The computational box is a 
rectangle of size $10a\times15a$. The density of the ambient fluid is set to $\rho_1^* = 
0.2$, and viscosity is $\mu_1^* = 0.1$. The ratio of the physical properties of the drop 
to the ambient fluid is set to $1$. 
The surface tension is $\sigma_0^* = 0.1$ at the reference temperature $T_R^* = 0$, with 
$\sigma_T^* = -0.1$. The temperature gradient is set to $\Delta T^* = 0.1\bar{3}$,
which is fixed by setting $T_1^*$ = 0 and $T_2^* = 1$. 
The drop is initially centered horizontally at $3a$ from the bottom wall. Figure 
\ref{fig:herrmann2D} shows the comparison of our method with the results in 
\citep{herrmann2008}, along with temporal convergence of our method.
Compared to the results by \citet{herrmann2008}, our results do not exhibit 
oscillations, which agrees with the asymptotic solution of constant rise velocity.
Another difference is that our terminal velocity converges to a smaller value with
decreasing time step. However, the timestep used in the results of 
\citet{herrmann2008} is not specified in their paper.

\begin{figure}[t]
\centering
  \includegraphics[width = 0.75\linewidth]{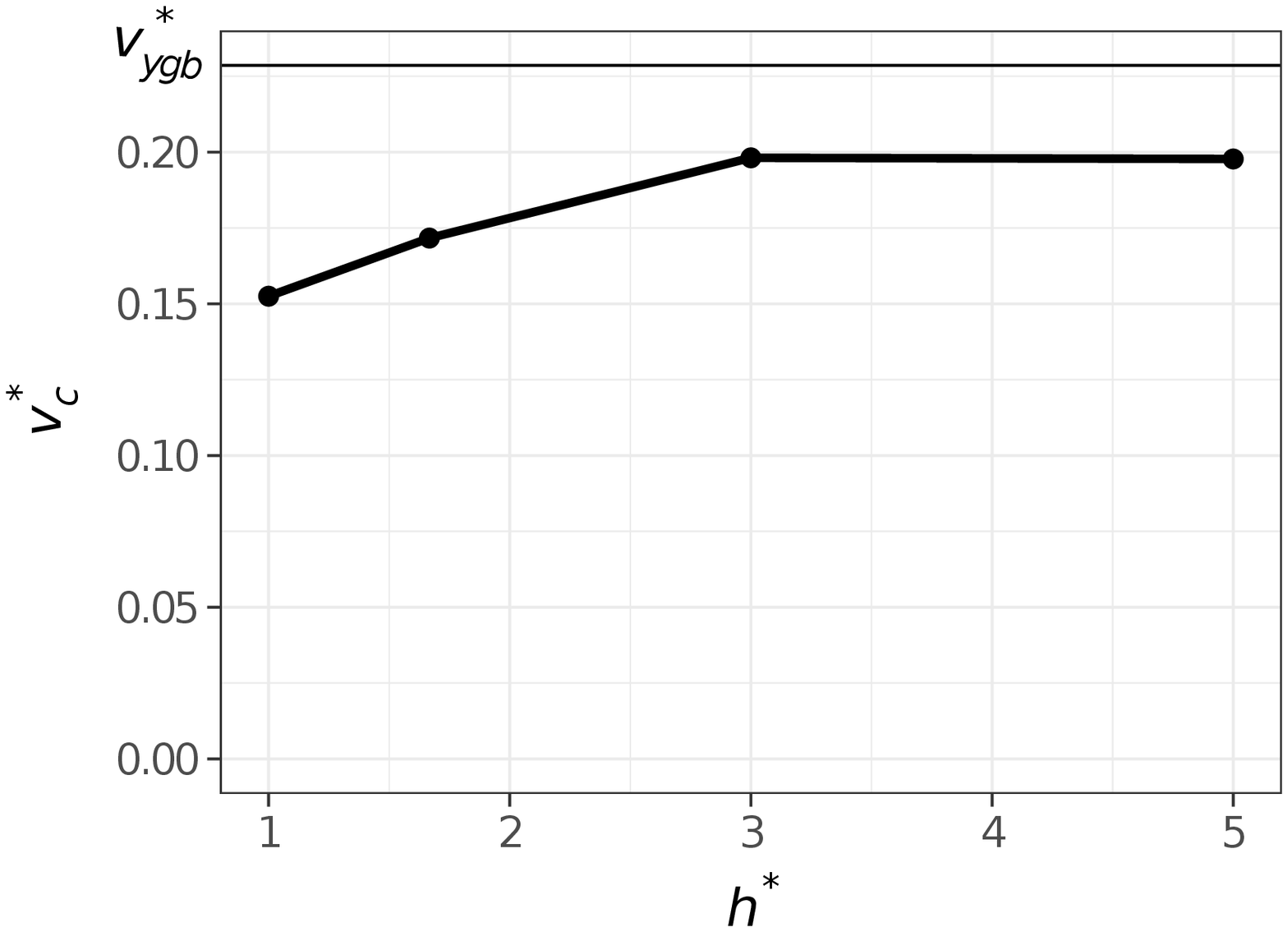}
  \caption{Convergence of the terminal velocity with increased distance from the
  wall, $h^*$, for $\text{Re} = \text{Ma} = 2.5\times10^{-3}$ and 
  $\text{Ca} = 1.25 \times 10^{-3} $ for 2D simulations.}
  \label{fig:wall_dist}
\end{figure}

We also test the convergence to the analytical solution obtained in the limit of  
Ma and Re approaching zero and in the unbounded domain, where the terminal 
velocity approaches $v^*_{ygb}$ value given in equation \eqref{eq:YGB}. 
Figure \ref{fig:wall_dist} shows the terminal velocity of a droplet for a 2D
simulation as a function of a distance from the wall for 
$\text{Re} = \text{Ma} = 2.5\times10^{-3}$ and $\text{Ca} = 1.25\times10^{-3} $.
The terminal velocity converges to a value lower than $v^*_{ygb}$ due to the 
difference in the geometry. We next show that our 3D result 
in fact converges to this analytical solution.

\begin{figure}[t]
\centering
 \includegraphics[width = 0.75\linewidth, trim=0 0 35mm 0, clip=true]{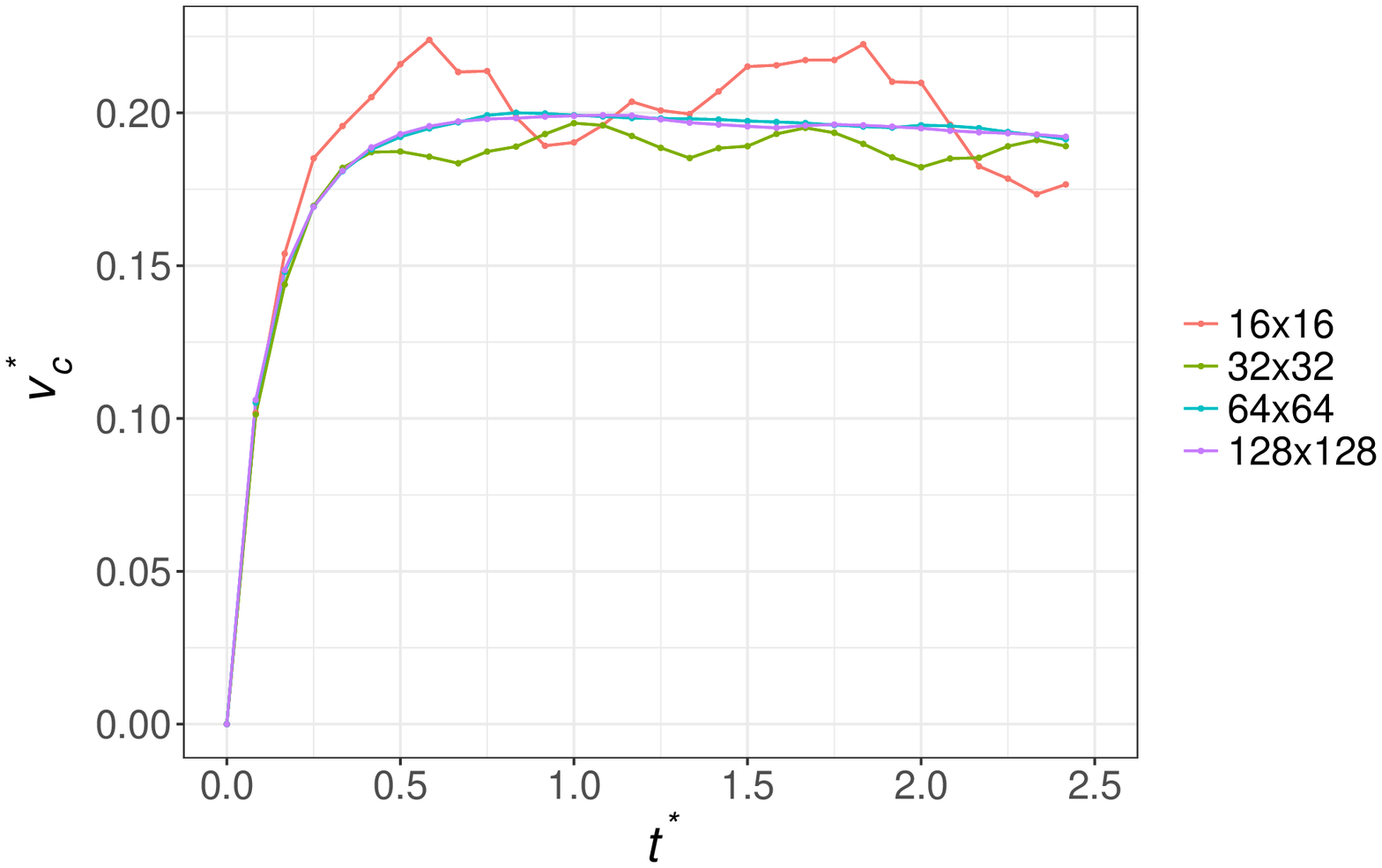}
  \caption{Convergence of the migration velocity in a 3D simulation with mesh refinement
   for $\Delta =$ $1/16$ ({\color{red} \solidrule}), $1/32$ ({\color{green}\solidrule}), 
   and $1/64$ ({\color{cyan}\solidrule}), and $1/128$ ({\color{purple}\solidrule});
   $\text{Re} = \text{Ma} = 0.72$ and $\text{Ca} = 0.0576 $.}
  \label{fig:3D_Bothe}
\end{figure}

\begin{figure}[t]
\centering
  \includegraphics[width = 0.75\linewidth]{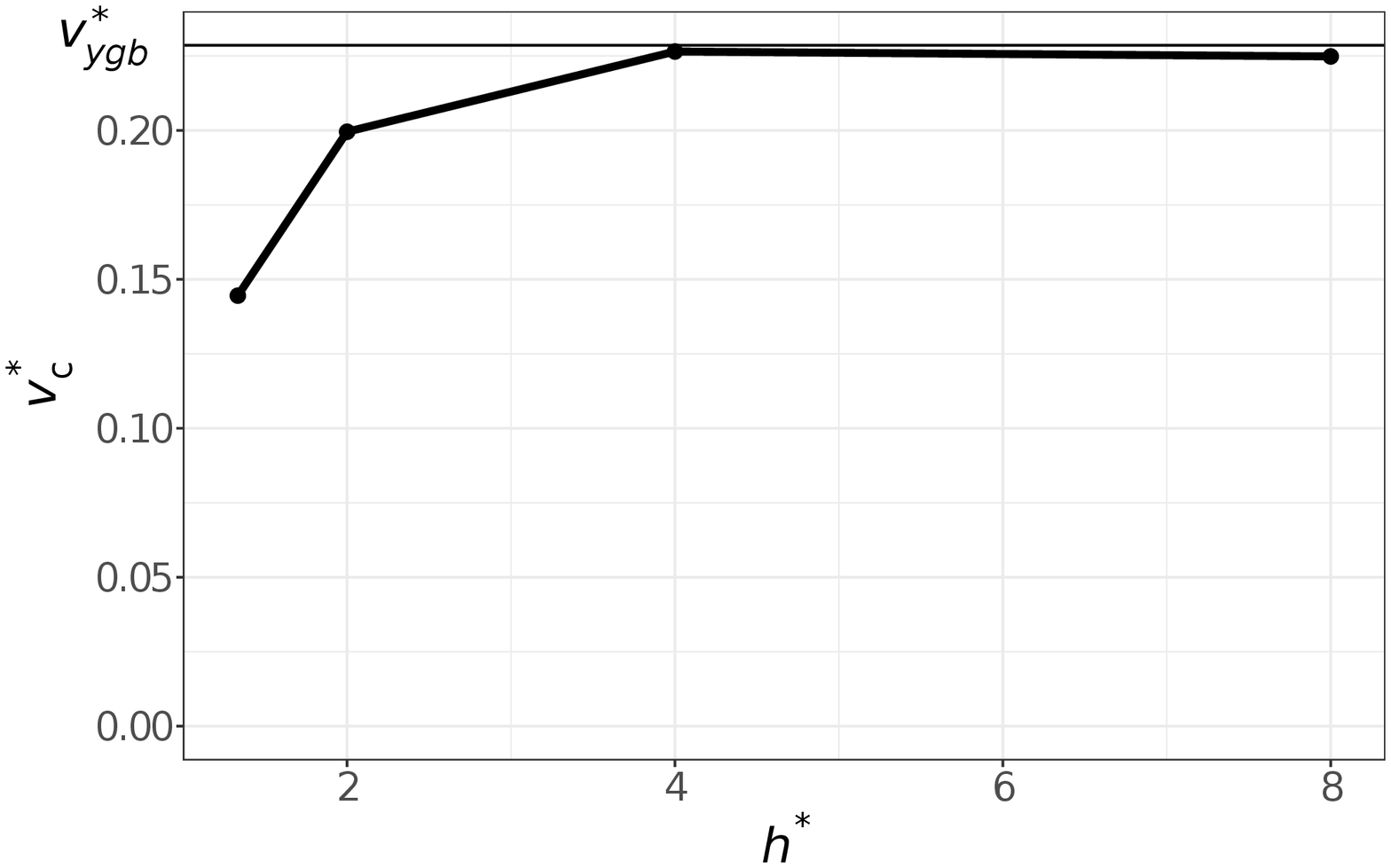}
  \caption{Convergence of the terminal velocity with increased distance from the 
  wall, $h^*$, for $\text{Re} = \text{Ma} = 2.5\times10^{-3}$ and $\text{Ca} = 10^{-3}$
  for 3D simulations. }
  \label{fig:3D_wall_dist}
\end{figure}

We perform similar tests for the 3D simulations. Figure \ref{fig:3D_Bothe} shows 
the migration velocity for $\text{Re} = \text{Ma} = 0.72$ and $\text{Ca} = 0.0576 $. 
The parameters and the domain size are equivalent to the simulation results shown 
in Figure \ref{fig:Bothe}. The results also show that the oscillations in the 
computed velocity decay with mesh refinement and the terminal velocity converges 
to a higher value compared to the 2D case. However, this value is still smaller than 
$v^*_{ygb}$ due to the small domain size and relatively large $\text{Re}$ and 
$\text{Ma}$ numbers. 
Figure \ref{fig:3D_wall_dist} shows the terminal velocity 
of a droplet for a 3D simulation as a function of a distance from the wall for
$\text{Re} = \text{Ma} = 2.5\times10^{-3}$ and $\text{Ca} = 10^{-3}$. 
As shown, the terminal velocity approaches the analytical value $v^*_{ygb}$.

In this section we have shown the comparison of our method with existing literature
and with a limiting analytical solution. Our method shows the convergence to the 
analytical value of the terminal velocity.   Furthermore, the trend of the solution 
as well as the time needed to reach the terminal velocity are consistent with the 
previously reported results.

\subsection{Coalescence and non-coalescence of sessile drops}
\label{sec:merge}

Next we demonstrate the performance of our numerical methods through an example 
of the coalescence behavior of sessile drops with different surface tension.   
We model the experiments of the coalescence of two droplets with different alcohol
concentrations by Karpitschka
et.~al.~\citep{Karpitschka2010,Karpitschka2012,KarpitschkaJFM}. In their experimental
study, they show three coalescence regimes depending on the surface tension 
difference between the two droplets: immediate coalescence, delayed coalescence, 
and non-coalescence. 
They identify a key parameter that governs the transition between the delayed and
non-coalescence regimes: specific Marangoni number M $=3\Delta \sigma / (2 {\bar \sigma}\theta^2)$ 
\citep{KarpitschkaJFM}, 
where $\Delta \sigma$ is the difference in the surface tension between the two 
drops and ${\bar \sigma}$ is the average of the surface tension of two drops. They 
determine a threshold Marangoni number M$_t\approx2\pm0.2$ experimentally for
the transition between the delayed coalescence and non-coalescence regimes.

Here we show that our 
numerical simulations also reveal the three regimes in agreement with the 
experimental observations in \citep{Karpitschka2010,Karpitschka2012}. From the 
numerical simulation point of view, this problem involves a level of difficulty:
unlike temperature, the concentration should remain strictly confined to the liquid phase 
and should not leak out to the ambient phase. To avoid this difficulty, we combine
our variable surface tension methodology with the numerical technique already 
implemented in the original version of {\sc gerris}\citep{Gerris} which prevents 
the concentration from leaking out of the liquid domain into the ambient phase.

\begin{figure}[t]
\centering
  \includegraphics[width = 0.8\linewidth]{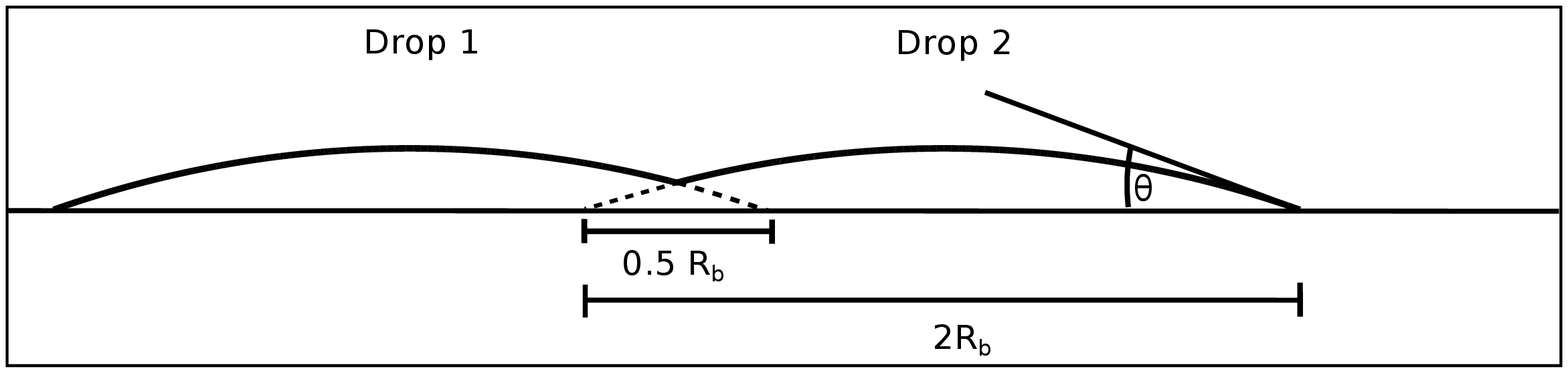}
    \caption{A schematic of the drop coalescence problem. }
    \label{fig:initial_coales} 
\end{figure}

\begin{figure}[t]
\centering
  \includegraphics[width = 0.49\linewidth]{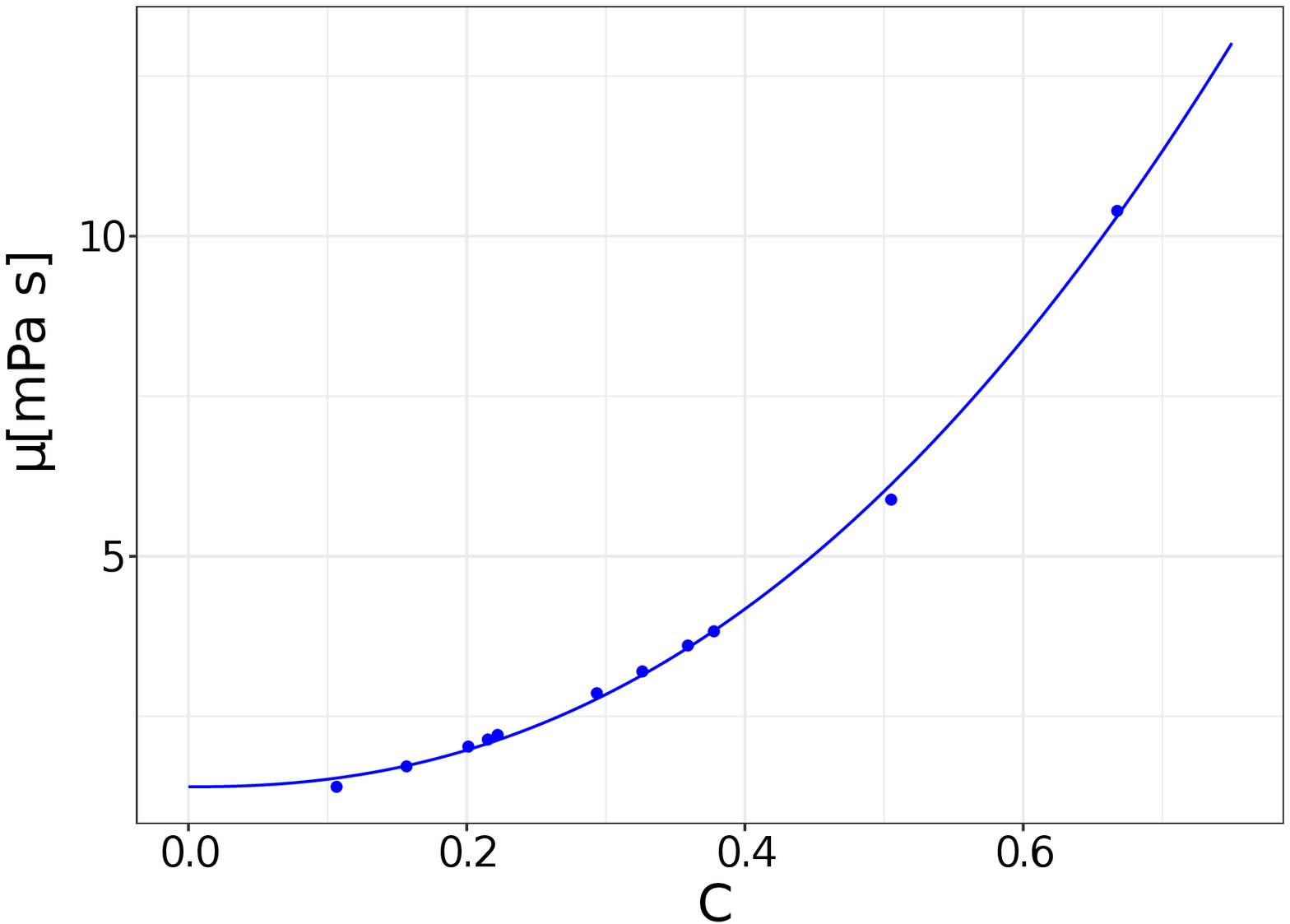}
  \includegraphics[width = 0.49\linewidth]{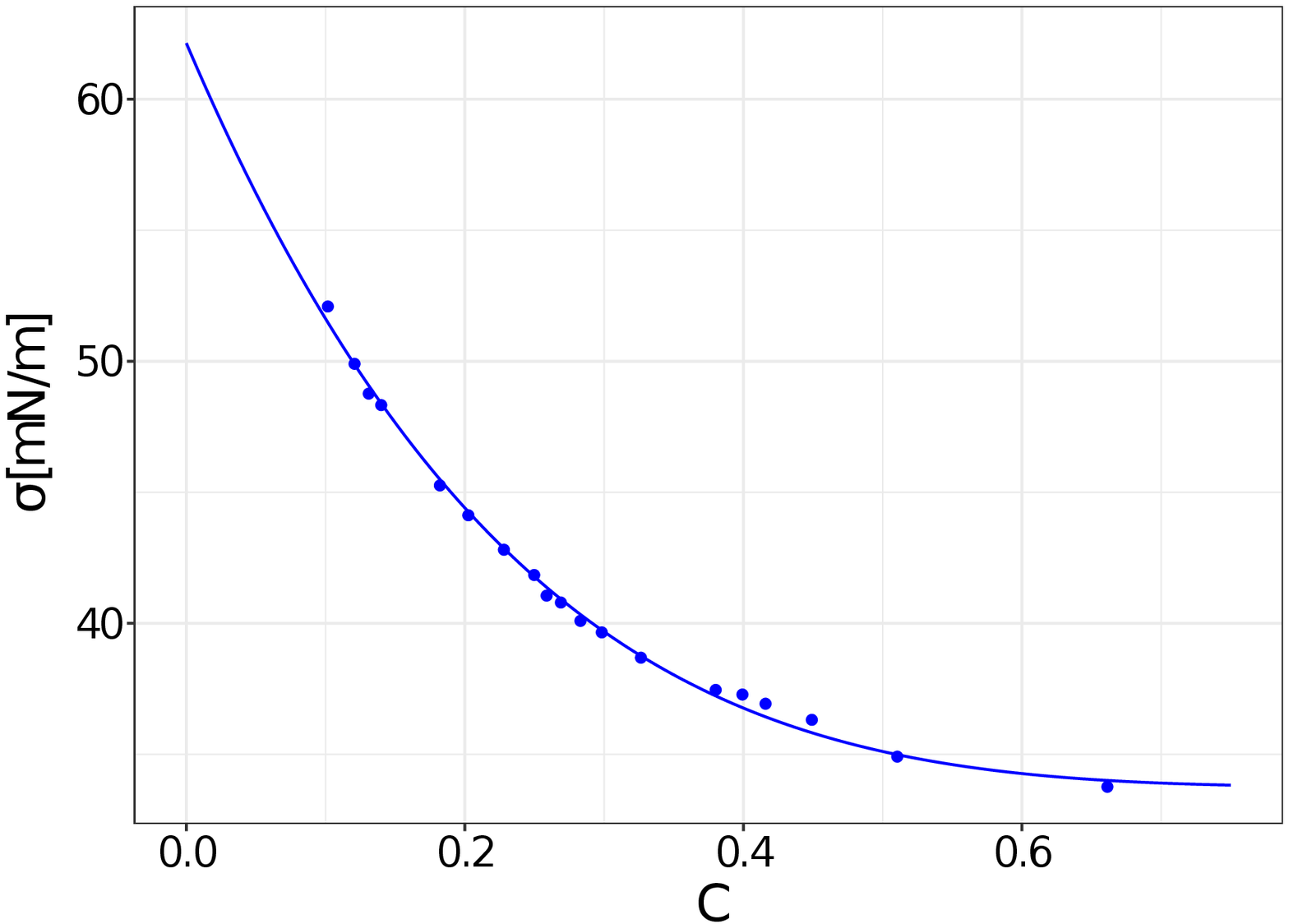}\\
  \hspace{7mm}(a)\hspace{55mm}(b)
    \caption{Viscosity (a) and  surface tension (a) of the mixture of 1,2-Butanediol 
    and water as a function of alcohol concentration. Points represent the data from
    \citet{Karpitschka2010}, and the lines show the fit of the points.}
    \label{fig:butanediol} 
\end{figure}

We model the 2D problem since the dominant flow dynamics in the problem is
in the region connecting the two droplets, where the surface tension gradient is 
the strongest, and in this region we can ignore the out of plane curvatures. 
Initially, the drops have the shape of a circular segment with the base radius $R_b$
and a contact angle $\theta$, and are connected by an overlap of $0.25R_b$
(see Figure \ref{fig:initial_coales}). The drops have equal base radius $R_b$ and
we assume that their densities are equal. The viscosity depends on the alcohol
concentration $C$, where we use a nonlinear fit to the data given in
\citep{Karpitschka2010} of the form 
\begin{equation}
    \mu \left( C \right) = \mu_1 + a_\mu \left(\mu_2 - \mu_1 \right)
    \left( 1 - C \right)^{n_\mu},
\end{equation}
shown in Figure \ref{fig:butanediol}(a). Drops are composed of the mixture of the
1,2-Butanediol and water, but they differ in the concentrations of alcohol. 
Figure \ref{fig:butanediol}(b) shows the surface tension dependence on the 
concentration of 1,2-Butanediol in water. Similarly as for the viscosity, we fit 
this data to a function of the form
\begin{equation}
    \sigma \left( C \right) = \sigma_1 + a_\sigma \left(\sigma_2 - \sigma_1 \right)
    \left( 1 - C \right)^{n_\sigma},
\end{equation}
Parameters $a$ and $n$ are determined from the fit. 

\begin{figure}[t]
\centering
  \includegraphics[width = 1\textwidth,trim={0 0 1.81in 0},clip]{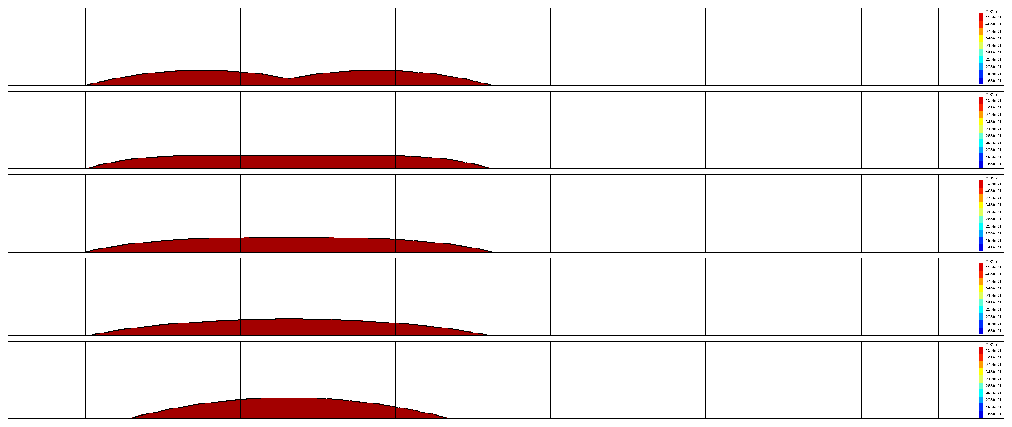}
    \caption{ Evolution of two drops with equal alcohol concentration, i.e.~no surface 
    tension difference, at times $t = 0\, s$, $t = 0.02\, s$, $t = 0.04\, s$, 
    $t = 0.1\, s$ and $t = 1\, s$ from top to bottom. The color shows the 
    concentration of alcohol. Each box is equivalent to $2\, mm$.}
    \label{fig:immediate}
\end{figure}

We first show a simulation of two drops, with equal surface tension.
We consider the case where the concentration of alcohol is $45$\%, and the base 
radii of the circular segments are both $R_b=3\, mm$. Along with a no-slip boundary 
condition at the substrate, we also impose a $\theta=15^\circ$ contact angle. 
For the contact angle implementation in {\sc gerris} and related numerical discussion
the reader is referred to \citep{Afkhami2009a,Afkhami2009}. Figure \ref{fig:immediate} 
shows the evolution of the interface at different times. The droplets coalesce 
immediately, fully merge after $0.1\, s$, and assume an equilibrium shape of 
one large circular segment at a later time. The color represents the concentration 
of alcohol, which is contained inside of the fluid and zero in the surrounding.

\begin{figure}[t]
\centering
  \includegraphics[width = 0.9\textwidth,trim={0 0 0.23in 0},clip]{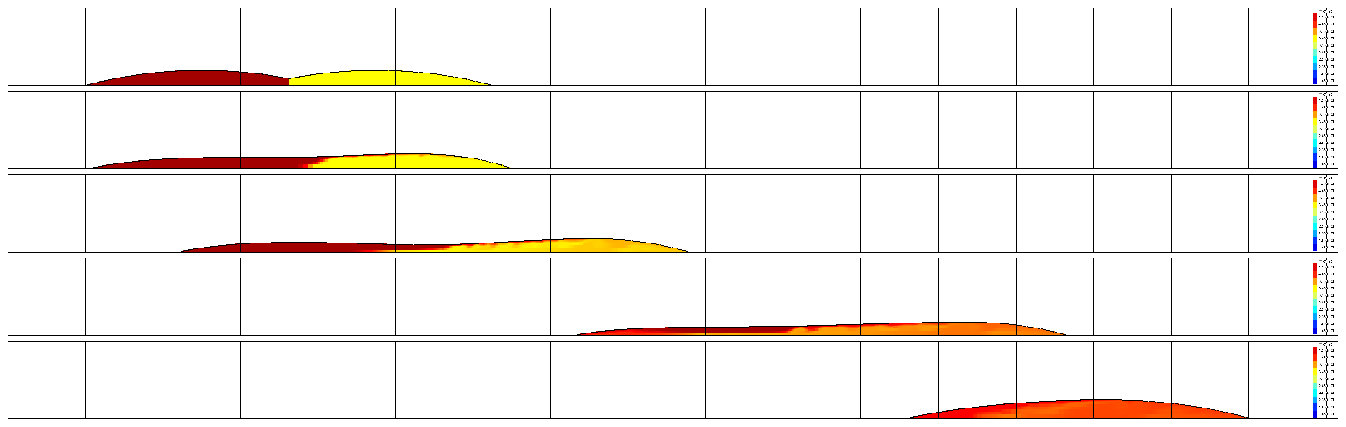}
  \includegraphics[width = 0.09\textwidth,trim={0 2.28in 0 0},clip]{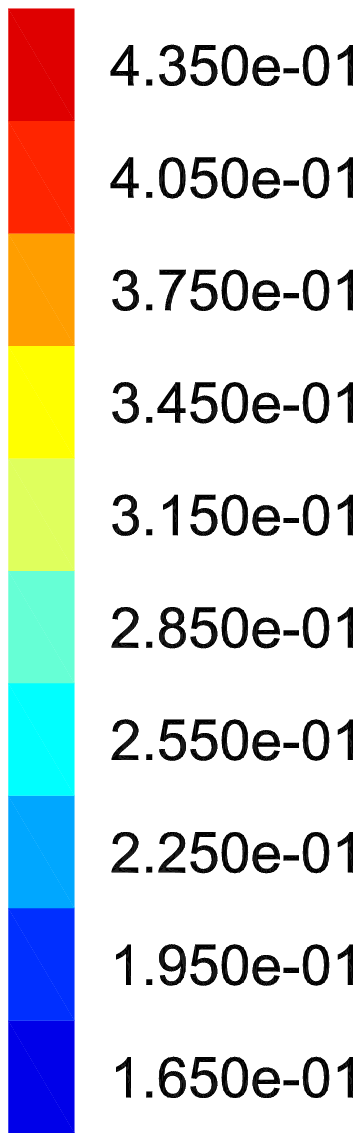}
    \caption{Evolution of two drops with small surface tension difference at 
    times $t = 0\, s$, $t = 0.1\, s$, $t = 1\, s$, $t = 5\, s$, and $t = 10\, s$
    from top to bottom.  The color shows the concentration of alcohol. Each box 
    is equivalent to $2\, mm$.}
    \label{fig:intermed} 
\end{figure}

\begin{figure}[t]
\centering
  \includegraphics[width = 0.275\textwidth]{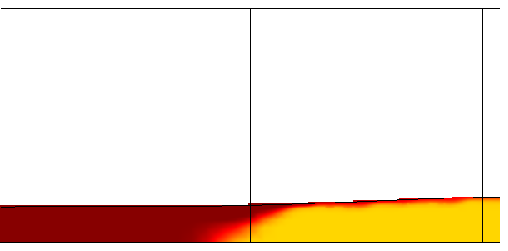}
  \includegraphics[width = 0.275\textwidth]{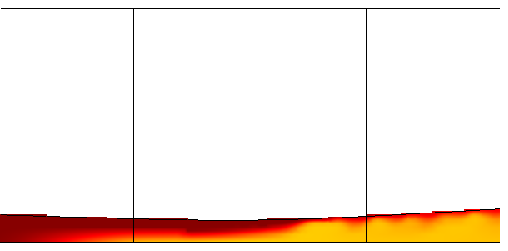}
  \includegraphics[width = 0.275\textwidth]{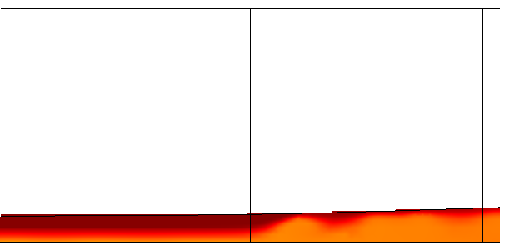}
  \includegraphics[width = 0.09\textwidth,trim={0 2.28in 0 0},clip]{scale.eps}
    \caption{Closeup of the neck region between the two drops shown in Figure
    \ref{fig:intermed}, at times $t = 0.1,\ s$, $t = 1\, s$, and $t = 5,\ s$ 
    from left to right. The color shows the concentration of alcohol.}
    \label{fig:intermed_zoom} 
\end{figure}

Next we examine the case where $\text{M}\approx1.2 < \text{M}_t$.
Figure \ref{fig:intermed} shows the simulations of this
intermediate regime where droplets coalescence is delayed. Here, we set drop $1$
to $45$\% and drop $2$ to $35$\% of alcohol. 
The connected drops move toward higher surface tension due to the Marangoni induced 
flow until the concentrations are mixed, resulting in a smaller gradient in the 
surface tension. 
Figure \ref{fig:intermed_zoom} shows closeup images of the neck region between the
two drops corresponding to the three panels in the middle shown in Figure
\ref{fig:intermed}. In this figure, we show
the flow mixing dynamics which leads to the decrease of the surface tension difference 
in the neck region, resulting in a consequent full coalescence of the two drops.

\begin{figure}[t]
\centering
  \includegraphics[width = 0.9\textwidth,trim={0 0 0.23in 0},clip]{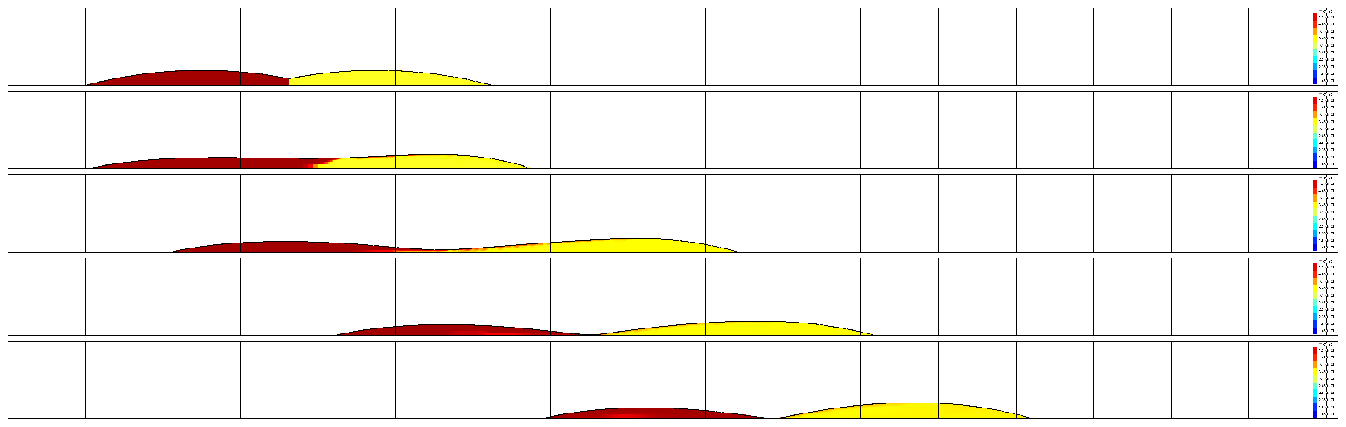}
    \includegraphics[width = 0.09\textwidth,trim={0 2.28in 0 0},clip]{scale.eps}
    \caption{Non-coalescence of drops at times $t = 0 \,s$, $t = 0.1\, s$, $t = 1\, s$,
    $t = 2\, s$, and $t = 6\, s$ from top to bottom. 
    The color shows the concentration of alcohol. Each box is equivalent to $2\, mm$.}
    \label{fig:nocoal} 
\end{figure}

\begin{figure}[t]
\centering
  \includegraphics[width = 0.275\textwidth]{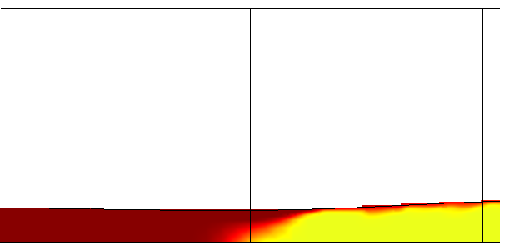}
  \includegraphics[width = 0.275\textwidth]{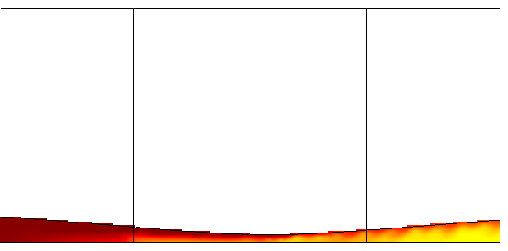}
  \includegraphics[width = 0.275\textwidth]{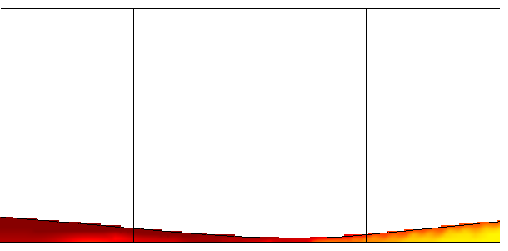}
  \includegraphics[width = 0.09\textwidth,trim={0 2.28in 0 0},clip]{scale.eps}
    \caption{Closeup of the neck region between the two drops shown in Figure \ref{fig:nocoal}, 
    at times $t = 0.1\, s$, $t = 1\, s$, and $t = 2\, s$ from left to right.  
    The color shows the concentration of alcohol.}
    \label{fig:nocoal_zoom} 
\end{figure}

Next we consider a case in the non-coalescence regime. We set drop $1$ 
to $45$\% and drop $2$ to $33$\% of alcohol. Figure \ref{fig:nocoal} shows the 
simulation results for $\text{M} \approx1.8\approx \text{M}_t$. In this case, the 
Marangoni induced flow initially pushes the fluid from drop $1$ towards drop $2$. 
However, this results in the thinning of the connecting neck between the drops 
(at $t = 1\,s$), and the fluid cannot pass from drop $1$ to drop $2$ anymore. 
Figure \ref{fig:nocoal_zoom} shows closeup images of the neck region between the 
two drops corresponding to the middle three panels shown in Figure \ref{fig:nocoal}. 
Compared to the previous case where droplets coalescence is delayed (M $\approx1.2$), 
the behavior of the mixing of the fluids in the neck region is prevented by the
thinning of the neck. Hence these droplets do not coalesce, but instead they move 
together with a constant velocity $u_d$ on the substrate in the direction of the 
higher surface gradient.

\begin{figure}[htb]
\centering
\begin{subfigure}{0.5\textwidth}
  \includegraphics[width = \textwidth,trim={0 0 2.1in 0},clip]{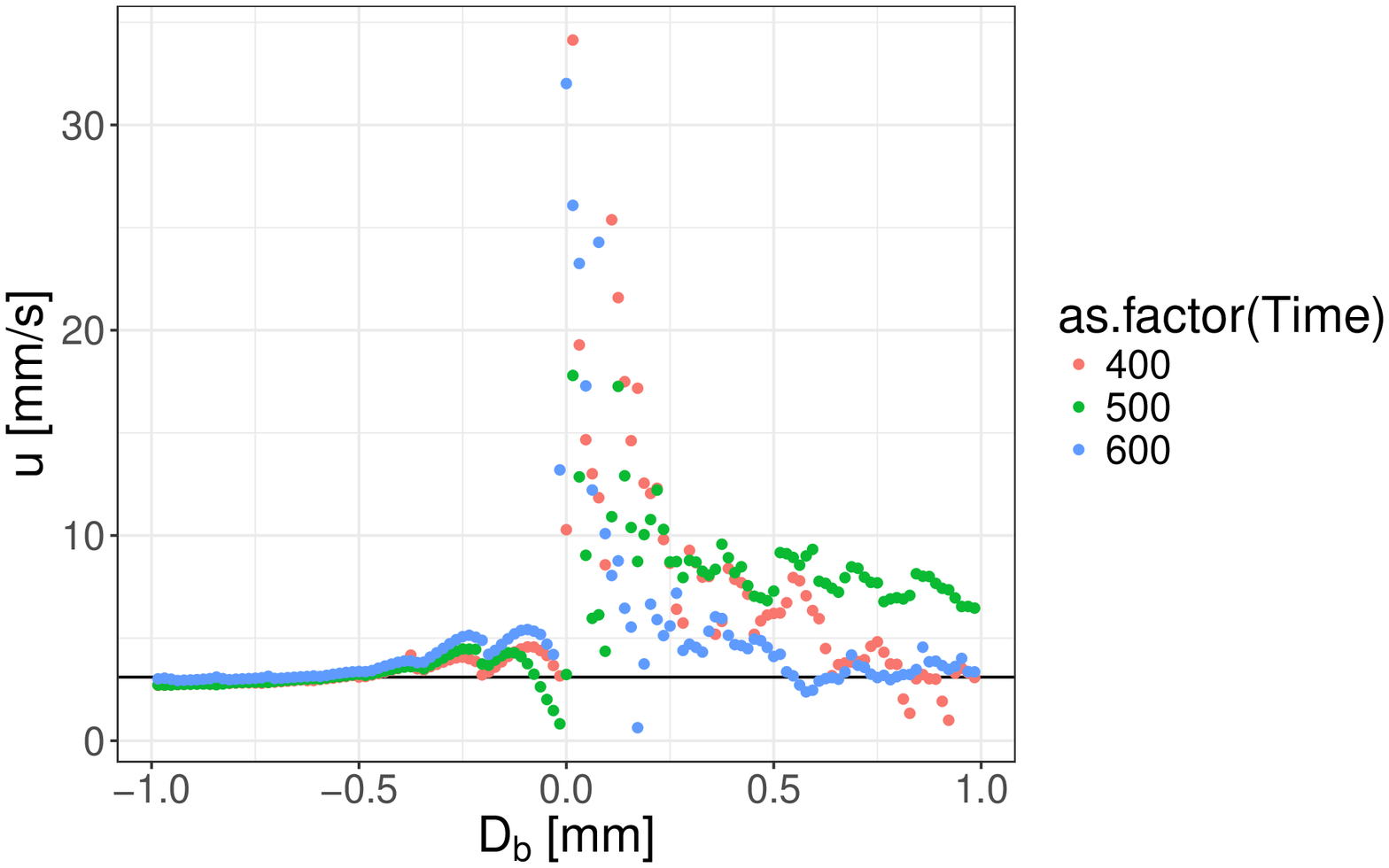}
  \caption{}
\end{subfigure}
\begin{subfigure}{0.5\textwidth}
  \includegraphics[width = \textwidth,trim={0 0 2.1in 0},clip]{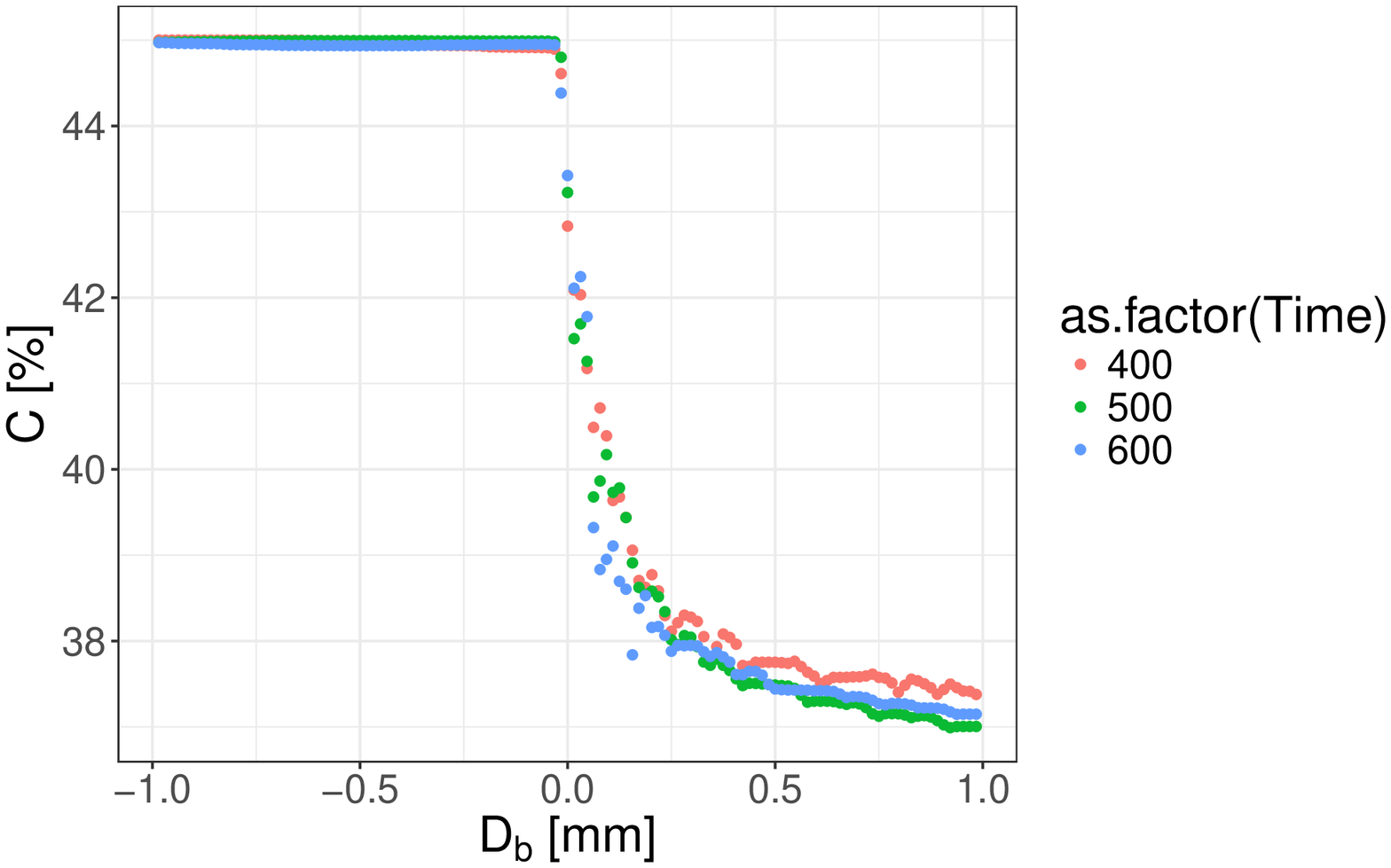}
  \caption{}
\end{subfigure}
    \caption{The $x$ component of the velocity field at the interface interfacial 
    points (a) and the concentration of alcohol at the interfacial points (b) 
    for the non-coalescent drops for the example shown in Figure \ref{fig:nocoal}.
    The color shows different times, with red, green and blue being 
    $t = 4\, s$, $t = 5\, s$, and $t = 6\, s$, respectively. }
    \label{fig:nocoal_U}
\end{figure}

This quasi-steady behavior is also observed in the experiments by \citet{Karpitschka2012}.  
Figure \ref{fig:nocoal_U}(a) shows the velocity of the points at the interface 
after the quasi-steady state is reached as a function of the distance from the 
bridge region, $D_b$. The points to the left of the bridge region have a velocity
$\approx 2u_d$ (solid line). At the bridge region the interface is close to the solid
substrate and the velocity becomes close to zero due to the no-slip boundary condition. 
In the region close to the bridge in drop $2$, the velocity has a jump and reaches
the maximum value due to the
Marangoni effect resulting from a high surface tension gradient at the neck region. 
Away from the bridge, the velocity is again comparable to $u_d$. This behavior is
in qualitative agreement with the experimental observation by 
\citet{Karpitschka2012,KarpitschkaJFM}.
To provide more insight into the flow through the neck region, in Figure 
\ref{fig:nocoal_U}(b) we present the alcohol concentration at the interface as a 
function of the distance from the bridge region, $D_b$. As shown, a localized and 
steady state surface tension gradient is established through the neck 
region. 
This Marangoni effect can counteract the capillary effect that would otherwise 
result in the coalescence and can therefore sustain the non-coalescence and the 
movement of drops temporarily.

\section{Conclusions}

We have developed a new numerical methodology for including variable surface tension 
in a VOF based Navier-Stokes solver. The method handles both temperature or 
concentration dependent surface tension variations. We employ a height function 
inspired formulation to compute surface gradients and the resulting stresses at 
the interface (Marangoni forces) in a more general numerical framework.
We show the robustness and accuracy of our developed method by studying the convergence 
of the computation of the surface gradient for multiple geometries and the 
convergence of the terminal velocity for the classical problem of the drop migration
with an imposed constant temperature gradient. The drop migration simulation results
are in agreement with the available theoretical and numerical results. We also show
that our method produces results consistent with experimental data in the case of
concentration dependent surface tension. Our numerical implementation extends to
adaptively refined meshes which improves the computational efficiency for Marangoni induced 
flows that require a high resolution around the interface.   

The presented approach represents a first attempt for implementing a general 
variable surface tension in the VOF method.  As presented here, our method can 
subsequently be used directly for surface tension 
dependence on the surfactant concentration. This includes implementing the solution 
to the surfactant transport equation for soluble and insoluble surfactants.  
Our methodology can provide tools for developing more robust and accurate numerical
simulations for two-phase flows with surfactants.  Surfactant flows have many applications, 
e.g.\ in chemical industry, pharmaceuticals  and technology \citep{rosen2012surfactants}, and their 
understanding will have  far reaching effects in many areas. 

The numerical verifications and validations with available literature demonstrate
the efficiency and applicability of our methodology. Our numerical approach is 
implemented in an adaptive mesh refinement framework, which now makes detailed 
numerical simulations that incorporate the effects of tangential (Marangoni) 
stresses feasible.   This is particularly relevant for a number of flow problems 
where Marangoni effect may play a crucial role, such as the evolution of thin films 
on nanoscale, where Marangoni effects may result either from concentration 
gradients (mixture of two fluids) or thermal gradients due to internal or external 
sources.  Our future research will continue in this direction.

\section*{Acknowledgements}
This work was partially supported by the NSF grants No. DMS-$1320037$ (S. A.) and No. CBET-$1604351$ (L. K., S. A.).

\appendix
\section{Appendix}
\subsection{Height functions}
\label{sec:HF}

\begin{figure}[t]
 \begin{center}
  \includegraphics[width=0.4\textwidth]{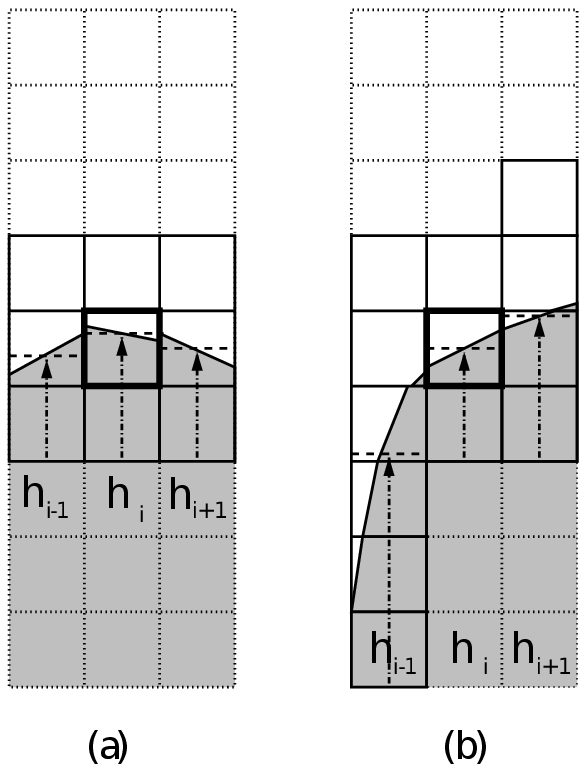}
  \caption{ Height function computation with compact stencil constructed 
  independently for each column. (a) Stencil for a slightly curved interface. 
(b) Asymmetric stencil, where each column requires different number of cells. 
From \citep{Popinet2009a}. }
\label{fig:PopHF}
 \end{center}
\end{figure}
An accurate computation of interface normals and curvature in the VOF method can
be achieved using the height function method \citep{Cummins2005}.
In this method, height functions are defined in each interfacial cell as a sum of the volume 
fractions in fluid column constructed of interfacial cells in vertical or
horizontal orientation. 
Figure \ref{fig:PopHF} shows columns with vertical orientation used for calculating
height functions implemented in {\sc Gerris} solver \citep{Popinet2009a}; the solid 
lines show the stencil size required to calculate the curvature in the cell 
marked by the bold lines. The height function for the the column $i$ is
\begin{equation}
 \label{eq:HF}
 h_i = \sum\limits_{k} \chi_{i,k},
\end{equation}
where the index $k$ includes all interfacial cells in the columns. 
The number of cells used for construction of the columns is optimized based on the 
each column, i.e.\ there is no fixed stencil size, as illustrated in Figure \ref{fig:PopHF}. 
Figure \ref{fig:PopHF}(a) shows a symmetric stencil where only three cells are 
required for calculating height function in each column. 
Figure \ref{fig:PopHF}(b) shows an asymmetric stencil where each column requires 
including different number of cells. 

In 2D, the height function can be calculated in either $x$ or $y$ direction, 
depending on the orientation of the interface. Here, we will present the discretization 
for height function for columns in the $y$ direction (as in Figure \ref{fig:PopHF}). For 
heights collected in the $x$ direction the equations are equivalent with derivatives 
of height function with respect to $x$ replaced by the derivatives with 
respect to $y$. The curvature of the interface is calculated from the height 
functions as
\begin{equation}
 \label{eq:kappaHF}
 \kappa = \frac{h_{xx}}{\left(1 + h_x^2 \right)^{3/2}},
\end{equation}
where the derivatives of the height functions in equation \eqref{eq:kappaHF} 
are calculated using a second order central difference
\begin{gather}
\label{eq:HFderivatives}
 h_x = \frac{h_{i+1} - h_{i-1}}{2\Delta}, \\
 h_{xx} = \frac{h_{i+1} - 2 h_{i} + h_{i-1}}{\Delta^2},
\end{gather}
where $\Delta$ is the cell size. 
In 3D, e.g.\ if the columns are computed in $z$ direction, the curvature is 
\begin{equation}
\label{eq:kappaHF_3D}
    \kappa = \frac{h_{xx} + h_{yy} + h_{xx} h_y^2 + h_{yy}h_x^2 - 2h_{xy}h_x h_y }{\left(1 + h_x^2 + h_y^2 \right)^{3/2}}.
\end{equation}

\subsection{Algorithm}
\label{sec:alg}

Algorithm \ref{alg:sigma} shows the pseudocode for approximating the interfacial
value of the surface tension $\tilde{\sigma}$ in each column. 

\begin{algorithm}[H]
\For{Each component $c \in \{x, y, z\} \equiv \{0, 1, 2\}$ }{
 \For{All interfacial cells $\mathcal{C}$}{
  Set 
  $V \leftarrow \chi \left(\mathcal{C} \right) $,
  $\tilde{\sigma}^c \leftarrow V \sigma \left(\mathcal{C} \right) $\;
  \For{Each direction $d \in \{2c, 2c+1 \}$}{
    Set $\mathcal{N} \leftarrow$ neighboring cell of $\mathcal{C}$ in direction $d$\;
    \While{$0< \chi \left( \mathcal{N} \right) < 1$}{
      Set 
      $v_d \leftarrow \chi \left(\mathcal{C} \right)$, 
      $\tilde{\sigma}^c \leftarrow \tilde{\sigma}^c + v_d \sigma \left(\mathcal{N} \right) $,
      $V \leftarrow V + v_d $ \;
      Set $\mathcal{N} \leftarrow$ neighboring cell of $\mathcal{N}$ in direction $d$\;
    }
   }
   \eIf{ $\{v_{2c}, v_{2c+1} \} \supset \{ 0, 1 \}$ }{
    Set $\tilde{\sigma}^c  \left(\mathcal{C} \right) \leftarrow \tilde{\sigma}^c/V $\;
   }{
    Return inconsistent $\tilde{\sigma}^c  \left(\mathcal{C} \right)$ \;
   }
  }
 }
 \caption{Defining interfacial values $\tilde{\sigma}$ for each column direction.
 \label{alg:sigma}}
\end{algorithm}

\bibliographystyle{model1-num-names}  
\bibliography{Marangoni_Vof_clean}

\begin{thebibliography}{40}
\expandafter\ifx\csname natexlab\endcsname\relax\def\natexlab#1{#1}\fi
\providecommand{\url}[1]{\texttt{#1}}
\providecommand{\href}[2]{#2}
\providecommand{\path}[1]{#1}
\providecommand{\DOIprefix}{doi:}
\providecommand{\ArXivprefix}{arXiv:}
\providecommand{\URLprefix}{URL: }
\providecommand{\Pubmedprefix}{pmid:}
\providecommand{\doi}[1]{\href{http://dx.doi.org/#1}{\path{#1}}}
\providecommand{\Pubmed}[1]{\href{pmid:#1}{\path{#1}}}
\providecommand{\bibinfo}[2]{#2}
\ifx\xfnm\relax \def\xfnm[#1]{\unskip,\space#1}\fi
\bibitem[{Scriven and Sternling(1960)}]{scriven1960}
\bibinfo{author}{L.~E. Scriven}, \bibinfo{author}{C.~V. Sternling},
\newblock \bibinfo{title}{{The Marangoni effects}},
\newblock \bibinfo{journal}{Nature} \bibinfo{volume}{187}
  (\bibinfo{year}{1960}) \bibinfo{pages}{186--188}.
\bibitem[{Farahi et~al.(2004)Farahi, Passian, Ferrell, and
  Thundat}]{Farahi2004}
\bibinfo{author}{R.~H. Farahi}, \bibinfo{author}{A.~Passian},
  \bibinfo{author}{T.~L. Ferrell}, \bibinfo{author}{T.~Thundat},
\newblock \bibinfo{title}{{Microfluidic manipulation via Marangoni forces}},
\newblock \bibinfo{journal}{Appl. Phys. Lett.} \bibinfo{volume}{85}
  (\bibinfo{year}{2004}) \bibinfo{pages}{4237--4239}.
\bibitem[{Kundan et~al.(2015)Kundan, Plawsky, Wayner, Chao, Sicker, Motil,
  Lorik, Chestney, Eustace, and Zoldak}]{Kundan2015}
\bibinfo{author}{A.~Kundan}, \bibinfo{author}{J.~L. Plawsky},
  \bibinfo{author}{P.~C. Wayner}, \bibinfo{author}{D.~F. Chao},
  \bibinfo{author}{R.~J. Sicker}, \bibinfo{author}{B.~J. Motil},
  \bibinfo{author}{T.~Lorik}, \bibinfo{author}{L.~Chestney},
  \bibinfo{author}{J.~Eustace}, \bibinfo{author}{J.~Zoldak},
\newblock \bibinfo{title}{{Thermocapillary phenomena and performance
  limitations of a wickless heat pipe in microgravity}},
\newblock \bibinfo{journal}{Phys. Rev. Lett.} \bibinfo{volume}{114}
  (\bibinfo{year}{2015}) \bibinfo{pages}{146105--5}.
\bibitem[{Subramanian et~al.(2002)Subramanian, Balasubramaniam, and
  Wozniak}]{subramanian2002}
\bibinfo{author}{R.~S. Subramanian}, \bibinfo{author}{R.~Balasubramaniam},
  \bibinfo{author}{G.~Wozniak},
\newblock \bibinfo{title}{{Fluid mechanics of bubbles and drops}},
\newblock in: \bibinfo{editor}{R.~Monti} (Ed.), \bibinfo{booktitle}{Physics of
  Fluids in Microgravity}, \bibinfo{publisher}{Taylor and Francis},
  \bibinfo{year}{2002}, pp. \bibinfo{pages}{149--177}.
\bibitem[{Trice et~al.(2008)Trice, Thomas, Favazza, Sureshkumar, and
  Kalyanaraman}]{trice_prl08}
\bibinfo{author}{J.~Trice}, \bibinfo{author}{D.~Thomas},
  \bibinfo{author}{C.~Favazza}, \bibinfo{author}{R.~Sureshkumar},
  \bibinfo{author}{R.~Kalyanaraman},
\newblock \bibinfo{title}{{Novel Self-Organization Mechanism in Ultrathin
  Liquid Films: Theory and Experiment}},
\newblock \bibinfo{journal}{Phys. Rev. Lett.} \bibinfo{volume}{101}
  (\bibinfo{year}{2008}) \bibinfo{pages}{017802--4}.
\bibitem[{Dong and Kondic(2016)}]{dong2016}
\bibinfo{author}{N.~Dong}, \bibinfo{author}{L.~Kondic},
\newblock \bibinfo{title}{Instability of nanometric fluid films on a thermally
  conductive substrate},
\newblock \bibinfo{journal}{Phys. Rev. Fluids} \bibinfo{volume}{1}
  (\bibinfo{year}{2016}) \bibinfo{pages}{063901--16}.
\bibitem[{Davis(1987)}]{Davis1987}
\bibinfo{author}{S.~Davis},
\newblock \bibinfo{title}{{Thermocapillary instabilities}},
\newblock \bibinfo{journal}{Ann. Rev. Fluid Mech.} \bibinfo{volume}{19}
  (\bibinfo{year}{1987}) \bibinfo{pages}{403--435}.
\bibitem[{Craster and Matar(2009)}]{cm_rmp09}
\bibinfo{author}{R.~Craster}, \bibinfo{author}{O.~Matar},
\newblock \bibinfo{title}{Dynamics and stability of thin liquid films},
\newblock \bibinfo{journal}{Rev. Mod. Phys.} \bibinfo{volume}{81}
  (\bibinfo{year}{2009}) \bibinfo{pages}{1131--1198}.
\bibitem[{Muradoglu and Tryggvason(2014)}]{Muradoglu2014}
\bibinfo{author}{M.~Muradoglu}, \bibinfo{author}{G.~Tryggvason},
\newblock \bibinfo{title}{{Simulations of soluble surfactants in 3D multiphase
  flow}},
\newblock \bibinfo{journal}{J. Comput. Phys.} \bibinfo{volume}{274}
  (\bibinfo{year}{2014}) \bibinfo{pages}{737--757}.
\bibitem[{Xu et~al.(2006)Xu, Li, Lowengrub, and Zhao}]{Xu2006}
\bibinfo{author}{J.-J. Xu}, \bibinfo{author}{Z.~Li},
  \bibinfo{author}{J.~Lowengrub}, \bibinfo{author}{H.~Zhao},
\newblock \bibinfo{title}{{A level-set method for interfacial flows with
  surfactant}},
\newblock \bibinfo{journal}{J. Comput. Phys.} \bibinfo{volume}{212}
  (\bibinfo{year}{2006}) \bibinfo{pages}{590--616}.
\bibitem[{Teigen et~al.(2011)Teigen, Song, Lowengrub, and Voigt}]{Teigen2011}
\bibinfo{author}{K.~E. Teigen}, \bibinfo{author}{P.~Song},
  \bibinfo{author}{J.~Lowengrub}, \bibinfo{author}{A.~Voigt},
\newblock \bibinfo{title}{A diffuse-interface method for two-phase flows with
  soluble surfactants},
\newblock \bibinfo{journal}{J. Comput. Phys.} \bibinfo{volume}{230}
  (\bibinfo{year}{2011}) \bibinfo{pages}{375 -- 393}.
\bibitem[{Blanchette et~al.(2009)Blanchette, Messio, and Bush}]{Blanchette2009}
\bibinfo{author}{F.~Blanchette}, \bibinfo{author}{L.~Messio},
  \bibinfo{author}{J.~W.~M. Bush},
\newblock \bibinfo{title}{{The influence of surface tension gradients on drop
  coalescence}},
\newblock \bibinfo{journal}{{Phys. Fluids}} \bibinfo{volume}{21}
  (\bibinfo{year}{2009}) \bibinfo{pages}{072107--10}.
\bibitem[{Blanchette and Shapiro(2012)}]{Blanchette2012}
\bibinfo{author}{F.~Blanchette}, \bibinfo{author}{A.~M. Shapiro},
\newblock \bibinfo{title}{{Drops settling in sharp stratification with and
  without Marangoni effects}},
\newblock \bibinfo{journal}{{Phys. Fluids}} \bibinfo{volume}{24}
  (\bibinfo{year}{2012}) \bibinfo{pages}{042104--17}.
\bibitem[{Lai et~al.(2008)Lai, Tseng, and Huang}]{Lai2008}
\bibinfo{author}{M.-C. Lai}, \bibinfo{author}{Y.-H. Tseng},
  \bibinfo{author}{H.~Huang},
\newblock \bibinfo{title}{{An immersed boundary method for interfacial flows
  with insoluble surfactant}},
\newblock \bibinfo{journal}{J. Comput. Phys.} \bibinfo{volume}{227}
  (\bibinfo{year}{2008}) \bibinfo{pages}{7279--7293}.
\bibitem[{Booty and Siegel(2010)}]{Booty2010}
\bibinfo{author}{M.~R. Booty}, \bibinfo{author}{M.~Siegel},
\newblock \bibinfo{title}{{A hybrid numerical method for interfacial fluid flow
  with soluble surfactant}},
\newblock \bibinfo{journal}{J. Comput. Phys.} \bibinfo{volume}{229}
  (\bibinfo{year}{2010}) \bibinfo{pages}{3864--3883}.
\bibitem[{Schranner and Adams(2016)}]{Schranner2016}
\bibinfo{author}{F.~S. Schranner}, \bibinfo{author}{N.~A. Adams},
\newblock \bibinfo{title}{A conservative interface-interaction model with
  insoluble surfactant},
\newblock \bibinfo{journal}{J. Comput. Phys.} \bibinfo{volume}{327}
  (\bibinfo{year}{2016}) \bibinfo{pages}{653--677}.
\bibitem[{Drumright-Clarke and Renardy(2004)}]{Renardy2004}
\bibinfo{author}{M.~A. Drumright-Clarke}, \bibinfo{author}{Y.~Renardy},
\newblock \bibinfo{title}{The effect of insoluble surfactant at dilute
  concentration on drop breakup under shear with inertia},
\newblock \bibinfo{journal}{Phys. Fluids} \bibinfo{volume}{16}
  (\bibinfo{year}{2004}) \bibinfo{pages}{14--21}.
\bibitem[{James and Lowengrub(2004)}]{James2004}
\bibinfo{author}{A.~J. James}, \bibinfo{author}{J.~Lowengrub},
\newblock \bibinfo{title}{{A surfactant-conserving volume-of-fluid method for
  interfacial flows with insoluble surfactant}},
\newblock \bibinfo{journal}{J. Comput. Phys.} \bibinfo{volume}{201}
  (\bibinfo{year}{2004}) \bibinfo{pages}{685--722}.
\bibitem[{Ma and Bothe(2011)}]{Ma2011}
\bibinfo{author}{C.~Ma}, \bibinfo{author}{D.~Bothe},
\newblock \bibinfo{title}{{Direct numerical simulation of thermocapillary flow
  based on the Volume of Fluid method}},
\newblock \bibinfo{journal}{Int. J. Multiphase Flow} \bibinfo{volume}{37}
  (\bibinfo{year}{2011}) \bibinfo{pages}{1045--1058}.
\bibitem[{Francois and Swartz(2010)}]{Francois2010}
\bibinfo{author}{M.~M. Francois}, \bibinfo{author}{B.~K. Swartz},
\newblock \bibinfo{title}{{Interface curvature via volume fractions, heights,
  and mean values on nonuniform rectangular grids}},
\newblock \bibinfo{journal}{J. Comput. Phys.} \bibinfo{volume}{229}
  (\bibinfo{year}{2010}) \bibinfo{pages}{527--540}.
\bibitem[{L\'{o}pez and Hern\'{a}ndez(2010)}]{Lopez2010}
\bibinfo{author}{J.~L\'{o}pez}, \bibinfo{author}{J.~Hern\'{a}ndez},
\newblock \bibinfo{title}{{On reducing interface curvature computation errors
  in the height function technique}},
\newblock \bibinfo{journal}{J. Comput. Phys.} \bibinfo{volume}{229}
  (\bibinfo{year}{2010}) \bibinfo{pages}{4855--4868}.
\bibitem[{Alexeev et~al.(2005)Alexeev, Gambaryan-Roisman, and
  Stephan}]{Alexeev2005}
\bibinfo{author}{A.~Alexeev}, \bibinfo{author}{T.~Gambaryan-Roisman},
  \bibinfo{author}{P.~Stephan},
\newblock \bibinfo{title}{{Marangoni convection and heat transfer in thin
  liquid films on heated walls with topography: Experiments and numerical
  study}},
\newblock \bibinfo{journal}{{Phys. Fluids}} \bibinfo{volume}{17}
  (\bibinfo{year}{2005}) \bibinfo{pages}{062106--13}.
\bibitem[{Popinet(2009)}]{Popinet2009a}
\bibinfo{author}{S.~Popinet},
\newblock \bibinfo{title}{{An accurate adaptive solver for
  surface-tension-driven interfacial flows}},
\newblock \bibinfo{journal}{J. Comput. Phys.} \bibinfo{volume}{228}
  (\bibinfo{year}{2009}) \bibinfo{pages}{5838--5866}.
\bibitem[{Popinet(2003)}]{Popinet2003}
\bibinfo{author}{S.~Popinet},
\newblock \bibinfo{title}{{Gerris: a tree-based adaptive solver for the
  incompressible {E}uler equations in complex geometries}},
\newblock \bibinfo{journal}{J. Comput. Phys.} \bibinfo{volume}{190}
  (\bibinfo{year}{2003}) \bibinfo{pages}{572--600}.
\bibitem[{Brackbill et~al.(1992)Brackbill, Kothe, and Zemach}]{Brackbill1992}
\bibinfo{author}{J.~U. Brackbill}, \bibinfo{author}{D.~B. Kothe},
  \bibinfo{author}{C.~Zemach},
\newblock \bibinfo{title}{{A continuum method for modeling surface tension}},
\newblock \bibinfo{journal}{J. Comput. Phys.} \bibinfo{volume}{100}
  (\bibinfo{year}{1992}) \bibinfo{pages}{335--354}.
\bibitem[{Landau and Lifshitz(1987)}]{landau1987}
\bibinfo{author}{L.~D. Landau}, \bibinfo{author}{E.~M. Lifshitz},
  \bibinfo{title}{{Fluid mechanics, 2nd}}, volume~\bibinfo{volume}{6},
  \bibinfo{publisher}{Pergamon Press, Oxford}, \bibinfo{year}{1987}.
\bibitem[{Levich and Krylov(1969)}]{levich1969}
\bibinfo{author}{V.~G. Levich}, \bibinfo{author}{V.~S. Krylov},
\newblock \bibinfo{title}{{Surface-tension-driven phenomena}},
\newblock \bibinfo{journal}{Annu. Rev. Fluid Mech.} \bibinfo{volume}{1}
  (\bibinfo{year}{1969}) \bibinfo{pages}{293--316}.
\bibitem[{Popinet(1999)}]{Gerris}
\bibinfo{author}{S.~Popinet}, \bibinfo{title}{{Gerris Flow Solver}},
  \bibinfo{year}{1999}. \URLprefix
  \url{http://gfs.sourceforge.net/wiki/index.php}.
\bibitem[{Afkhami and Bussmann(2008)}]{Afkhami2009a}
\bibinfo{author}{S.~Afkhami}, \bibinfo{author}{M.~Bussmann},
\newblock \bibinfo{title}{{Height functions for applying contact angles to 2{D}
  {VOF} simulations}},
\newblock \bibinfo{journal}{Int. J. Numer. Meth. Fluids} \bibinfo{volume}{57}
  (\bibinfo{year}{2008}) \bibinfo{pages}{453--472}.
\bibitem[{Aulisa et~al.(2007)Aulisa, Manservisi, Scardovelli, and
  Zaleski}]{aulisa2007}
\bibinfo{author}{E.~Aulisa}, \bibinfo{author}{S.~Manservisi},
  \bibinfo{author}{R.~Scardovelli}, \bibinfo{author}{S.~Zaleski},
\newblock \bibinfo{title}{{Interface reconstruction with least-squares fit and
  split advection in three-dimensional Cartesian geometry}},
\newblock \bibinfo{journal}{J. Comput. Phys.} \bibinfo{volume}{225}
  (\bibinfo{year}{2007}) \bibinfo{pages}{2301--2319}.
\bibitem[{Wozniak et~al.(1988)Wozniak, Siekmann, and Srulijes}]{wozniak1988}
\bibinfo{author}{G.~Wozniak}, \bibinfo{author}{J.~Siekmann},
  \bibinfo{author}{J.~Srulijes},
\newblock \bibinfo{title}{{Thermocapillary bubble and drop dynamics under
  reduced gravity-survey and prospects}},
\newblock \bibinfo{journal}{Zeitschrift f{\"u}r Flugwissenschaften und
  Weltraumforschung} \bibinfo{volume}{12} (\bibinfo{year}{1988})
  \bibinfo{pages}{137--144}.
\bibitem[{Nas and Tryggvason(2003)}]{nas2003}
\bibinfo{author}{S.~Nas}, \bibinfo{author}{G.~Tryggvason},
\newblock \bibinfo{title}{{Thermocapillary interaction of two bubbles or
  drops}},
\newblock \bibinfo{journal}{Int. J. Multiphase Flow} \bibinfo{volume}{29}
  (\bibinfo{year}{2003}) \bibinfo{pages}{1117--1135}.
\bibitem[{Herrmann et~al.(2008)Herrmann, Lopez, Brady, and
  Raessi}]{herrmann2008}
\bibinfo{author}{M.~Herrmann}, \bibinfo{author}{J.~M. Lopez},
  \bibinfo{author}{P.~Brady}, \bibinfo{author}{M.~Raessi},
\newblock \bibinfo{title}{{Thermocapillary motion of deformable drops and
  bubbles}},
\newblock in: \bibinfo{booktitle}{Proceedings of the Summer program},
  \bibinfo{address}{Stanford University, Center for Turbulence Research},
  \bibinfo{year}{2008}, pp. \bibinfo{pages}{155--170}.
\bibitem[{Young et~al.(1959)Young, Goldstein, and Block}]{Young1959}
\bibinfo{author}{N.~O. Young}, \bibinfo{author}{J.~S. Goldstein},
  \bibinfo{author}{M.~J. Block},
\newblock \bibinfo{title}{The motion of bubbles in a vertical temperature
  gradient},
\newblock \bibinfo{journal}{{J. Fluid Mech.}} \bibinfo{volume}{6}
  (\bibinfo{year}{1959}) \bibinfo{pages}{350--356}.
\bibitem[{Karpitschka and Riegler(2010)}]{Karpitschka2010}
\bibinfo{author}{S.~Karpitschka}, \bibinfo{author}{H.~Riegler},
\newblock \bibinfo{title}{{Quantitative experimental study on the transition
  between fast and delayed coalescence of sessile droplets with different but
  completely miscible liquids}},
\newblock \bibinfo{journal}{Langmuir} \bibinfo{volume}{26}
  (\bibinfo{year}{2010}) \bibinfo{pages}{11823--11829}.
\bibitem[{Karpitschka and Riegler(2012)}]{Karpitschka2012}
\bibinfo{author}{S.~Karpitschka}, \bibinfo{author}{H.~Riegler},
\newblock \bibinfo{title}{{Noncoalescence of sessile drops from different but
  miscible liquids: Hydrodynamic analysis of the twin drop contour as a
  self-stabilizing traveling wave}},
\newblock \bibinfo{journal}{Phys. Rev. Lett.} \bibinfo{volume}{109}
  (\bibinfo{year}{2012}) \bibinfo{pages}{066103--5}.
\bibitem[{Karpitschka and Riegler(2014)}]{KarpitschkaJFM}
\bibinfo{author}{S.~Karpitschka}, \bibinfo{author}{H.~Riegler},
\newblock \bibinfo{title}{{Sharp transition between coalescence and
  non-coalescence of sessile drops}},
\newblock \bibinfo{journal}{{J. Fluid Mech.}} \bibinfo{volume}{743}
  (\bibinfo{year}{2014}) \bibinfo{pages}{R1}.
\bibitem[{Afkhami et~al.(2009)Afkhami, Zaleski, and Bussmann}]{Afkhami2009}
\bibinfo{author}{S.~Afkhami}, \bibinfo{author}{S.~Zaleski},
  \bibinfo{author}{M.~Bussmann},
\newblock \bibinfo{title}{{A Mesh-Dependent Model for Applying Dynamic Contact
  Angles to {VOF} Simulations}},
\newblock \bibinfo{journal}{J. Comput. Phys.} \bibinfo{volume}{228}
  (\bibinfo{year}{2009}) \bibinfo{pages}{5370--5389}.
\bibitem[{Rosen and Kunjappu(2012)}]{rosen2012surfactants}
\bibinfo{author}{M.~J. Rosen}, \bibinfo{author}{J.~T. Kunjappu},
  \bibinfo{title}{Surfactants and interfacial phenomena},
  \bibinfo{publisher}{John Wiley \& Sons}, \bibinfo{year}{2012}.
\bibitem[{Cummins et~al.(2005)Cummins, Francois, and Kothe}]{Cummins2005}
\bibinfo{author}{S.~J. Cummins}, \bibinfo{author}{M.~M. Francois},
  \bibinfo{author}{D.~B. Kothe},
\newblock \bibinfo{title}{{Estimating curvature from volume fractions}},
\newblock \bibinfo{journal}{Comput. Struct.} \bibinfo{volume}{83}
  (\bibinfo{year}{2005}) \bibinfo{pages}{425--434}.

\end{thebibliography}

\end{document}